\documentclass[12pt]{article}
\pdfoutput = 1
\usepackage{cite}
\usepackage{array}
\usepackage{epsfig}
\usepackage{amssymb}
\usepackage{graphics,graphpap}
\usepackage{amssymb}
\usepackage{amsmath}
\usepackage{slashed}
\usepackage{dsfont}
\newcommand{\model}{\overline{331}}

\def\eps{\varepsilon}
\def\epe{\varepsilon'/\varepsilon}

\newcommand{\tev}{\, {\rm TeV}}
\newcommand{\gev}{\, {\rm GeV}}
\newcommand{\mev}{\, {\rm MeV}}

\newcommand{\Heff}{{\cal H}_\text{ eff}}

\newcommand{\be}{\begin{equation}}
\newcommand{\ee}{\end{equation}}
\newcommand{\bea}{\begin{eqnarray}}
\newcommand{\eea}{\end{eqnarray}}

\newcommand{\bi}{\begin{itemize}}
\newcommand{\ei}{\end{itemize}}
\newcommand{\ord}{{\cal O}}

\newcommand{\vcb}{|V_{cb}|}
\newcommand{\vtd}{|V_{td}|}
\newcommand{\vub}{|V_{ub}|}
\newcommand{\vts}{|V_{ts}|}
\newcommand{\vus}{|V_{us}|}

\def\kpn{K^+\rightarrow\pi^+\nu\bar\nu}

\def\klpn{K_{L}\rightarrow\pi^0\nu\bar\nu}

\usepackage{graphicx}

\medskipamount % besser als explizite Angabe in pt; etwas Platz nach neuem Absatz

\usepackage{fancyhdr}
\advance \headheight by 3.0truept       % for 12pt mandatory...
\newlength{\textlength}
\newlength{\overlinelength}

 \def\s#1{\setbox0=\hbox{$#1$}%
   \rlap{\ifdim\wd0>.7em\kern.22\wd0\else\kern.1\wd0\fi /}#1}


\begin{document}

%%%%%%%%%% Title page
\begin{titlepage}
\begin{flushright}
%\begin{tabular}{l}
{FLAVOUR(267104)-ERC-38}\\
 {BARI-TH/13-671}\\
Nikhef-2013-008\\
UT-13-09
%\end{tabular}
\end{flushright}
\vskip0.7cm
\begin{center}
{\Large \bf \boldmath The Anatomy of Neutral Scalars
with FCNCs in the Flavour Precision Era}
\vskip0.5cm
{\bf Andrzej~J.~Buras$^{a,b}$, Fulvia~De~Fazio$^{c}$,
    Jennifer Girrbach$^{a,b}$ \\  Robert Knegjens$^{d}$ and Minoru Nagai$^{e}$
 \\[0.4 cm]}
{\small
$^a$TUM Institute for Advanced Study, Lichtenbergstr. 2a, D-85747 Garching, Germany\\
$^b$Physik Department, TUM, James-Franck-Stra{\ss}e, D-85747 Garching, Germany\\
$^c$Istituto Nazionale di Fisica Nucleare, Sezione di Bari, Via Orabona 4,
I-70126 Bari, Italy\\
$^d$ Nikhef, Science Park 105, NL-1098 XG Amsterdam, The Netherlands\\
$^e$ Department of Physics, University of Tokyo, Tokyo 113-0033, Japan
}
\vskip0.71cm

%{\em Version of \today}

%\vskip0.1cm

{\large\bf Abstract\\[10pt]} \parbox[t]{\textwidth}{\small
In many extensions of the Standard Model (SM) flavour changing neutral
current (FCNC) processes can be mediated by tree-level
heavy neutral scalars and/or
 pseudo-scalars $H^0(A^0)$. This
generally introduces new sources of flavour violation and CP violation
as well as left-handed (LH) and  right-handed (RH) {\it scalar}
($1\mp\gamma_5$)
currents.
These new physics (NP) contributions imply
a pattern of deviations from SM expectations for FCNC processes that depends
only on the couplings of $H^0(A^0)$ to fermions  and on their  masses. In
situations in which a single $H^0$ or $A^0$ dominates NP contributions
stringent correlations between $\Delta F=2$ and $\Delta F=1$ observables exist.
 Anticipating the Flavour Precision Era (FPE) ahead
of us we illustrate this by searching for allowed oases in the
landscape of a given model assuming significantly smaller uncertainties in CKM and hadronic
parameters than presently available. To this end we  analyze
$\Delta F=2$ observables in
$B^0_{s,d}-\bar B^0_{s,d}$ and $K^0-\bar K^0$ systems and rare $B$ and $K$ decays with charged leptons in the final state including
both left-handed and right-handed scalar couplings of $H^0$ and $A^0$
to quarks in various
combinations.  
We identify a number of correlations
between various flavour observables that could test and distinguish these different scenarios. The prominent
role of the decays $B_{s,d}\to \mu^+\mu^-$ in these
studies is emphasized. Imposing the existing flavour constraints, a rich
pattern of deviations from the SM expectations in rare $B_{s,d}$ decays emerges provided $M_{H}\le 1\tev$. NP effects in rare $K$ decays,
except for  $K_L \to\mu^+\mu^-$, turn out 
 to be very small. In $K_L \to\mu^+\mu^-$ they can be as large as the SM contributions but due to hadronic uncertainties this is still
insufficient to 
learn much about new scalars from this decay in the context of models considered here.
Flavour violating SM Higgs contributions to rare $B_d$ and $K$ decays turn out to be
negligible once the constraints from $\Delta F=2$ processes are taken into
account. But $\mathcal{B}(B_s\to\mu^+\mu^-)$  can still be enhanced 
up to $8\%$. Finally, we point out striking differences between the correlations found here and in  scenarios in which tree-level 
FCNC are mediated by a new neutral gauge boson $Z'$.
}

\vfill
\end{center}
\end{titlepage}

\setcounter{footnote}{0}

\clearpage
\pagestyle{empty}
\tableofcontents
\clearpage
\pagestyle{fancyplain}
\newpage

\section{Introduction}
\label{sec:1}
The recent discovery of a scalar particle with a mass of $126\gev$ opened
the gate to the unexplored world of scalar particles which could be
elementary or composite. While we will surely learn a lot about the properties
of these new objects through collider experiments like ATLAS and CMS,
also low energy processes, in particular flavour violating transitions, will teach us
about their nature. In the Standard Model (SM) and in many of its extensions
there are no fundamental flavour-violating couplings of scalars\footnote{Unless otherwise specified we will use the name {\it scalar} for
both scalars and pseudo-scalars.} to quarks and leptons but such couplings can be generated through loop corrections leading
in the case of $\Delta F=1$ transitions to Higgs-Penguins (HP) and in $\Delta F=2$ transitions
to double Higgs-Penguins (DHP). However, when the masses of the scalar particles are significantly lower than the heavy new particles exchanged in the loops, the HP and DHP look at the electroweak scale as flavour violating tree diagrams.
Beyond the SM such diagrams can also be present at the fundamental level, an important example being the left-right symmetric models. From the point of view of low
energy theory there is no distinction between these possibilities as long as
the vertices involving heavy particles in a Higgs-Penguin cannot be resolved
and to first approximation what really matters is the mass of the exchanged
scalar and its flavour violating couplings, either fundamental or generated
at one-loop level. While all this can be formulated with the help of
effective field theories and spurion technology, we find it more transparent
to study directly tree diagrams with heavy particle exchanges.

In a recent paper \cite{Buras:2012jb}  an anatomy of neutral gauge boson ($Z'$ and $Z$) couplings to quark
flavour changing neutral currents (FCNC) has been presented.
 Anticipating the Flavour Precision Era (FPE) ahead
of us  and consequently assuming significantly smaller
uncertainties in CKM and hadronic
parameters than presently available, it was possible to find
 allowed oases in the landscape of new parameters in these models and
to uncover stringent correlations between $\Delta F=2$ and $\Delta F=1$ observables characteristic for such NP scenarios.

The goal of the present paper is to perform a similar analysis for scalar neutral
particles and to investigate whether the patterns of flavour violation in
these two different NP scenarios (gauge bosons and scalars) can be
distinguished through correlations between quark flavour observables. Already
at this stage it is useful to note the following differences in NP
contributions to quark flavour observables in these two scenarios:
\begin{itemize}
\item
While the lower bounds on masses of $Z'$ gauge bosons from collider
experiments are at least $1-2\tev$, new neutral scalars with masses as low as a few
hundred $\gev$ are not excluded.
\item
While in the $Z'$ scenarios in addition to new operators also SM operators
with modified Wilson coefficients can be present, in the case of tree-level
scalar exchanges all effective low energy operators are new.
\item
While there is some overlap between operators contributing to $\Delta F=2$
processes in $Z'$ and scalar cases after the inclusion of QCD corrections, their Wilson coefficients
are very different. Moreover, in $\Delta F=1$ transitions there is no
overlap with the operators present in $Z'$ models.
\item
Concerning flavour violating couplings of $Z$ and the SM Higgs $h$, in the
case of the $Z$ boson large NP effects, in particular in rare $K$ decays, are still allowed
but then its effects in $\Delta F=2$ processes turn out to be very small
\cite{Buras:2012jb}. { In the Higgs case, the smallness of the Higgs coupling
to muons and electrons precludes any visible effects from tree-level Higgs exchanges in rare $K$ and $B_d$ decays with muon or electron pair in the final state
once constraints from $\Delta F=2$ processes are taken into account. 
The corresponding effects in $B_s\to\mu^+\mu^-$ are small but can still be at 
the level of $8\%$.}
Simultaneously tree-level Higgs contributions to $\Delta F=2$ transitions can still
provide in principle solutions to possible tensions within the SM.

\item
At first sight the couplings of scalars to neutrinos look totally negligible
but if the masses of neutrinos are generated by a different mechanism than
coupling to scalars, like in the case of the see-saw mechanism, it is
not a priori obvious that such couplings in some NP scenarios could be
 measurable. Our working assumption in the present paper will be that this is
not the case. Consequently NP
effects of scalars in $\kpn$, $\klpn$ and $b\to s\nu\bar\nu$ transitions
will be assumed to be negligible in contrast to
$Z'$ models, where NP effects in these decays could be very important \cite{Buras:2012jb}. As we will see, scalar contributions to  
$K_L\to\pi^0\ell^+\ell^-$ although in principle larger than for 
 $\kpn$, $\klpn$ and $b\to s\nu\bar\nu$ transitions, are found to be small.
In $K_L \to\mu^+\mu^-$ they can be as large as the SM contribution but due to hadronic uncertainties this is still insufficient to 
learn much about scalars from this decay, at least in the context of models 
considered by us.
\end{itemize}

In order to have an easy comparison with the anatomy of FCNCs mediated by a
neutral gauge
bosons presented in \cite{Buras:2012jb} the structure of the present work
will be similar to the structure of the latter paper but not identical, as
rare $K$ decays play in this paper a subleading role so that emphasis
will be put on $B_s$ and $B_d$ systems.
In Section~\ref{sec:2} we describe
 our strategy by defining the relevant couplings and
listing processes to be considered. Our analysis
 will only involve processes which are theoretically clean and have
 simple structure. Here we will also introduce a number of
 different scenarios for the scalar couplings to quarks thereby
reducing the number
 of free parameters.
In Section~\ref{DF2} we will first present a compendium of formulae relevant
for the study of $\Delta F=2$ processes mediated by tree-level neutral scalar
exchanges
including for
the first time NLO QCD corrections to these NP contributions.
In Section~\ref{RareB} we discuss rare $B$ decays, in particular
$B_{s,d}\to\mu^+\mu^-$.  In Section~\ref{RareK}
 rare $K$
decays are
considered. In Section~\ref{sec:3a}
we present a general qualitative view on NP contributions to flavour
observables stressing analytic correlations between $\Delta F=2$ and
$\Delta F=1$ observables.
In Section~\ref{sec:4} we present our strategy for the numerical analysis
and in Section~\ref{sec:Excursion} we execute our strategy for the determination of scalar
couplings in the $B_s$ and $B_d$ systems. We discuss several scenarios for them
and identify stringent correlations between various observables.
We also investigate what the imposition of the $U(2)^3$ flavour
symmetry on scalar couplings would imply.
In Section~\ref{sec:U(2)} we present the results for rare $K$ decays, where
NP effects are found to be small. { In Section~\ref{sec:ZSM} we demonstrate that the contributions of
the SM Higgs with induced flavour violating couplings, even if in principle relevant for $\Delta F=2$ transitions, are irrelevant for rare $K$ and $B_{d}$ decays with small but still visible effects in $B_s\to\mu^+\mu^-$.}  A summary of our main results and a brief outlook for the future
 are given in  Section~\ref{sec:5}.

\section{Strategy}
\label{sec:2}
\subsection{Basic Model Assumptions}

Our paper is dominated
by tree-level contributions to FCNC processes mediated
by a heavy neutral scalar or pseudoscalar.  We use a common
name, $H^0$, for them unless otherwise specified. When a distinction will have
to be made, we will either use $H^0$ and $A^0$ for scalar and pseudoscalar, respectively or in order to distinguish SM Higgs from additional
spin 0 particles we
will use the familiar 2HDM and MSSM notation: $(H,A,h)$.

Our main goal is to consider the simplest extension of the SM in which the only
new particle in the low energy effective theory
is a single neutral particle with spin $0$ and the question arises whether this is
possible from the point of view of an underlying original theory. If the
scalar in question is not a $SU(2)$ singlet, then it must be placed in a complete
$SU(2)$ multiplet, e.g. a second doublet as is the case of 2HDM or the MSSM. However,
this implies the existence of its $SU(2)$ partners in a given multiplet with
masses close to the masses of our scalar. In fact in the decoupling regime
in 2HDM and MSSM the masses of $(H^\pm,H^0,A^0)$ are approximately degenerate.
While $SU(2)$ breaking effects in the Higgs potential allow for mass splittings, they must be of $\ord(v)$ at most and consequently the case of
the dominance of a single scalar is rather unlikely.

It follows then that our scalar should be a $SU(2)$ singlet. In this case,
the scalar-quark couplings of $H$ come from the following low energy effective
operator
\be
L=\lambda_L^{ij}\frac{H^0}{\Lambda}\bar q_R^i q_L^j h_{\rm SM}+h.c.
\ee
with $\Lambda$ denoting the cut-off scale of the low energy theory. After the
spontaneous breakdown of $SU(2)$ the scalar left-handed coupling is given by
\be
\Delta_L^{ij}(H^0)=\frac{1}{\sqrt{2}}\frac{v}{\Lambda}\lambda_L^{ij},
\ee
with analogous expression for the right-handed coupling.

\begin{figure}[!htb]
\includegraphics[width = 0.6\textwidth]{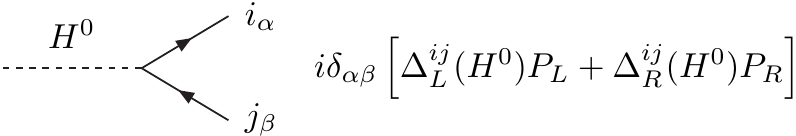}
\caption{\it Feynman rules for  a  neutral colourless scalar particle $H^0$
with mass $M_H$, where $i,\,j$ denote different
quark flavours and $\alpha,\,\beta$ the colours. $P_{L,R}=(1\mp\gamma_5)/2$.}\label{neutral}
\end{figure}

Now, in the case of $\Delta F=2$
transitions
the scalar  contributions are governed only by the couplings $\Delta_{L,R}^{ij}(H^0)$
to quarks and the corresponding Feynman rule has
been shown in Fig~\ref{neutral}.
Here $(i,j)$ denote quark flavours. Note the following important property
\be\label{basic}
\Delta_L^{ij}(H^0)= [\Delta_R^{ji}(H^0)]^*
\ee
that distinguishes it from the corresponding gauge couplings in which there
is no chirality flip.

The couplings $\Delta_{L,R}^{ij}(H^0)$ are dimensionless quantities but
as these are scalar and not gauge couplings they can involve ratios
of quark masses and the electroweak vacuum expectation value $v$ or other
mass scales. While from the SM, 2HDM and MSSM we are used to having scalar couplings
proportional to the masses of the participating quarks, it should be emphasized
that this is not a general property. It applies only if the scalar and the SM
Higgs, responsible for $SU(2)$ breakdown, are in the same $SU(2)$ multiplet
or a multiplet of a larger gauge group $G$. Then after the breakdown of $G$
to $SU(2)$, the scalar appears as a singlet of $SU(2)$ symmetry, with
couplings to quarks involving their masses after $SU(2)$ breakdown. While
this is the case in several models, in our simple extension of the SM, it is more natural to think that the involved scalar couplings are unrelated to the
generation of quark masses.

In spite of the last statement 
is useful to recall how the quark masses could enter the scalar couplings.
 Which
quark masses are involved depends on the model. Considering for definiteness the
$B_s$ system let us just list a few cases encountered in the literature:
\begin{itemize}
\item
In models with MFV in which the scalar couplings are just Yukawa couplings
one has
\be\label{basic1}
\Delta_L^{bs}(H^0)\propto \frac{m_b}{v}, \qquad \Delta_R^{bs}(H^0)\propto \frac{m_s}{v} \qquad {\rm (MFV)}
\ee
implying that $\Delta_L^{bs}$ dominate in these scenarios. Note however
that using (\ref{basic}) these relations also give
\be\label{basic2}
\Delta_L^{sb}(H^0)\propto \frac{m_s}{v}, \qquad \Delta_R^{sb}(H^0)\propto \frac{m_b}{v} \qquad {\rm (MFV)}
\ee
which implies  some care when stating whether LH or RH scalar couplings are dominant.
 Below we will use the ordering $bq$ for
$\Delta F=2$ operators in $B_q$ $(q=s,d)$  systems while $qb$ in the case of rare
$B_q$ decays. In the $K$ system $sd$ will be used for both $\Delta F=2$ and
$\Delta F=1$ couplings.
\item
In non-MFV scenarios the mass dependence in scalar couplings can be
 reversed
\be\label{basic3}
\Delta_L^{bs}(H^0)\propto \frac{m_s}{v}, \qquad \Delta_R^{bs}(H^0)\propto \frac{m_b}{v} \qquad {\rm (non-MFV)}
\ee
implying that $\Delta_R^{bs}$ dominate in these scenarios. Correspondingly
(\ref{basic2}) is changed to
\be\label{basic4}
\Delta_L^{sb}(H^0)\propto \frac{m_b}{v}, \qquad \Delta_R^{sb}(H^0)\propto \frac{m_s}{v} \qquad {\rm (non-MFV)}
\ee
promoting the so-called primed operators in $\Delta F=1$ decays.
\item
There exist also models in which flavour violating  neutral scalar couplings
do not involve the masses of external quarks. This is the case for the neutral
heavy Higgs in the left-right symmetric models analysed in
\cite{Blanke:2011ry} where the scalar down quark couplings are proportional
to up-quark masses, in particular $m_t/v$. In the case of a manifest left-right
symmetry with the right-handed mixing matrix being equal to the CKM matrix
one finds
\be\label{basic5}
\Delta_L^{bs}(H^0)=\Delta_R^{bs}(H^0).
\ee
Even if in the concrete model analysed in \cite{Blanke:2011ry} the
right-handed mixing matrix equal  to the CKM matrix is ruled out by the data, there
could be other model constructions in which (\ref{basic5}) could be satisfied.
Also the LH and RH couplings differing by sign could in principle be possible.
\end{itemize}

\subsection{Scenarios for Scalar Couplings}
In order to take these different possibilities into account and having
also in mind that scalar couplings could be independent of quark masses,
we  consider the following four scenarios for their couplings to quarks
keeping the pair $(i,j)$ fixed:
\begin{enumerate}
\item
Left-handed Scenario (LHS) with complex $\Delta_L^{bq}\not=0$  and $\Delta_R^{bq}=0$,
\item
Right-handed Scenario (RHS) with complex $\Delta_R^{bq}\not=0$  and $\Delta_L^{bq}=0$,
\item
Left-Right symmetric Scenario (LRS) with complex
$\Delta_L^{bq}=\Delta_R^{bq}\not=0$,
\item
Left-Right asymmetric Scenario (ALRS) with complex
$\Delta_L^{bq}=-\Delta_R^{bq}\not=0$,
\end{enumerate}
with analogous scenarios for the pair $(s,d)$. For rare $B_q$ decays in which
the ordering $qb$ is used, the rule (\ref{basic}) has to be applied to each
scenario. For $K$ physics this is not required. In the course of our
paper we will list specific examples of models that share the properties
of these different scenarios.
We will see that these simple cases will give us a profound insight
into the flavour structure of models in which NP is dominated by left-handed
scalar
currents or right-handed scalar currents or left-handed and right-handed
scalar currents
of the same size.  We will also consider a model in which both a scalar and a pseudoscalar with  approximately the same mass couple equally
to quarks and leptons.
Moreover we will study a scenario with underlying
flavour $U(2)^3$ symmetry
which will imply relations between $\Delta_L^{bd}$  and $\Delta_L^{bs}$
couplings and interesting phenomenological consequences.

The idea of looking at NP scenarios  with the dominance of certain quark couplings to neutral gauge bosons or neutral scalars
is not new and has been motivated by detailed studies
in concrete models like supersymmetric flavour models
\cite{Altmannshofer:2009ne}, LHT model with T-parity \cite{Blanke:2006eb,Blanke:2009am} or Randall-Sundrum scenario with custodial protection (RSc)
\cite{Blanke:2008yr}. See also \cite{Altmannshofer:2011gn,Altmannshofer:2012ir}. Also our recent analysis of tree-level FCNCs mediated by $Z'$ and $Z$
in \cite{Buras:2012jb}
demonstrates this type of NP in
a transparent manner.

\subsection{Scalar vs Pseudoscalar}
It will turn out to be useful to exhibit the differences between the
scalar and pseudoscalar spin 0 particles, although one should emphasize
that in the presence of CP violation, the mass eigenstate $H^0$  propagating in  a tree-diagram is not necessarily a CP eigenstate. Therefore, generally
the coupling to $\mu^+\mu^-$ appearing at many places in our paper can have
the general structure
\be
L=\frac{1}{2}\bar\mu (\Delta_S^{\mu\bar\mu}(H^0)+\gamma_5\Delta_P^{\mu\bar\mu}(H^0))H\mu
\ee
where generalizing the Feynman rule in Fig.~\ref{neutral} to charged lepton couplings we have introduced:
\begin{align}\begin{split}\label{equ:mumuSPLR}
 &\Delta_S^{\mu\bar\mu}(H)= \Delta_R^{\mu\bar\mu}(H)+\Delta_L^{\mu\bar\mu}(H),\\
&\Delta_P^{\mu\bar\mu}(H)= \Delta_R^{\mu\bar\mu}(H)-\Delta_L^{\mu\bar\mu}(H).\end{split}
\end{align}
$\Delta^{\mu\bar\mu}_{S}$ is real and $\Delta^{\mu\bar\mu}_{P}$ purely
imaginary as required by the hermiticity of the Hamiltonian which can be
verified by means of (\ref{basic}).

The expressions for various observables will be first given in terms of the
couplings $\Delta_{L,R}^{ij}(H)$ and $\Delta_{S,P}^{\mu\bar\mu}(H)$ and can be
directly used in the case of the scalar particle being CP-even eigenstate, like
$(H^0,h)$ in 2HDM or MSSM setting  $\Delta_{P}^{\mu\bar\mu}(H)=0$. However,
when the mass eigenstate is a pseudoscalar $A$, implying $\Delta_S^{\mu\bar\mu} = 0$, it will be useful to exhibit
the {\it i} which we illustrate here for the $B_s^0$ system:
\be\label{PSEUDO}
 \Delta_L^{bs}(A)= -i  \tilde\Delta_L^{bs}(A), \qquad
                \Delta_R^{bs}(A)= +i  \tilde\Delta_R^{bs}(A), \qquad
\Delta_P^{\mu\bar\mu}(A)=i\tilde\Delta_P^{\mu\bar\mu}(A).
\ee
Here the flavour violating couplings $\tilde\Delta_{L,R}^{bs}(A)$ are still
complex, while $\tilde\Delta_P^{\mu\bar\mu}(A)$ is real.

The following useful relations follow from (\ref{basic}) and (\ref{PSEUDO}):
\be
\Delta_R^{sb}(A)=i[\tilde\Delta^{bs}_L(A)]^*,\qquad
\Delta_L^{sb}(A)=-i[\tilde\Delta^{bs}_R(A)]^*.
\ee

As far as $\Delta F=2$ transitions are concerned this distinction between
  scalar and pseudoscalar mass eigenstate is only relevant in a concrete
model in which the relevant couplings are given in terms of fundamental
parameters. However, as in our numerical analysis we will treat the flavour
violating quark-scalar
couplings as arbitrary complex numbers to be bounded by $\Delta F=2$
observables it will not be possible
to distinguish a scalar and pseudoscalar boson on the basis of $\Delta F=2$
transitions alone. On the other hand, when rare decays, in particular $B_{s,d}\to\mu^+\mu^-$, are considered there is a difference  between these two cases as
the pseudoscalar contributions interfere with SM contribution, while the
scalar ones do not. Consequently the allowed values for $\tilde\Delta_P^{\mu\bar\mu}(A)$ and $\Delta_S^{\mu\bar\mu}(A)$ will differ from each other and we will
find other differences. Finally, if both scalar and pseudoscalar
contribute to tree-level decays and have approximately the same mass as
well as couplings related by symmetries, also their contributions to
$\Delta F=2$ processes differ. We will consider a simple example in the
course of our presentation.

\subsection{Steps}
Let us then outline our strategy for the determination of flavour violating $H$
couplings to quarks and for finding correlations between flavour observables in
the context of the simple scenarios listed above.
Our strategy will only be fully effective in the second half of this
decade, when hadronic uncertainties will be reduced and the data on
various observables significantly improved. It involves ten steps including
a number of working assumptions:

{\bf Step 1:}

Determination of CKM parameters by means of tree-level decays
and of the necessary non-perturbative parameters by means of lattice calculations. This step will provide the results for all observables considered below
within the SM as well as all non-perturbative parameters entering the NP
contributions. As $\vub$ is presently poorly known, it will be interesting in
the spirit of our recent papers \cite{Buras:2012jb,Buras:2012sd,Blanke:2011ry}
to investigate how the outcome of this step
depends on the value of $\vub$ with direct implications for the necessary
size of NP contributions which will be different in different observables.

{\bf Step 2:}

We will assume that the ratios
\be\label{leptonic}
\frac{\Delta_{S,P}^{\mu\bar\mu}(H)}{ M_{H}}
\ee
for scalar and pseudoscalar bosons
have been determined in pure leptonic processes and that the scalar couplings
to neutrinos are negligible. The properties of these couplings have
been discussed above.
In principle these ratios
can be determined up to the sign from quark flavour violating processes and
in fact we will be able to bound them from the present data
on $B_s\to\mu^+\mu^-$
but
their independent knowledge increases predictive power of our analysis. In particular
the knowledge of their signs allows us to remove certain discrete ambiguities
and is crucial for the distinction between LHS and RHS scenarios in
$B_{s,d}\to\mu^+\mu^-$ decays. Of course, in concrete models like 2HDM or
supersymmetric models these couplings depend on the fundamental
parameters of a given model.

{\bf Step 3:}

Here we will consider the $B_s^0$ system and the observables
 \be\label{Step3}
\Delta M_s, \quad S_{\psi \phi}, \quad \mathcal{B}(B_s\to\mu^+\mu^-),\quad
\mathcal{A}_{\Delta\Gamma}^{\mu^+\mu^-},\quad
S^s_{\mu^+\mu^-},
\ee
where
$\mathcal{A}_{\Delta\Gamma}^{\mu^+\mu^-}$ and
$S^s_{\mu^+\mu^-}$ can be extracted from the time-dependent $B_s\to\mu^+\mu^-$ rate~\cite{deBruyn:2012wj,deBruyn:2012wk}.
Explicit expressions for these observables in terms of the
relevant couplings can be found in  Sections~\ref{sec:3} and \ref{RareB}.

Concentrating in this step on the LHS scenario, NP contributions to these
three observables are fully described by
\be\label{Step3a}
\frac{\Delta_L^{bs}(H)}{M_{H}}=-\frac{\tilde s_{23}}{M_{H}}
e^{-i\delta_{23}},\quad \frac{\Delta_{S,P}^{\mu\bar\mu}(H)}{ M_{H}},
\ee
with the second ratio known from Step 2. Here $\tilde s_{23}\ge 0$
and it is found to be below unity but it does not represent any mixing
parameter as in \cite{Buras:2012dp}. The minus sign is introduced
to cancel the minus sign in $V_{ts}$ in the phenomenological formulae listed
in the
next section.

Thus we have five observables
to our disposal and two parameters in the quark sector to determine. This
allows to remove certain discrete ambiguities, determine all parameters
uniquely for a given $M_H$
and predict correlations between these five observables  that
are characteristic for this scenario.

{\bf Step 4:}

Repeating this exercise in the $B_d^0$ system we have to
our disposal
 \be\label{Step4}
\Delta M_d, \quad S_{\psi K_S}, \quad \mathcal{B}(B_d\to\mu^+\mu^-), \quad
S^d_{\mu^+\mu^-}.
\ee
 Explicit expressions for these observables in terms of the
relevant couplings can be found in  Sections~\ref{sec:3} and \ref{RareB}.

Now NP contributions  to these
three observables are fully described by
\be\label{Step4a}
\frac{\Delta_L^{bd}(H)}{M_{H}}=\frac{\tilde s_{13}}{M_{H}}
e^{-i\delta_{13}},\quad \frac{\Delta_{S,P}^{\mu\bar\mu}(H)}{ M_{H}},
\ee
with the last one known from Step 2 and bounded in Step 3. Again we can determine all the couplings uniquely for a given $M_H$. Our notations and sign conventions are as in Step 3 with
 $\tilde s_{13}\ge 0$  but no minus sign as $V_{td}$ has no such sign.

{\bf Step 5:}

Moving to the $K$ system we have to our disposal
\be\label{Step5}
\varepsilon_K, \quad K_L\to\pi^0\ell^+\ell^-,\quad K_L\to\mu^+\mu^-,
\ee
where in view of hadronic uncertainties the last decay on this list will
only be used to make sure that the existing rough bound on its short distance
branching ratio
is satisfied.  Unfortunately tree-level neutral Higgs contributions to $\kpn$ and $\klpn$ are expected to be negligible, but this fact by itself offers
an important test and distinction from tree level neutral gauge boson
exchanges where these decays could still be significantly affected  \cite{Buras:2012jb}. Also the decays  $K_L\to\pi^0\ell^+\ell^-$ are subject to considerable hadronic uncertainties and their measurements are not expected in this decade. Yet, as they are known to be sensitive to NP effects it is of
interest to consider them as well and compare the scalar case
with the case of $Z'$ models \cite{Buras:2012jb}.

In the present paper we do not study the ratio $\epe$, which
is rather accurately measured but presently subject to much larger hadronic uncertainties
than observables listed in (\ref{Step5}).
Yet, it should be emphasized that $\epe$
is important for the tests of $H$ FCNC
scenarios as it is very sensitive to any NP contribution \cite{Buras:1998ed,Buras:1999da,Blanke:2007wr}.

Explicit expressions for the  observables in the $K$ system
in terms of the
relevant couplings can be found in   Sections~\ref{sec:3} and \ref{RareK}.

Now NP contributions  to these
observables are fully described by
\be\label{Step5a}
\frac{\Delta_L^{sd}(H)}{M_{H}}=-\frac{\tilde s_{12}}{M_{H}}
e^{-i\delta_{12}}, \quad
\frac{\Delta_{S,P}^{\mu\bar\mu}(H)}{ M_{H}}
\ee
The
ratios involving muon couplings are already constrained or determined in
previous steps.
Consequently,
we can bound
 quark couplings involved by using the data on the observables in (\ref{Step5}).
Moreover we identify certain correlations characteristic for LHS
scenario.
$\tilde s_{12}\ge 0$ and the minus sign is chosen to cancel the one of
$V_{ts}$.

We can already announce at this stage that the results in $K$ physics turned out
    to be much less interesting than in the $B_s$ and $B_d$ systems and
     we summarize them separately in Section~\ref{sec:U(2)}.

{\bf Step 6:}

As all parameters of LHS scenario have been fixed in the first five steps we
are in the position to make predictions for the following processes
\be\label{Step6b}
B\to X_s\ell^+\ell^-, \quad B\to K\ell^+\ell^-, \quad B\to K^*\ell^+\ell^-
\ee
and test whether they provide additional constraints on the couplings.
Again as in the case of $\kpn$ and $\klpn$ also the $b\to s\nu\bar\nu$
transitions are expected to be SM-like which provides a distinction from
the gauge boson mediated tree-level transitions  \cite{Buras:2012jb}.

{\bf Step 7:}

We repeat Steps 3-6 for the case of RHS. We will see that in
               view of the change of the sign of NP contributions to
               $B_{s,d}\to\mu^+\mu^-$ and $ K_L\to\mu^+\mu^-$ decays
                the structure of the correlations between various observables
                will distinguish this scenario from the LHS one. Yet,
as we will find out, by going from LHS to RHS scenario we can keep
results of Steps 3-5 unchanged by interchanging
simultaneously two {\it big oases} in the parameter space  that we
encountered already in our study of the $\model$ model \cite{Buras:2012dp} and
$Z'$ models  \cite{Buras:2012jb}.
 This LH-RH invariance present in Steps 3-5 can be broken by the
$b\to s \ell^+\ell^-$ transition in  (\ref{Step6b}).
They
 allow us to
distinguish the physics of RH scalar currents from LH ones.
As only
RH couplings are present in the NP contributions in this scenario, we can use the parametrization of these couplings as in (\ref{Step3a}), (\ref{Step4a}) and
(\ref{Step5a}) keeping in mind that now RH couplings are involved.

{\bf Step 8:}

We repeat Steps 3-6 for the case of LRS. In the case of tree-level gauge boson 
contributions the new features relative to the previous
scenarios is
enhanced NP contributions due to the presence of LR operators in  $\Delta F=2$ transitions.
Yet, in the scalar case, the matrix elements of SLL and SRR operators
 present in previous scenarios are  also significant larger
 than the SM ones and the addition of LR operators has
a more modest effect than in the gauge boson case. However, one of the important 
new feature is the vanishing of NP contributions
               to $B_{s,d}\to\mu^+\mu^-$ and $ K_L\to\mu^+\mu^-$ decays. As the LH and RH couplings
               are equal we can again  use the
parametrization of these couplings as in (\ref{Step3a}), (\ref{Step4a}) and
(\ref{Step5a}) but their values will change due to different constraints
from $\Delta F=2$ transitions. Also in this step 
$b\to s \ell^+\ell^-$
transitions can play an important role.

{\bf Step 9:}

We repeat Steps 3-6 for the case of ALRS. Here the new feature relatively
 to LRS are non-vanishing NP contributions to
               $B_{s,d}\to\mu^+\mu^-$, including $S^{d,s}_{\mu^+\mu^-}$
               CP asymmetries. Again the $b\to s \ell^+\ell^-$ transitions
                    will exhibit their strength in testing the theory in
                 a different environment:
                NP contributions to $\Delta F=2$ observables
               due to the presence of LR operators. As the LH and RH couplings
                differ only by a sign we can again  use the
parametrization of these couplings as in (\ref{Step3a}), (\ref{Step4a}) and
(\ref{Step5a}) but their values will change due to different constraints
from $\Delta F=2$ transitions.

{\bf Step 10:}

One can consider next the case of simultaneous LH and RH
               couplings that are unrelated to each other. This step is more challenging as one has more free parameters and in order to reach clear cut
conclusions one would need a concrete model for $H$ couplings or a very
involved numerical analysis \cite{Altmannshofer:2011gn,Altmannshofer:2012ir,Becirevic:2012fy}.
 A simple model in which both a scalar and a pseudoscalar with 
approximately the same mass couple equally to quarks and leptons has been 
recently presented in \cite{Buras:2013uqa} showing that the structure of 
correlations can be quite rich. We refer to this paper for details.

Once this analysis of $H$ contributions is completed it will be straightforward to apply it to the case of the SM Higgs boson with flavour
violating couplings. Yet, we will see that this case is less interesting than
the case of $Z$  with flavour violating couplings.

\boldmath
 \section{$\Delta F=2$ Processes}\label{DF2}
\unboldmath
 \label{sec:3}
\subsection{Preliminaries}

In the SM the dominant top quark contributions to $\Delta F=2$
processes are described by {\it flavour universal real valued} function
given as follows
($x_t=m_t^2/M_W^2$):
\begin{align}
S_0(x_t)  = \frac{4x_t - 11 x_t^2 + x_t^3}{4(1-x_t)^2}-\frac{3 x_t^2\log x_t}{2
(1-x_t)^3}~.
\end{align}
In other CMFV models $S_0(x_t)$ is replaced by a different function which is
 still flavour universal
and is
real valued. This implies very stringent relations between various
observables in three meson system in question
which have been reviewed in \cite{Buras:2003jf}.

In the presence of $H$ tree-level contributions the
flavour universality is generally broken and one needs three different functions
\be\label{FS}
S(K), \qquad  S(B_d), \qquad S(B_s),
\ee
to describe $K^0-\bar K^0$ and $B^0_{s,d}-\bar B^0_{s,d}$ systems. Moreover,
they all become complex quantities. Therefore CMFV relations are generally
broken. In introducing these functions we will
include in their definitions the contributions
of operators with $LL$, $RR$ and $LR$ Dirac structures.

The derivation of the formulae listed below
is so simple that we will not present it here.
In any case, the compendium of relevant formulae given below and in
next sections is self-contained as far as numerical
analysis is concerned.

\boldmath
\subsection{Master Functions Including $H$ Contributions}
\unboldmath
 Calculating the contributions of $H$ to $\Delta F=2$ transitions it is straightforward
 to write down the expressions for the master functions $S(M)$ in (\ref{FS})
in terms of the
 couplings defined in Fig.~\ref{neutral}.

We define first the relevant CKM factors
\be
\lambda_i^{(K)} =V_{is}^*V_{id},\qquad
\lambda_t^{(d)} =V_{tb}^*V_{td}, \qquad \lambda_t^{(s)} =V_{tb}^*V_{ts},
\ee
and introduce
\be\label{gsm}
g_{\text{SM}}^2=4\frac{G_F}{\sqrt 2}\frac{\alpha}{2\pi\sin^2\theta_W}=1.78137\times 10^{-7} \gev^{-2}\,.
\ee

The $\Delta F=2$ master functions for $M=K,B_q$ are then given as follows
\begin{equation}\label{Seff}
S(M)=S_0(x_t)+\Delta S(M)\equiv|S(M)|e^{i\theta_S^M}
\end{equation}
with $\Delta S(M)$ receiving contributions from various operators so that
it is useful to write
\begin{equation}\label{sum}
\Delta S(M)=[\Delta S(M)]_{\rm SLL}+[\Delta S(M)]_{\rm SRR}+[\Delta S(M)]_{\rm LR}.
\end{equation}

The contributing new operators
are defined for the $K$ system as follows
\cite{Buras:2001ra,Buras:2012fs}
\begin{subequations}\label{equ:operatorsZ}
\bea
{Q}_1^\text{LR}&=&\left(\bar s\gamma_\mu P_L d\right)\left(\bar s\gamma^\mu P_R d\right)\,,\\
{Q}_2^\text{LR}&=&\left(\bar s P_L d\right)\left(\bar s P_R d\right)\,.
\eea
\end{subequations}

{\allowdisplaybreaks
\begin{subequations}\label{equ:operatorsHiggs}
 \bea
{Q}_1^\text{SLL}&=&\left(\bar s P_L d\right)\left(\bar s P_L d\right)\,,\\
{Q}_1^\text{SRR}&=&\left(\bar s P_R d\right)\left(\bar s P_R d\right)\,,\\
{Q}_2^\text{SLL}&=&\left(\bar s \sigma_{\mu\nu} P_L d\right)\left(\bar s\sigma^{\mu\nu}  P_L d\right)\,,\\
{Q}_2^\text{SRR}&=&\left(\bar s \sigma_{\mu\nu}  P_R d\right)\left(\bar s \sigma^{\mu\nu} P_R d\right)\,,
\eea
\end{subequations}}%
where $P_{R,L}=(1\pm\gamma_5)/2$ and
we suppressed colour indices as they are summed up in each factor. For instance $\bar s\gamma_\mu P_L d$ stands for $\bar s_\alpha\gamma_\mu P_L d_\alpha$ and similarly for other factors.
For $B_q^0-\bar B_q^0$ mixing our conventions for new operators are:
\begin{subequations}\label{equ:operatorsZb}
\bea
{Q}_1^\text{LR}&=&\left(\bar b\gamma_\mu P_L q\right)\left(\bar b\gamma^\mu P_R q\right)\,,\\
{Q}_2^\text{LR}&=&\left(\bar b P_L q\right)\left(\bar b P_R q\right)\,,
\eea
\end{subequations}
{\allowdisplaybreaks
\begin{subequations}\label{equ:operatorsHiggsb}
 \bea
{Q}_1^\text{SLL}&=&\left(\bar b P_L q\right)\left(\bar b P_L q\right)\,,\\
{Q}_1^\text{SRR}&=&\left(\bar b P_R q\right)\left(\bar b P_R q\right)\,,\\
{Q}_2^\text{SLL}&=&\left(\bar b \sigma_{\mu\nu} P_L q\right)\left(\bar b\sigma^{\mu\nu}  P_L q\right)\,,\\
{Q}_2^\text{SRR}&=&\left(\bar b \sigma_{\mu\nu}  P_R q\right)\left(\bar b \sigma^{\mu\nu} P_R q\right)\,.
\eea
\end{subequations}}%

In order to calculate the SLL, SRR and LR contributions to $\Delta S(M)$
we introduce
quantities familiar from SM expressions for mixing amplitudes
\be
T(B_q)=\frac{G_F^2}{12\pi^2}F_{B_q}^2\hat B_{B_q}m_{B_q}M_{W}^2
\left(\lambda_t^{(q)}\right)^2\eta_B,
\label{eq:3.6}
\ee
\be
T(K)=\frac{G_F^2}{12\pi^2}F_{K}^2\hat B_{K}m_{K}M_{W}^2
\left(\lambda_t^{(K)}\right)^2\eta_2
\label{eq:3.7},
\ee
where $\eta_i$ are QCD corrections and $\hat B_i$ known SM non-perturbative
factors.

Then
\be\label{DSKSLL}
T(K)[\Delta S(K)]_{\rm SLL}=
-\frac{(\Delta_L^{sd}(H))^2}{2M_H^2}\left[C_1^\text{SLL}(\mu_H)
 \langle Q_1^\text{SLL}(\mu_H,K)\rangle +C_2^\text{SLL}(\mu_H)\langle Q_2^\text{SLL}(\mu_H,K)\rangle
 \right]
\ee
with the SRR contribution obtained by replacing L by R. Note that this
replacement  only affects the coupling $\Delta_L^{sd}(H)$ as the hadronic matrix elements being evaluated in QCD remain unchanged and the Wilson coefficients have been so defined that they also remain unchanged.
For LR contributions
we find
\be\label{DSKLR}
T(K)[\Delta S(K)]_{\rm LR}=
-\frac{\Delta_L^{sd}(H)\Delta_R^{sd}(H)}{
 M_H^2} \left [ C_1^\text{LR}(\mu_H) \langle Q_1^\text{LR}(\mu_H,K)\rangle +
 C_2^\text{LR}(\mu_H) \langle Q_2^\text{LR}(\mu_H,K)\rangle \right]\,.
\ee

Including NLO QCD corrections \cite{Buras:2012fs} the Wilson coefficients of
the involved
operators are given by
\begin{align}\label{equ:WilsonH}
C_1^\text{SLL}(\mu)= C_1^\text{SRR}(\mu)&=
1+\frac{\alpha_s}{4\pi}\left(-3\log\frac{M_H^2}{\mu^2}+\frac{9}{2}\right)\,,\\
\begin{split}
C_2^\text{SLL}(\mu) =
C_2^\text{SRR}(\mu)
&=\frac{\alpha_s}{4\pi}\left(-\frac{1}{12}\log\frac{M_H^2}{\mu^2}+\frac{1}{8}
\right)\,,\end{split}\\
 C_1^\text{LR}(\mu)& =-\frac{3}{2}\frac{\alpha_s}{4\pi}\,,\\
C_2^\text{LR}(\mu) &=  1-\frac{\alpha_s}{4\pi}\frac{3}{N}=  1-\frac{\alpha_s}{4\pi}\,.
\end{align}

Next
\be
\langle Q^a_i(\mu_{H},K)\rangle \equiv \frac{m_K F_K^2}{3} P^a_i(\mu_{H},K)
\ee
are the matrix elements of operators evaluated at the matching scale
$\mu_{H}=\ord(M_{H})$
and
 $P^a_i$ are  the coefficients introduced in \cite{Buras:2001ra}.
{ The $\mu_{H}$ dependence of  $P^a_i(\mu_H)$ cancels the one of $\Delta_{L,R}(H)$ and of $C^a_i(\mu_{H})$ so that $S(K)$ does 
not depend  on $\mu_{H}$. It should 
be emphasized at this point that in contrast to gauge boson couplings the 
couplings $\Delta_{L,R}(H)$ are scale dependent and consistently with the 
NLO calculation in \cite{Buras:2012fs} they are defined here at $\mu_{H}=\ord(M_{H})$. In our numerical calculations we will simply set $\mu_{H}=M_{H}$.}

Similarly for $B_q$ systems we have
\be\label{BSLL}
T(B_q)[\Delta S(B_q)]_{\rm SLL}=
-\frac{(\Delta_L^{bq}(H))^2}{2M_H^2}\left[C_1^\text{SLL}(\mu_H)
 \langle Q_1^\text{SLL}(\mu_H,B_q)\rangle +C_2^\text{SLL}(\mu_H)\langle Q_2^\text{SLL}(\mu_H,B_q)\rangle\right]
\ee
\be\label{BLR}
T(B_q)[\Delta S(B_q)]_{\rm LR}=
-\frac{\Delta_L^{bq}(H)\Delta_R^{bq}(H)}{
 M_H^2} \left [ C_1^\text{LR}(\mu_H) \langle Q_1^\text{LR}(\mu_H,B_q)\rangle +
 C_2^\text{LR}(\mu_H) \langle Q_2^\text{LR}(\mu_H,B_q)\rangle \right]\,,
\ee
where the Wilson coefficients   $C^a_i(\mu_{H})$  are as in the $K$ system
and the matrix elements are given by
\be
\langle Q^a_i(\mu_{H},B_q)\rangle \equiv \frac{m_{B_q} F_{B_q}^2}{3} P^a_i(\mu_{H},B_q).
\ee
For SRR contributions one proceeds as in the $K$ system.

Finally, we collect in Table~\ref{tab:Qi}
central values of  $\langle Q^a_i(\mu_{H})\rangle$. They are given in the
$\overline{\text{MS}}$-NDR scheme and are based on lattice calculations
in \cite{Boyle:2012qb,Bertone:2012cu} for $K^0-\bar K^0$ system and  in
\cite{Bouchard:2011xj} for
$B_{d,s}^0-\bar B^0_{d,s}$ systems. For the $K^0-\bar K^0$ system we have just
used the average of the results in \cite{Boyle:2012qb,Bertone:2012cu} that
are consistent with each other.
As the values of the relevant $B_i$ parameters in these papers have been
evaluated at $\mu=3\gev$ and $4.2\gev$, respectively, we have used the
formulae in  \cite{Buras:2001ra} to obtain the values of the matrix
elements in question at $\mu_{H}$. For simplicity we choose this scale to
be $M_{H}$ but any scale of this order would give the same results for
the physical quantities up to NNLO QCD corrections that are negligible
at these high scales.  The renormalization scheme dependence of the
matrix elements is canceled by the one of the Wilson coefficients.

In the case of tree-level SM Higgs exchanges we evaluate the matrix elements
at $m_t(m_t)$ as the inclusion of NLO QCD corrections allows us to choose
any scale of $\ord(M_H)$ without changing physical results. Then
in the formulae above
one should replace $M_{H}$ by the SM Higgs mass and $\mu_{H}$ by $m_t(m_t)$. 
{ This also means that the flavour violating couplings of SM Higgs are 
defined here at $m_t(m_t)$. } The values
of hadronic matrix elements at $m_t(m_t)$ in the
$\overline{\text{MS}}$-NDR scheme
are given in Table~\ref{tab:Qi1}.

%%%%%%%%%%%%%%%%%%%%%%%%%%%%%%%%%%%%%%%%%%%%%%%%%%%%%%%%%%%%%%%%%%%%%%%%%%%%%%%%

\begin{table}[!ht]
{\renewcommand{\arraystretch}{1.3}
\begin{center}
\begin{tabular}{|c||c|c|c|c|}
\hline
&$\langle Q_1^\text{LR}(\mu_H)\rangle$& $\langle Q_2^\text{LR}(\mu_H)\rangle$&$\langle Q_1^\text{SLL}(\mu_H)\rangle$&$\langle
Q_2^\text{SLL}(\mu_H)\rangle$\\
\hline
\hline
$K^0$-$\bar K^0$ &$-0.14$ &$0.22$ & $-0.074$ & $-0.128$\\
\hline
$B_d^0$-$\bar B_d^0$& $-0.25$ &$0.34$  &$-0.11$ &$-0.22 $\\
\hline
$B_s^0$-$\bar B_s^0$& $-0.37$ &$ 0.51$ &$-0.17$  & $-0.33$\\
\hline
\end{tabular}
\end{center}}
\caption{\it Hadronic matrix elements $\langle Q_i^a(\mu_H)\rangle$  in units of GeV$^3$ at $\mu_H=1\tev$.
\label{tab:Qi}}
\end{table}

\begin{table}[!ht]
{\renewcommand{\arraystretch}{1.3}
\begin{center}
\begin{tabular}{|c||c|c|c|c|}
\hline
&$\langle Q_1^\text{LR}(m_t)\rangle$& $\langle Q_2^\text{LR}(m_t)\rangle$&$\langle Q_1^\text{SLL}(m_t)\rangle$&$\langle
Q_2^\text{SLL}(m_t)\rangle$\\
\hline
\hline
$K^0$-$\bar K^0$ &$-0.11$ &$0.18$  & $-0.064$ & $-0.107$\\
\hline
$B_d^0$-$\bar B_d^0$& $-0.21$ &$0.27$  &$-0.095$ &$-0.191 $\\
\hline
$B_s^0$-$\bar B_s^0$& $-0.30$ &$ 0.40$ &$-0.14$  & $-0.29$\\
\hline
\end{tabular}
\end{center}}
\caption{\it Hadronic matrix elements $\langle Q_i^a(\mu_t)\rangle$  in units of GeV$^3$ at $m_t(m_t)$.
\label{tab:Qi1}}
\end{table}

\boldmath
 \subsection{Basic Formulae for $\Delta F=2$ Observables}
\unboldmath

The $\Delta B=2$ mass differences are given as follows:
\be\label{DMd}
\Delta M_d=\frac{G_F^2}{6 \pi^2}M_W^2 m_{B_d}|\lambda_t^{(d)}|^2   F_{B_d}^2\hat B_{B_d} \eta_B |S(B_d)|\,,
\ee
\be\label{DMs}
\Delta M_s =\frac{G_F^2}{6 \pi^2}M_W^2 m_{B_s}|\lambda_t^{(s)}|^2   F_{B_s}^2\hat B_{B_s} \eta_B |S(B_s)|\,.
\ee
The corresponding mixing induced CP-asymmetries are then given
by
\begin{equation}
S_{\psi K_S} = \sin(2\beta+2\varphi_{B_d})\,, \qquad
S_{\psi\phi} =  \sin(2|\beta_s|-2\varphi_{B_s})\,,
\label{eq:3.44}
\end{equation}
where the phases $\beta$ and $\beta_s$ are defined by
\be\label{vtdvts}
V_{td}=\vtd e^{-i\beta}, \qquad V_{ts}=-\vts e^{-i\beta_s}.
\ee
$\beta_s\simeq -1^\circ\,$.
 The new phases $\varphi_{B_q}$  are directly related to the phases of the functions
$S(B_q)$:
\be
2\varphi_{B_q}=-\theta_S^{B_q}.
\ee

Our phase conventions are as in \cite{Buras:2012jb} and our previous papers
quoted in this work.  Consequently $S_{\psi\phi}^{\rm SM}\approx 0.04$. On
the other hand the experimental results are usually given for the phase
\be
\phi_s=2\beta_s+\phi^{\rm NP}
\ee
so that
\be
S_{\psi\phi}=-\sin(\phi_s), \qquad  2\varphi_{B_s}=\phi^{\rm NP}.
\ee
Using this dictionary the most recent result  for $\phi_s$  from the 
LHCb analysis of CP-violation in
$B_s\to \psi \phi$ decay implies \cite{Raven:2012fb}
\be
2|\beta_s|-2\varphi_{B_s} = 0.001\pm 0.104,
\ee
that is close to its SM value. But the uncertainties are still sufficiently large so that it is of interest to investigate correlations of $S_{\psi\phi}$
with other observables in the $B_s$ system.

For the CP-violating parameter $\varepsilon_K$  and $\Delta M_K$ we have
respectively
\be
\varepsilon_K=\frac{\kappa_\eps e^{i\varphi_\eps}}{\sqrt{2}(\Delta M_K)_\text{exp}}\left[\Im\left(M_{12}^K\right)\right]\,,\qquad \Delta M_K=2\Re\left(M_{12}^K\right),
\label{eq:3.35}
\ee
where
\be\label{eq:3.4}
\left(M_{12}^K\right)^*=\frac{G_F^2}{12\pi^2}F_K^2\hat
B_K m_K M_{W}^2\left[
\lambda_c^{2}\eta_1x_c +\lambda_t^{2}\eta_2S(K) +
2\lambda_c\lambda_t\eta_3S_0(x_c,x_t)
\right]\,.
\ee
Here, $S_0(x_c,x_t)$
is a {\it real valued} one-loop box function for which explicit expression is given e.\,g.~in \cite{Blanke:2006sb}. The
factors $\eta_i$ are QCD {corrections} evaluated at the NLO level in
\cite{Herrlich:1993yv,Herrlich:1995hh,Herrlich:1996vf,Buras:1990fn,Urban:1997gw}. For $\eta_1$ and $\eta_3$ also NNLO corrections are known \cite{Brod:2010mj,Brod:2011ty}.
Next $\varphi_\eps = (43.51\pm0.05)^\circ$ and $\kappa_\eps=0.94\pm0.02$ \cite{Buras:2008nn,Buras:2010pza} takes into account
that $\varphi_\eps\ne \tfrac{\pi}{4}$ and includes long distance  effects in $\Im( \Gamma_{12})$ and $\Im (M_{12})$.

In the rest of the paper, unless otherwise stated, we will assume that all four parameters in the CKM
matrix have been determined through tree-level decays without any NP pollution
and pollution from QCD-penguin diagrams so that their values can be used
universally  in
all NP models considered by us.

\boldmath
\section{Rare B Decays}\label{RareB}
\unboldmath
\subsection{Preliminaries}
These decays played already for many years a significant role in constraining
NP models. In particular $B_s\to\mu^+\mu^-$ was instrumental in bounding
scalar contributions in the framework of supersymmetric
models and two Higgs doublet models (2HDM). Recently a very detailed analysis
of the decay $B_s\to\mu^+\mu^-$ including the observables involved in
the time-dependent rate has been presented \cite{Buras:2013uqa}.
Below, after recalling the
relevant effective Hamiltonian that can be used for other $b\to s\ell^+\ell^-$
transitions, we will summarize the final formulae for the most important
observables in  $B_s\to\mu^+\mu^-$ that have been derived and discussed in
more detail in \cite{Buras:2013uqa} and in particular earlier in
\cite{deBruyn:2012wk}.
While our analysis of $B_s\to\mu^+\mu^-$ is less detailed than the one in \cite{Buras:2013uqa}, our main goal here is to discuss the
correlations of  $B_s\to\mu^+\mu^-$ observables with $\Delta F=2$ observables, in particular $S_{\psi\phi}$,
which were not presented there.
Moreover, we analyze here similar correlations involving
$B_d\to\mu^+\mu^-$ observables and $S_{\psi K_S}$.
\boldmath
\subsection{Effective Hamiltonian for $b\to s\ell^+\ell^-$}\label{sec:bqll}
\unboldmath
For our discussion of $B_{d,s} \to \mu^+ \mu^-$ and for the imposition of the
constraints from other
$b\to s\ell^+\ell^-$ transitions, like $B\to K^*\ell^+\ell^-$, $B\to K\ell^+\ell^-$ and $B\to X_s\ell^+\ell^-$,
we will need the corresponding effective
Hamiltonian which is
a generalization of the SM one:
\be\label{eq:Heffqll}
 \Heff(b\to s \ell\bar\ell)
= \Heff(b\to s\gamma)
-  \frac{4 G_{\rm F}}{\sqrt{2}} \frac{\alpha}{4\pi}V_{ts}^* V_{tb} \sum_{i = 9,10,S,P} [C_i(\mu)Q_i(\mu)+C^\prime_i(\mu)Q^\prime_i(\mu)]
\end{equation}
where
\begin{subequations}
\begin{align}
Q_9 & = (\bar s\gamma_\mu P_L b)(\bar \ell\gamma^\mu\ell),&  &Q_9^\prime  =  (\bar s\gamma_\mu P_R b)(\bar \ell\gamma^\mu\ell),\\
Q_{10} & = (\bar s\gamma_\mu P_L b)(\bar \ell\gamma^\mu\gamma_5\ell),&  &Q_{10}^\prime  =  (\bar s\gamma_\mu P_R b)(\bar \ell\gamma^\mu\gamma_5\ell),\\
Q_S &= m_b(\bar s P_R b)(\bar \ell\ell),& & Q_S^\prime = m_b(\bar s P_L b)(\bar \ell\ell),\\
Q_P & = m_b(\bar s P_R b)(\bar \ell\gamma_5\ell),& & Q_P^\prime = m_b(\bar s P_L b)(\bar \ell\gamma_5\ell).
\end{align}
\end{subequations}
Including the factors of $m_b$ into the definition of scalar operators makes
their matrix elements and their Wilson coefficients scale independent.
$\Heff(b\to s\gamma)$ stands for the effective Hamiltonian for the
$b\to s\gamma$ transition that involves
the dipole operators. We will not discuss $b\to s\gamma$ in this paper as
it appears first at one-loop level and a neutral scalar contribution would
only be of relevance in the presence 
of flavor-conserving scalar couplings to down-type quarks which would 
introduce new parameters without any impact on our results. 

Note the
difference of ordering of flavours relatively to $\Delta F=2$ as already stressed in Section~\ref{sec:2}. Therefore the unprimed operators $Q_S$ and $Q_P$
represent the LHS scenario and the primed ones  $Q_S^\prime$ and $Q_P^\prime$
the RHS scenario. We neglect effects proportional to $m_s$ in each case but keep $m_s$ and $m_d$ different from zero when they are shown explicitly.

The  Wilson coefficients $C_9$ and $C_{10}$  do not receive any new
contributions from scalar exchanges and take SM values
\begin{align}\label{C9SM}
 \sin^2\theta_W C^{\rm SM}_9 &=[\eta_Y Y_0(x_t)-4\sin^2\theta_W Z_0(x_t)],\\
   \sin^2\theta_W C^{\rm SM}_{10} &= -\eta_Y Y_0(x_t).
 \end{align}
On the other hand with $m_s\ll m_b$ we have $C_9^\prime=C_{10}^\prime=0$.
Here  $Y_0(x_t)$ and  $Z_0(x_t)$ are SM one-loop functions given by
\be\label{YSM}
Y_0(x_t)=\frac{x_t}{8}\left(\frac{x_t-4}{x_t-1} + \frac{3 x_t \log x_t}{(x_t-1)^2}\right),
\ee
\begin{align}
  Z_0 (x) & = -\frac{1}{9} \log x + \frac{18 x^4 - 163 x^3 + 259 x^2 - 108 x}{144 (x-1)^3} + \frac{32 x^4 - 38 x^3 - 15 x^2 + 18 x}{72
(x-1)^4}\log x \,.
\end{align}
The coefficient $\eta_{Y}$ is a QCD factor which for $m_t=m_t(m_t)$ is close to unity:  $\eta_{Y}=1.012$
\cite{Buchalla:1998ba,Misiak:1999yg}.

For the coefficients of scalar operators we find
\begin{align}
 m_b(\mu_H)\sin^2\theta_W C_S &= \frac{1}{g_{\text{SM}}^2}\frac{1}{ M_H^2}\frac{\Delta_R^{sb}(H)\Delta_S^{\mu\bar\mu}(H)}{V_{ts}^* V_{tb}},\\
 m_b(\mu_H)\sin^2\theta_W C_S^\prime &= \frac{1}{g_{\text{SM}}^2}\frac{1}{ M_H^2}\frac{\Delta_L^{sb}(H)\Delta_S^{\mu\bar\mu}(H)}{V_{ts}^* V_{tb}},\\
 m_b(\mu_H)\sin^2\theta_W C_P &= \frac{1}{g_{\text{SM}}^2}\frac{1}{ M_H^2}\frac{\Delta_R^{sb}(H)\Delta_P^{\mu\bar\mu}(H)}{V_{ts}^* V_{tb}},\\
 m_b(\mu_H)\sin^2\theta_W C_P^\prime &= \frac{1}{g_{\text{SM}}^2}\frac{1}{ M_H^2}\frac{\Delta_L^{sb}(H)\Delta_P^{\mu\bar\mu}(H)}{V_{ts}^* V_{tb}},
\end{align}
where $\Delta_{S,P}^{\mu\bar\mu}(H)$ are defined in (\ref{equ:mumuSPLR}). { It should be emphasized at this point that the 
couplings $\Delta_{L,R}^{sb}(H)$ 
extracted from $\Delta M_s$ and $S_{\psi\phi}$ are defined at $\mu_H=M_H$, therefore, as shown explicitly, $m_b$ has to be 
evaluated also at this scale in order 
to keep these coefficients scale independent. In the case of the SM Higgs $m_b$
has to be evaluated at $m_t(m_t)$ as at this scale the flavour violating SM Higgs couplings in $\Delta F=2$ processes are defined. In what follows we will not 
show this dependence explicitly.} For $m_b$ at $1~$TeV and at $125~$GeV we use the values
\begin{align}
 & m_b(1~\text{TeV}) = 2.54~\text{GeV}\,,\qquad m_b(163~\text{GeV}) =  2.81~\text{GeV}\,.
\end{align}

Next we recall that in terms of the couplings used in the analysis of $B_{s,d}^0-\bar B_{s,d}^0$ mixings we have
\be\label{dictionary}
\Delta_R^{sb}(H)=[\Delta_L^{bs}(H)]^*,\qquad  \Delta_L^{sb}(H)=[\Delta_R^{bs}(H)]^*,
\ee
which should be kept in mind when studying correlations between $\Delta F=1$
and $\Delta F=2$ transitions. These relations can be directly used in
the case of $C_S$ and $C_S^\prime$ but in the case of $C_P$ and $C_P^\prime$,
as discussed in Section~\ref{sec:2},
it is useful to use in this context the following relations:
\be
\Delta_R^{sb} \Delta_P^{\mu\bar\mu}=-[\tilde\Delta_L^{bs}]^* \tilde\Delta_P^{\mu\bar\mu}, \qquad
\Delta_L^{sb} \Delta_P^{\mu\bar\mu}=[\tilde\Delta_R^{bs}]^* \tilde\Delta_P^{\mu\bar\mu}
\ee
with $\Delta_P^{\mu\bar\mu}$ being imaginary but $\tilde \Delta_P^{\mu\bar\mu}$ real.

\boldmath
\subsection{Observables for $B_{s}\to \mu^+ \mu^-$}
\unboldmath
In the general analysis of  $B_{s}\to \mu^+ \mu^-$ in \cite{Buras:2013uqa}, which goes beyond the
NP scenario considered here, the basic four observables are
\be\label{trio}
\overline{R}, \qquad \mathcal{A}^{\mu\mu}_{\Delta\Gamma}, \qquad S^s_{\mu\mu},\quad  S_{\psi\phi}.
\ee
 Here, the observable $\overline{R}$, defined in (\ref{Rdef}),  is
just the ratio of the branching ratio that includes
$\Delta\Gamma_s$ effects and of the SM prediction for the branching ratio that
also includes them. The relation of $\overline{R}$ to $R$ introduced 
in \cite{deBruyn:2012wk} is given below. Following \cite{Buras:2013uqa} we 
will denote branching ratios containing $\Delta\Gamma_s$ effects with a {\it bar} while those without these effects without it.

The next two observables, ${\cal A}^{\mu\mu}_{\Delta\Gamma}$ and ${\cal S}_{\mu\mu}$ can be extracted from flavour untagged and tagged time-dependent measurements of  $B_{s}\to \mu^+ \mu^-$, respectively.
As these three observables depend also on the new phase $\varphi_{B_s}$ in
the $B_s^0-\bar B_s^0$ mixing, also the mixing induced CP-asymmetry $S_{\psi\phi}$ is involved here.

In order to calculate these observables one
 introduces
\begin{align}
P&\equiv \frac{C_{10}-C_{10}^\prime}{C_{10}^{\rm SM}}+
\frac{m^2_{B_s}}{2m_\mu}\frac{m_b}{m_b+m_s} \frac{C_P-C_P^\prime}{C_{10}^{\rm SM}}
\equiv |P|e^{i\varphi_P}\label{PP}\\
S&\equiv \sqrt{1-\frac{4m_\mu^2}{m_{B_s}^2}}\frac{m^2_{B_s}}{2m_\mu}
\frac{m_b}{m_b+m_s}
\frac{C_S-C_S^\prime}{C_{10}^{\rm SM}}
\equiv |S|e^{i\varphi_S}.
\label{SS}
\end{align}

One finds then three basic formulae \cite{deBruyn:2012wk,Fleischer:2012fy,Buras:2013uqa}
\begin{align}\label{Rdef}
	\overline{R} &\equiv \frac{\overline{\mathcal{B}}(B_{s}\to\mu^+\mu^-)}{\overline{\mathcal{B}}(B_{s}\to\mu^+\mu^-)_{\rm SM}}
	= \left[\frac{1+{\cal A}^{\mu\mu}_{\Delta\Gamma}\,y_s}{1+y_s} \right] \times (|P|^2 + |S|^2)\notag\\
	&= \left[\frac{1+y_s\cos(2\varphi_P-2\varphi_{B_s})}{1+y_s} \right] |P|^2 + \left[\frac{1-y_s\cos(2\varphi_S-2\varphi_{B_s})}{1+y_s} \right] |S|^2,
\end{align}
\begin{align}
{\cal A}^{\mu\mu}_{\Delta\Gamma} &= \frac{|P|^2\cos(2\varphi_P-2\varphi_{B_s}) - |S|^2\cos(2\varphi_S-2\varphi_{B_s})}{|P|^2 + |S|^2},\label{ADG}\\
	S^s_{\mu\mu}
	&=\frac{|P|^2\sin(2\varphi_P-2\varphi_{B_s})-|S|^2\sin(2\varphi_S-2\varphi_{B_s})}{|P|^2+|S|^2}.
	\label{Ssmu}
\end{align}
Here \cite{LHCb-Mor-12}
\begin{equation}\label{defys}
	y_s\equiv\tau_{B_s}\frac{\Delta\Gamma_s}{2}
=0.088\pm0.014.
\end{equation}

 The ratio $R$ of \cite{deBruyn:2012wk}, which did not include $\Delta\Gamma_s$ effects in the SM result and $\overline{R}$ which
includes them are related by
\be
\overline{R}= (1-y_s) R.
\ee
The advantage of $\overline{R}$ over $R$ is that in the SM it is equal to unity 
and its departure from unity summarizes total NP effects present both in the decay and mixing.

Another useful variable encountered in this discussion is
\be\label{rys}
r(y_s)\equiv\frac{1-y_s^2}{1+\mathcal{A}^{\mu^+\mu^-}_{\Delta\Gamma} y_s}.
\ee
It is the correction factor that one has to introduce in any model in
order to compare the branching ratio calculated in this model without 
$\Delta\Gamma_s$ effects and the branching ratio which
includes them \cite{DescotesGenon:2011pb,deBruyn:2012wj,deBruyn:2012wk}
\be
\label{Fleischer1}
\mathcal{B}(B_{s}\to\mu^+\mu^-) =
r(y_s)~\overline{\mathcal{B}}(B_{s}\to\mu^+\mu^-).
\ee

It should be emphasized that presently only $\overline{\mathcal{B}}(B_{s}\to\mu^+\mu^-)$ is known experimentally but once
$\mathcal{A}^{\mu\mu}_{\Delta\Gamma}$ will be extracted from time-dependent measurements, we will be able 
to obtain $\mathcal{B}(B_{s}\to\mu^+\mu^-)$ directly from experiment as well.
Evidently, in any model the branching ratios without $\Delta\Gamma_s$ 
effect are related to the corresponding SM branching ratio through
\be\label{THBr}
\mathcal{B}(B_{s}\to\mu^+\mu^-)=\mathcal{B}(B_{s}\to\mu^+\mu^-)_{\rm SM}(|P|^2 + |S|^2).
\ee

As $F_{B_s}$ cancels out in the evaluation of $\mathcal{A}^{\mu\mu}_{\Delta\Gamma}$ and ${\cal S}_{\mu\mu}$, these are theoretically
clean observables and offer new ways to test NP models. Indeed, as seen 
in (\ref{ADG}) and (\ref{Ssmu}), both observables depend on NP contributions 
and this is also the case of the conversion factor $r(y_s)$.
In the SM and CMFV models $S=0$ and $\varphi_P=\varphi_{B_s}=0$ so that
\be
\mathcal{A}^{\mu\mu}_{\Delta\Gamma}=1, \quad S_{\mu\mu}^s=0,
\quad r(y_s)=0.912\pm0.014 \qquad  ({\rm SM,~CMFV}), 
\ee
independently of NP parameters present in the whole class of CMFV models.

As $\mathcal{A}^{\mu\mu}_{\Delta\Gamma}$ does not rely on flavour tagging, which is difficult for a rare decay, it will be easier to determine than
$S_{\mu\mu}^s$.
Given limited statistics, experiments may first measure the $B_s\to\mu^+\mu^-$ effective lifetime,
a single exponential fit to the untagged rate, from which $\mathcal{A}^{\mu\mu}_{\Delta\Gamma}$ can also be deduced~\cite{deBruyn:2012wk}. See also
\cite{Buras:2013uqa} for discussion.

While $\Delta\Gamma_d$ is very small and $y_d$ can be set to zero,
in the case of $B_d\to\mu^+\mu^-$ one can still consider the CP asymmetry
$S_{\mu\mu}^d$ \cite{Fleischer:2012fy}, for which one can use all expressions
given above with the flavour index ``$s$'' replaced by ``$d$''.

\boldmath
\subsection{Present Data}
\unboldmath

The most recent  results from LHCb read \cite{Aaij:2012ac,LHCbBsmumu}
\be\label{LHCb2}
\overline{\mathcal{B}}(B_{s}\to\mu^+\mu^-) = (3.2^{+1.5}_{-1.2}) \times 10^{-9}, \quad
\mathcal{B}(B_{s}\to\mu^+\mu^-)_{\rm SM}=(3.25\pm0.17)\times 10^{-9},
\ee
\be\label{LHCb3}
\mathcal{B}(B_{d}\to\mu^+\mu^-) \le  9.4\times 10^{-10}, \quad
\mathcal{B}(B_{d}\to\mu^+\mu^-)_{\rm SM}=(1.05\pm0.07)\times 10^{-10}.
\ee
We have shown here SM predictions for these
observables that do not include the correction
${r(y_s)}$.   As $r(y_d)=1$ to an excellent approximation, the result for 
$B_{d}\to\mu^+\mu^-$
can be directly compared with experiment. In order to obtain these results we have used the parametric formulae of \cite{Buras:2012ru} and updated the lattice QCD values of
$F_{B_{s,d}}$ \cite{Dowdall:2013tga} and the life-times $\tau_{B_{s,d}}$ \cite{Amhis:2012bh}. 
Details can be found in \cite{Buras:2013uqa}.

If the correction factor ${r(y_s)}$ is taken into account the SM result
in (\ref{LHCb2}) changes to \cite{Buras:2013uqa}
\be
\label{FleischerSM}
\overline{\mathcal{B}}(B_{s}\to\mu^+\mu^-)_{\rm SM}= (3.56\pm0.18)\cdot 10^{-9}.
\ee
It is this branching that should be compared in such a case
with the results of LHCb given above.
For the latest discussions of these issues see
\cite{deBruyn:2012wj,deBruyn:2012wk,Buras:2012ru,Fleischer:2012fy}.
As discussed in \cite{Buras:2012ru,Misiak:2011bf} complete NLO electroweak corrections are still missing in this estimate. This result
should be available in the near future.\footnote{Martin Gorbahn, private communication.}

In our numerical results we will use $\overline{\mathcal{B}}(B_{s}\to\mu^+\mu^-)$ in (\ref{Fleischer1})
with $\mathcal{B}(B_{s}\to\mu^+\mu^-)$ given by (\ref{THBr}) and 
$r(y_s)$ by (\ref{rys}) with $\mathcal{A}^{\mu\mu}_{\Delta\Gamma}$ also 
affected by NP effects.

Combining the experimental and theoretical results quoted above gives
\begin{equation}
	\overline{R}_{\rm LHCb} = 0.90^{+0.42}_{-0.34} \in [0.30,1.80] \,  (95\%\ {\rm C.L}).
    \label{RexpValues}
\end{equation}
This range should be compared with its SM value, corresponding to
$P=1$, $S=0$ and $\phi_s^{\rm NP}=0$:
\be
\overline{R}_{\rm SM}=1.
\ee

\boldmath
\subsection{Scenarios for $S$ and $P$}\label{SPscenarios}
\unboldmath
Clearly, the outcome of the results for the observables in question depends
on the values of the muon couplings and whether a scalar or pseudoscalar
boson is involved. Moreover, as we stressed already in Section~\ref{sec:2} the
exchanged mass eigenstate does not have to be a CP eigenstate and can have
both scalar and pseudoscalar couplings to leptons. In \cite{Buras:2013uqa} a detailed classification of various
possibilities beyond the dynamical model considered here has been made and the
related purely phenomenological numerical analysis has been performed. Here we will make a
classification that is particularly suited for the dynamical model considered
by us.

{\bf Pseudoscalar Scenario:}

In this scenario $S=0$ and $P$ can be arbitrary complex number. We find then
\begin{align}\label{FA1}
	{\cal A}^{\mu\mu}_{\Delta\Gamma} = \cos(2\varphi_P - 2\varphi_{B_s}),\quad
	 S^s_{\mu\mu} = \sin(2\varphi_P - 2\varphi_{B_s}).
\end{align}
The branching ratio observable is given by
\begin{equation}\label{FA2}
	\overline{R} = |P|^2 \left[\frac{1+y_s\cos(2\varphi_P - 2\varphi_{B_s})}{1+y_s}\right].
\end{equation}
This scenario corresponds to scenario A in \cite{Buras:2013uqa}.

{\bf Scalar Scenario:}

In this scenario $P=1$ and $S$ can be arbitrary complex number. We find then
\begin{align}\label{SB}
	{\cal A}^{\mu\mu}_{\Delta\Gamma} &= \frac{\cos 2\varphi_{B_s} - |S|^2\cos(2\varphi_S -  2\varphi_{B_s})}{1+|S|^2},\notag\\
	 S^s_{\mu\mu} &= \frac{-\sin 2\varphi_{B_s} - |S|^2\sin(2\varphi_S -  2\varphi_{B_s})}{1+|S|^2},\notag\\
	\overline{R} &= \frac{1+y_s\cos 2\varphi_{B_s}}{1+y_s} + |S|^2 \left[\frac{1-y_s\cos(2\varphi_S -  2\varphi_{B_s})}{1+y_s}\right].
\end{align}
This scenario corresponds to scenario B in \cite{Buras:2013uqa}.

{\bf Mixed Scenario:}

We will consider a scenarios in which $P$ is modified from its SM value and
$S$ is non-zero. As we want to discuss the case of a single new particle 
with spin 0, this means that this particle
has both scalar and pseudoscalar couplings to muons.

A simple scenario with  both  a scalar ($H^0$) and pseudoscalar ($A^0$)
with approximately the same mass that couple equally to quarks and leptons up to the usual
{\it i} factor in the pseudoscalar coupling has been recently considered 
in \cite{Buras:2013uqa}.
This scenario can be realized as a special limit
in models like 2HDM and the MSSM with interesting consequences for
$\mathcal{A}^{\mu\mu}_{\Delta\Gamma}$. We refer to  \cite{Buras:2013uqa} 
for details.

\boldmath
\section{Rare $K$ Decays}\label{RareK}
\unboldmath

\boldmath
\subsection{Effective Hamiltonian for $d\to s\ell^+\ell^-$}\label{sec:dsll}
\unboldmath
For the study of $K_L\to\mu^+\mu^-$ and $K_L\to\pi^0\ell⁺\ell^-$ decays
we will need  the relevant effective Hamiltonian. It can be obtained
from the formulae of subsection~\ref{sec:bqll}. For completeness we
list here explicit formulae for operators and Wilson coefficients:
\begin{subequations}
\begin{align}
Q_9 & = (\bar s\gamma_\mu P_L d)(\bar \ell\gamma^\mu\ell),&  &Q_9^\prime  =  (\bar s\gamma_\mu P_R d)(\bar \ell\gamma^\mu\ell),\\
Q_{10} & = (\bar s\gamma_\mu P_L d)(\bar \ell\gamma^\mu\gamma_5\ell),&  &Q_{10}^\prime  =  (\bar s\gamma_\mu P_R d)(\bar \ell\gamma^\mu\gamma_5\ell),\\
Q_S &= m_s(\bar s P_L d)(\bar \ell\ell),& & Q_S^\prime = m_s(\bar s P_R d)(\bar \ell\ell),\\
Q_P & = m_s(\bar s P_L d)(\bar \ell\gamma_5\ell),& & Q_P^\prime = m_s(\bar s P_R d)(\bar \ell\gamma_5\ell).
\end{align}
\end{subequations}
 Note that because of the $sd$ ordering instead of $qb$ scalar operators have
$L$ and $R$ interchanged with respect to $b\to s,d$ transitions.

The  Wilson coefficients $C_9$ and $C_{10}$  do not receive any new
contributions from scalar exchange and take SM values as given in
(\ref{C9SM}). However,
in order to include charm component in $K_L\to \mu^+\mu^-$ we make replacement:
\be\label{tc}
\eta_Y Y_0(x_t) \longrightarrow \eta_Y Y_0(x_t)+
\frac{V_{cs}^* V_{cd}}{V_{ts}^* V_{td}} Y_{\rm NNL}
\ee
where at NNLO \cite{Gorbahn:2006bm}
\be
 Y_{\rm NNL}=\lambda^4 P_c(Y), \qquad P_c(Y)=0.113\pm 0.017~.
\ee

The coefficients of scalar operators are:
\begin{align}
 m_s\sin^2\theta_W C_S &= \frac{1}{g_{\text{SM}}^2}\frac{1}{ M_H^2}\frac{\Delta_L^{sd}(H)\Delta_S^{\mu\bar\mu}(H)}{V_{ts}^* V_{td}},\\
 m_s\sin^2\theta_W C_S^\prime &= \frac{1}{g_{\text{SM}}^2}\frac{1}{ M_H^2}\frac{\Delta_R^{sd}(H)\Delta_S^{\mu\bar\mu}(H)}{V_{ts}^* V_{td}},\\
 m_s\sin^2\theta_W C_P &= \frac{1}{g_{\text{SM}}^2}\frac{1}{ M_H^2}\frac{\Delta_L^{sd}(H)\Delta_P^{\mu\bar\mu}(H)}{V_{ts}^* V_{td}},\\
 m_s\sin^2\theta_W C_P^\prime &= \frac{1}{g_{\text{SM}}^2}\frac{1}{ M_H^2}\frac{\Delta_R^{sd}(H)\Delta_P^{\mu\bar\mu}(H)}{V_{ts}^* V_{td}}.
\end{align}

\boldmath
\subsection{$K_L\to\mu^+\mu^-$}\label{sec:KLmumu}
\unboldmath
Only the so-called short distance (SD)
part to a dispersive contribution
to $K_L\to\mu^+\mu^-$ can be reliably calculated. Therefore in what follows
this decay will be treated only as an additional constraint to be sure
that the rough upper bound given below is not violated.

The relevant branching ratio can be obtained by first introducing:
\be\label{PPK}
\hat P(K)\equiv {C_{10}-C_{10}^\prime}+
\frac{m^2_{K}}{2m_\mu}\frac{m_s}{m_d+m_s} (C_P-C_P^\prime)
\ee
\be\label{SSK}
\hat S(K)\equiv \sqrt{1-\frac{4m_\mu^2}{m_{K}^2}}\frac{m^2_{K}}{2m_\mu}
\frac{m_s}{m_d+m_s} (C_S-C_S^\prime)
\ee
We then find
\begin{align}\label{BRKLmumu}
 \mathcal{B}(K_L\to\mu^+\mu^-)_{\rm SD} =& \frac{G_F^4 M_W^4}{4 \pi^5}F_{K}^2 m_{K}
\tau_{K_L} m_\mu^2\sqrt{1-\frac{4m_\mu^2}{m_{K}^2}}\sin^4\theta_W \\
&\times\left\{\left[ \Re\left(V_{ts}^* V_{td} \hat P\right)\right]^2 +\left[\Im\left(V_{ts}^* V_{td} \hat S\right)\right]^2\right\} \nonumber
\end{align}
 and $\Re \leftrightarrow \Im$ for $K_S\to\mu^+\mu^-$ decay.
We recall that $C_{10}$ does not receives any contribution from
scalar exchanges  and
includes also SM charm contribution as given in (\ref{tc}). $C_{10}^\prime=0$
for scalar exchanges.

Equivalently we can write
\begin{equation}\label{BRKLmumu1}
 \mathcal{B}(K_L\to\mu^+\mu^-)_{\rm SD} =\kappa_\mu
\left\{\left[ \Re\left(V_{ts}^* V_{td} \hat P\right)\right]^2 +\left[\Im\left(V_{ts}^* V_{td} \hat S\right)\right]^2\right\},
\end{equation}
where
\be
\kappa_\mu=\frac{\alpha^2 \mathcal{B}(K^+\to\mu^+\nu)}{\lambda^2\pi^2}
\frac{\tau(K_L)}{\tau(K^+)}.
\ee

The extraction of the short distance
part from the data is subject to considerable uncertainties. The most recent
estimate gives \cite{Isidori:2003ts}
\be\label{eq:KLmm-bound}
\mathcal{B}(K_L\to\mu^+\mu^-)_{\rm SD} \le 2.5 \cdot 10^{-9}\,,
\ee
to be compared with $(0.8\pm0.1)\cdot 10^{-9}$ in the SM
\cite{Gorbahn:2006bm}.

\boldmath
\subsection{$K_L\to\pi^0\ell^+\ell^-$}
\unboldmath
The rare decays $K_L\to\pi^0e^+e^-$ and $K_L\to\pi^0\mu^+\mu^-$ are dominated
by CP-violating contributions. The indirect CP-violating
contributions are determined by the measured decays
$K_S\to\pi^0 \ell^+\ell^-$ and the parameter $\varepsilon_K$ in
a model independent manner. It is the dominant contribution within the SM
where one finds
\cite{Mescia:2006jd}
\begin{gather}
\mathcal{B}(K_L\to\pi^0e^+e^-)_\text{SM}=
3.54^{+0.98}_{-0.85}\left(1.56^{+0.62}_{-0.49}\right)\cdot 10^{-11}\,,\label{eq:KLpee}\\
\mathcal{B}(K_L\to\pi^0\mu^+\mu^-)_\text{SM}= 1.41^{+0.28}_{-0.26}\left(0.95^{+0.22}_{-0.21}\right)\cdot
10^{-11}\label{eq:KLpmm}\,,
\end{gather}
with the values in parentheses corresponding to the destructive interference
between directly and indirectly CP-violating contributions.
The last discussion  of the theoretical status of this interference
sign can be found in \cite{Prades:2007ud} where the results of \cite{Isidori:2004rb,Friot:2004yr,Bruno:1992za} are
critically analysed. From this discussion, constructive interference
seems to be  favoured though more work is necessary. In spite of significant
uncertainties in the SM prediction we will  investigate how large the
scalar contributions to these decays are still allowed by present constraints.
To this end we will confine our analysis
to the case of the constructive interference between the directly and
indirectly CP-violating contributions.

The present experimental bounds
\be
\mathcal{B}(K_L\to\pi^0e^+e^-)_\text{exp} <28\cdot10^{-11}\quad\text{\cite{AlaviHarati:2003mr}}\,,\qquad
\mathcal{B}(K_L\to\pi^0\mu^+\mu^-)_\text{exp} <38\cdot10^{-11}\quad\text{\cite{AlaviHarati:2000hs}}\,,
\ee
are still by one order of magnitude larger than the SM predictions, leaving
thereby large room for NP contributions. While in the case of $Z'$ models
large enhancements of  branching ratios were  not possible due
to constraints from data on $\kpn$ \cite{Buras:2012jb}, this constraint is
absent in the case of scalar contributions and it is of interest to see
by how much the branching ratios can be enhanced in the models considered
here still being consistent with all data, in particular with the bound
in (\ref{eq:KLmm-bound}).

In the LHT model the branching
ratios for both decays can be enhanced at most
by a factor of 1.5 \cite{Blanke:2006eb,Blanke:2009am}. Slightly larger
effects are still allowed in Randall-Sundrum models with custodial
protection (RSc) for left-handed couplings \cite{Blanke:2008yr}.
Even larger effects are found if the custodial protection
is absent \cite{Bauer:2009cf}.

Probably the most extensive model independent analysis of decays in question
has been performed in \cite{Mescia:2006jd}, where formulae for branching
ratios for both decays in the presence of new operators have been presented.
These formulae have been already used in \cite{Blanke:2006eb,Blanke:2008ac}
for the LHT model and in \cite{Blanke:2008yr} in the case of RSc.
In the LHT model, where only SM operators are present
the effects
of NP can be compactly summarized by generalization of the
real SM functions $Y_0(x_t)$ and $Z_0(x_t)$ to two complex functions $Y_K$ and
$Z_K$, respectively. As demonstrated in the context of the corresponding
analysis within RSc  \cite{Blanke:2008yr}, also in the presence of RH
currents two complex functions $Y_K$ and $Z_K$
are sufficient to describe jointly the SM and NP contributions.
Consequently the LHT formulae (8.1)--(8.8) of \cite{Blanke:2006eb} with
$Y_K$ and $Z_K$ given in (88) and (89) of \cite{Buras:2012jb} can be used
in the context of tree-level gauge boson exchanges.
The original papers behind these formulae can
be found in
\cite{Buchalla:2003sj,Isidori:2004rb,Friot:2004yr,Mescia:2006jd,Buras:1994qa}.

The case of scalar contributions is more involved.
In order to use the formulae of \cite{Mescia:2006jd} for scalar contributions
we introduce the following quantities:
\be
\omega_{7A}=-\frac{1}{2\pi}\frac{\eta_YY_0(x_t)}{\sin^2\theta_W}
\frac{\Im(\lambda_t^{(K)})}{1.4\cdot 10^{-4}},
\ee
\be
\bar y_P=\frac{y_P+y_P'}{2},\qquad \bar y_S=\frac{y_S+y_S'}{2}
\ee
with $y_i$ related to the Wilson coefficients in the present paper as
follows:
\be\label{conversion}
y_P= -\frac{M_W^2\sin^2\theta_W}{m_l}{V_{ts}^* V_{td}} C_P, \qquad
y_S= -\frac{M_W^2\sin^2\theta_W}{m_l}{V_{ts}^* V_{td}} C_S
\ee
with analogous formulae for primed coefficients. Here $m_l$ stands
for $m_e$ and $m_\mu$ as the authors of \cite{Mescia:2006jd}  anticipating
helicity suppression included these masses already in the effective
Hamiltonian.

Using \cite{Mescia:2006jd} we find then corrections from tree-level
$A^0$ and $H^0$ exchanges to
the branching ratios that should be added directly to SM results in (\ref{eq:KLpee}) and  (\ref{eq:KLpmm}):
\be
\Delta\mathcal{B}^{e^+e^-}_P=\left(1.9~\omega_{7A}\Im(\bar y_P)+0.038~(\Im(\bar y_P))^2\right)
                      \cdot 10^{-17},
\ee

\be
\Delta\mathcal{B}^{\mu^+\mu^-}_P=\left(0.26~\omega_{7A}\Im(\bar y_P)+0.0085~(\Im(\bar y_P))^2\right)
                      \cdot 10^{-12},
\ee

\be
\Delta\mathcal{B}^{e^+e^-}_S=\left(1.5~\Re(\bar y_S)+0.0039~(\Re(\bar y_S))^2\right)
                      \cdot 10^{-16},
\ee

\be
\Delta\mathcal{B}^{\mu^+\mu^-}_S=\left(0.04~\Re(\bar y_S)+0.0041~(\Re(\bar y_S))^2\right)
                      \cdot 10^{-12}.
\ee

Note that in the absence of helicity suppression the large suppression factors
above are canceled by the conversion factors in (\ref{conversion}).

The numerical results for these new  contributions are given in
 Section~\ref{sec:U(2)}.

\section{General Structure of New Physics Contributions}\label{sec:3a}
\subsection{Preliminaries}
We have seen in Section~\ref{sec:2} that the small number of free parameters in each of LHS, RHS, LRS and ALRS scenarios allows to expect definite correlations between  flavour observables in each step of the strategy outlined there. These
expectations will be confirmed through the numerical analysis below but it is instructive to develop first  a qualitative general view on NP
contributions in different scenarios before entering the details.

First, it should be realized that the confrontation of correlations in question
with future precise data will not only depend on the size of theoretical, parametric and experimental uncertainties, but also in an important manner on the
size of allowed deviations from SM expectations. The latter deviations are
presently constrained dominantly by $\Delta F=2$ observables and $B\to X_s\gamma$ decay. But as already demonstrated in \cite{Altmannshofer:2011gn,Altmannshofer:2012ir,Buras:2012dp,Buras:2012jb} with the the new data from the LHCb, ATLAS and CMS at hand
 also the decays $B_{s,d}\to\mu^+\mu^-$ and $b\to s\ell^+ \ell^-$ begin
to play important roles in this context. We will see their impact on our
analysis as well.

Now, in general NP scenarios in which there are many free parameters, it is
possible with the help of some amount of fine-tuning to satisfy constraints
from $\Delta F=2$ processes without a large impact on the size of NP contributions to $\Delta F=1$ processes. However, in the case at hand
in which NP in both $\Delta F=2$ and $\Delta F=1$ processes is governed by
flavour changing tree-diagrams, the situation is different.
Indeed, due to the property of {\it factorization} of decay
amplitudes into vertices and the propagator at the tree-level, the same quark flavour violating couplings and the same mass $M_{H}$ enter  $\Delta F=2$ and $\Delta F=1$ processes undisturbed by the presence of fermions entering the
usual box and penguin diagrams. Let us exhibit these correlations in
explicit terms.
\boldmath
\subsection{$\Delta F=1$ vs. $\Delta F=2$ Correlations}\label{CORR}
\unboldmath
In order to obtain transparent expressions we rewrite various contributions
$[\Delta S(K)]_{\rm AB}$ and $[\Delta S(B_q)]_{\rm AB}$ with $A,B=L,R$ to
$\Delta F=2$ amplitudes as
follows
\be\label{DSKa}
[\Delta S(K)]_{\rm AB}=\frac{ r^\text{AB}(K)}{M_{H}^2}
  \frac{\Delta_A^{sd}(H)\Delta_B^{sd}(H)}{[\lambda_t^{(K)}]^2}
 \ee
\be\label{DSBqa}
[\Delta S(B_q)]_{\rm AB}=\frac{ r^\text{AB}(B_q)}{M_{H}^2}
 \frac{\Delta_A^{bq}(H)\Delta_B^{bq}(H)}{[\lambda_t^{(q)}]^2}
 \ee
where the quantities $r^\text{AB}(M)$ can be found by comparing these expressions with (\ref{DSKSLL}), (\ref{DSKLR}), (\ref{BSLL}) and (\ref{BLR}) and analogous
expressions for the contributions of operators $Q_i^\text{SRR}$.
They depend on low energy
parameters, in particular on the meson system and logarithmically on $M_{H}$. The latter dependence can be
neglected for all practical purposes as long as  $M_{H}$ is above
 several hundreds of GeV and still in the reach of the LHC. We collect the values of $r^\text{AB}(M)$ in Table~\ref{tab:rABM}.

Defining ($i=b,s$)
\be
\tilde C_{S,P}^{(\prime)}=m_i\sin^2\theta_W  C_{S,P}^{(\prime)}
\ee
we can then derive the following relations between the Wilson coefficients
$\tilde C_{S,P}^{(\prime)}$  entering the $\Delta F=1$ processes
and the shifts  $[\Delta S(M)]_{\rm AB}$ in $\Delta F=2$ processes which are independent
of any parameters like $\tilde s_{ij}$ but depend sensitively on $M_{H}$ and on the couplings
$\Delta_{S,P}^{\mu\bar\mu}(H)$\footnote{Similar
relations have been derived in \cite{Buras:2012jb} in the context of
$Z'$ models.}. In particular
they do not depend explicitly on whether S1 or S2 scenarios for $\vub$ defined
below
are considered. This dependence is hidden in the allowed shifts in
$[\Delta S(K)]_{AB}$ and $[\Delta S(B_d)]_{AB}$ both in magnitudes and phases. We have then\footnote{The numerical values on the r.h.s of
these equations correspond to
$M_{H}=1\tev$.}

\be\label{REL1}
\frac{\tilde C_{S,P}(K)}{\sqrt{[\Delta S(K)]_{RR}}}=
\frac{\Delta_{S,P}^{\mu\bar\mu}(H)}{M_{H}g^2_{\rm SM}\sqrt{r^\text{RR}(K)}}=0.18\Delta_{S,P}^{\mu\bar\mu}(H) ,
\ee
\be\label{REL2}
\frac{\tilde C^\prime_{S,P}(K)}{\sqrt{[\Delta S(K)]_{LL}}}=
\frac{\Delta_{S,P}^{\mu\bar\mu}(H)}{M_{H}g^2_{\rm SM}\sqrt{r^\text{LL}(K)}}=0.18\Delta_{S,P}^{\mu\bar\mu}(H),
\ee
\be\label{REL3}
\frac{\tilde C_{S,P}(K)\tilde C^\prime_{S,P}(K)}{[\Delta S(K)]_{LR}}=
\frac{[\Delta_{S,P}^{\mu\bar\mu}(H)]^2}{M^2_{H}g^4_{\rm SM}r^\text{LR}(K)}=-0.005[\Delta_{S,P}^{\mu\bar\mu}(H)]^2.
\ee

For $B_q$ we have to make the following replacements in the formulae
above:
\be
[\Delta S(K)]_{AB}\longrightarrow [\Delta S(B_q)]_{AB}^*, \qquad
r^\text{AB}(K)\longrightarrow r^\text{AB}(B_q)
\ee
\be
\frac{\tilde C_{S,P}(B_q)}{\sqrt{[\Delta S(B_q)]_{RR}^\star}}=
\frac{\Delta_{S,P}^{\mu\bar\mu}(H)}{M_{H}g^2_{\rm SM}\sqrt{r^\text{RR}(B_q)}}=0.78\Delta_{S,P}^{\mu\bar\mu}(H) ,
\ee
\be
\frac{\tilde C^\prime_{S,P}(B_q)}{\sqrt{[\Delta S(B_q)]_{LL}^\star}}=
\frac{\Delta_{S,P}^{\mu\bar\mu}(H)}{M_{H}g^2_{\rm SM}\sqrt{r^\text{LL}(B_q)}}=0.78\Delta_{S,P}^{\mu\bar\mu}(H),
\ee
\be
\frac{\tilde C_{S,P}(B_q)\tilde C^\prime_{S,P}(B_q)}{[\Delta S(B_q)]_{LR}^\star}=
\frac{[\Delta_{S,P}^{\mu\bar\mu}(H)]^2}{M^2_{H}g^4_{\rm SM}r^\text{LR}(B_q)}=-0.1[\Delta_{S,P}^{\mu\bar\mu}(H)]^2.
\ee

\begin{table}[!tb]
\centering
\begin{tabular}{|c||c|c|}
\hline
 $r^\text{AB}(M)$& $LL/RR$ & $LR$  \\
\hline
\hline
  $K$ & $960$& $-5700$\\
$B_d$ & $51$& $-310$\\
$B_s$ & $50$& $-300$\\
\hline
\end{tabular}
\caption{\it $r^\text{AB}(M)$ in units of $\tev^2$ as defined in Eqs.~(\ref{DSKa}) and (\ref{DSBqa}) for $M_H = 1~$TeV.
}\label{tab:rABM}~\\[-2mm]\hrule
\end{table}

\subsection{Implications}
Inspecting these formulae we observe that
if the SM prediction for $\varepsilon_K$ is very close to its
experimental value, $\Delta S(K)$ cannot be large
and consequently at first sight
the values of the Wilson coefficients $C_{S,P}^{(\prime)}(K)$
cannot be large implying
suppressed NP contributions
to rare $K$ decays unless $H$ couplings to charged leptons in the final state are enhanced, although this enhancement can be bounded by rare $B_{s,d}$ decays.
Further  details depend on the value of $M_{H}$.
While in $Z'$ models
the present theoretical and parametric
uncertainties in $\varepsilon_K$ and $\Delta M_K$ still allow for large
effects in rare $K$ decays both in S1 and S2 scenarios, this turns out not to be the case in the models considered here.

Similarly in the $B_d$ and $B_s$ systems if the SM
predictions for $\Delta M_{s,d}$, $S_{\psi K_S}$ and $S_{\psi\phi}$ are very
close to the data, it is unlikely that large NP contributions to rare
$B_d$ and $B_s$ decays, in particular the asymmetries $S^{s,d}_{\mu^+\mu^-}$, will be found, unless again
$H$ couplings to charged leptons in
the final state are enhanced. Here the situation concerning theoretical
and parametric uncertainties is better than in the $K$ system and
the presence of several additional constraints from $b\to s$ transitions
allows to reach in the $B_s$ system clear cut conclusions.

In this context it is fortunate that within the SM there appears  to be a
tension between the values of $\varepsilon_K$ and
$S_{\psi K_S}$ \cite{Lunghi:2008aa,Buras:2008nn} so that some action from
NP is required. Moreover, parallel to this tension,
the values of $\vub$
extracted from inclusive and exclusive decays differ significantly
from each other. For a recent review see
\cite{Ricciardi:2012pf}.

If one does not average the inclusive and exclusive values of $\vub$ and
takes into account the tensions mentioned above, one is lead naturally to
two scenarios for NP:
 \begin{itemize}
\item
{\bf Exclusive (small) $\vub$ Scenario 1:}
$|\varepsilon_K|$ is smaller than its experimental determination,
while $S_{\psi K_S}$ is rather close to the central experimental value.
\item
{\bf Inclusive (large) $\vub$ Scenario 2:}
$|\varepsilon_K|$ is consistent with its experimental determination,
while $S_{\psi K_S}$ is significantly higher than its  experimental value.
\end{itemize}

Thus depending on which scenario is considered, we need either
{\it constructive} NP contributions to $|\varepsilon_K|$
(Scenario 1)
or {\it destructive} NP contributions to  $S_{\psi K_S}$ (Scenario 2).
However this  NP should not spoil the agreement with the data
for $S_{\psi K_S}$ (Scenario 1) and for $|\varepsilon_K|$ (Scenario 2).

While introducing these two scenarios, we should emphasize the following difference between them.
In Scenario 1, the central value of $|\varepsilon_K|$ is visibly smaller than
the very precise data  but the still  significant parametric uncertainty
due to $\vcb^4$ dependence in $|\varepsilon_K|$ and a large uncertainty
in the charm contribution found at the NNLO level in \cite{Brod:2011ty}
does not make this problem as pronounced as this is the case of
Scenario 2, where large $\vub$ implies definitely a value of $S_{\psi K_S}$
that is by $3\sigma$ above the data.

Our previous discussion allows to expect larger NP effects in rare
$B_d$ decays in scenario S2 than in S1. This will be indeed confirmed
by our numerical analysis. In the $K$ system one would expect larger NP
effects in scenario S1 than S2 but the present uncertainties in
$\varepsilon_K$ and $\Delta M_K$ do not allow to see this clearly.
The $B_s$ system is not affected by the choice of these scenarios and in fact
our results in S1 and S2 are basically indistinguishable from each other
as long as there is no correlation with the $B_d$ system. However, we
will demonstrate that the imposition of $U(2)^3$ symmetry on $H$ couplings
will introduce such correlation with interesting implications for the $B_s$
system.

We do not include $B⁺\to\tau^+\nu_\tau$ in this discussion as NP related
to this decay has nothing to do with neutral scalars,
at least at the tree-level.  Moreover, the disagreement
of the data with the SM in this case softened significantly with the new
data from Belle Collaboration \cite{BelleICHEP}. The
new world average provided by the UTfit collaboration of
$\mathcal{B}(B^+ \to \tau^+ \nu)_{\rm exp} = (0.99 \pm 0.25) \times 10^{-4}~$
\cite{Tarantino:2012mq}
is in perfect agreement with the SM in scenario S2 and only by $1.5\sigma$
above the SM value in scenario S1.

Evidently $\vub$ could be some average between the inclusive and exclusive
values, in which significant NP effects will be in principle allowed
simultaneously in $K$ and $B_d$ decays. This is in fact necessary in NP
scenarios in which NP effects to $\Delta F=2$ processes are negligible
and some optimal value for $\vub$, like $0.0037$ is chosen in order to
obtain rough agreement with the data. But then one should hope that
future data, while selecting this value of $\vub$, will also appropriately
imply a higher experimental value of $S_{\psi K_S}$  and new lattice results
will bring  modified non-perturbative
parameters in the remaining $\Delta F=2$ observables so that everything works.
This is the case of a recent analysis of FCNC processes within a model for
quark masses \cite{Buras:2013td}.
This discussion shows importantance of
the determination of the value of $\vub$ and of the non-perturbative parameters
in question (see article by A.~Buras in \cite{Varzielas:2012as}).

As already remarked above, the case of $B_s$ mesons is different as the
$B^0_s-\bar B_s^0$ system is not involved in the tensions discussed above.
Here the visible deviation of the $\Delta M_s$ in the SM from the data and the
asymmetry $S_{\psi\phi}$, still being not  accurately measured, govern the possible size of NP contributions in rare decays.

\boldmath
\subsection{Dependence on $M_{H}$}
\unboldmath
The correlations between $\Delta F=1$ and $\Delta F=2$ derived in subsection \ref{CORR} imply that when free NP parameters have been bounded by $\Delta F=2$
constraints, the Wilson coefficients of scalar operators are {\it inversely } proportional to  $M_{H}$. This means that in the case of NP contributions
significantly smaller than the SM contributions in $P$, the modifications of rare decay branching ratios due to NP will be governed
by the interference of
SM and NP contributions.  Consequently
such contributions to branching ratios will also be inversely  proportional to  $M_{H}$. If NP contribution to $P$ is of the size of the SM contributions than
this law will be modified and NP contributions will decrease faster with
increasing $M_H$. On the other hand in the absence of interference
between NP and SM contributions, as is the case of $S$,
the NP modifications of branching ratios will decrease as
$1/M^2_{H}$. Consequently, we expect that for sufficiently large $M_H$ only
NP contributions in $P$, as in $Z'$ scenarios, will matter unless the scalar
couplings are very much enhanced over pseudoscalar ones. Evidently,
for low values of  $M_H$ the $S$ contributions could be relevant.
Here in principle a SM Higgs, being a scalar, could play a prominent role, but
as we will demonstrate below this can only be the case for $\Delta F=2$
transitions.

While $M_H$ could still be as low as few hundreds of GeV,
in order to cover a large set of models,
we will choose as our nominal value $M_{H}=1\tev$. With the help of
the formulae in subsection \ref{CORR} it should be possible to
estimate approximately, how our results would change for other values of
$M_{H}$. In this context it should be noted that any change of $M_H$ can 
be compensated by the change in couplings $\Delta^{\mu\mu}_{S,P}$ unless 
these couplings are predicted in a given model or are known from other 
measurements.

With this general picture in mind we can now proceed to numerical analysis.

\boldmath
\section{Strategy for Numerical Analysis}\label{sec:4}
\unboldmath
\subsection{Preliminaries}
Similarly to our analyses in \cite{Buras:2012dp,Buras:2012jb} it is not the goal of the next  section to present a full-fledged numerical
analysis of all correlations including present theoretical, parametric and experimental
uncertainties as this would only wash out the effects we want to emphasize.
Yet, these uncertainties will be significantly
reduced in the coming years \cite{Antonelli:2009ws,Bediaga:2012py} and it is of interest to ask how
 the $H$ scenarios considered here
 would face precision flavour data and the reduction
of hadronic and CKM uncertainties. In this respect, as emphasized above,
correlations between
various observables are very important and we would like to exhibit these
correlations by assuming reduced uncertainties in question.

Therefore, in our numerical analysis we will  choose
as nominal values for three out of four CKM parameters:\vspace{1ex}
\be\label{fixed}
\vus=0.2252, \qquad \vcb=0.0406, \qquad \gamma=68^\circ,
\ee
and instead of taking into account their uncertainties directly, we will
take them effectively at a reduced level by increasing the experimental
uncertainties in $\Delta M_{s,d}$ and $\varepsilon_K$.
Here the values for
 $|V_{us}|$ and  $|V_{cb}|$ have been measured
 in tree level
decays. The value for $\gamma$ is consistent with CKM fits and as the
ratio $\Delta M_d/\Delta M_s$ in the SM agrees well with the data, this
choice is a legitimate one.
Other inputs are collected in
Table~\ref{tab:input}. For $\vub$ we will use as two values
\be\label{Vubrange}
|V_{ub}|=3.1\cdot 10^{-3}\qquad |V_{ub}|=4.0\cdot 10^{-3}
\ee
 that are in the ballpark of exclusive and inclusive determinations of
this CKM element
and representing thereby S1 and S2 scenarios, respectively.

\begin{table}[!tbh]
\center{\begin{tabular}{|l|l|}
\hline
$G_F = 1.16637(1)\times 10^{-5}\gev^{-2}$\hfill\cite{Nakamura:2010zzi} 	&  $m_{B_d}= 5279.5(3)\mev$\hfill\cite{Nakamura:2010zzi}\\
$M_W = 80.385(15) \gev$\hfill\cite{Nakamura:2010zzi}  								&	$m_{B_s} =
5366.3(6)\mev$\hfill\cite{Nakamura:2010zzi}\\
$\sin^2\theta_W = 0.23116(13)$\hfill\cite{Nakamura:2010zzi} 				& 	$F_{B_d} =
(188\pm4)\mev$\hfill \cite{Dowdall:2013tga}\\
$\alpha(M_Z) = 1/127.9$\hfill\cite{Nakamura:2010zzi}									& 	$F_{B_s} =
(225\pm3)\mev$\hfill \cite{Dowdall:2013tga}\\
$\alpha_s(M_Z)= 0.1184(7) $\hfill\cite{Nakamura:2010zzi}								&  $\hat B_{B_d} =
1.26(11)$\hfill\cite{Laiho:2009eu}\\\cline{1-1}
$m_u(2\gev)=(2.1\pm0.1)\mev $ 	\hfill\cite{Laiho:2009eu}						&  $\hat B_{B_s} =
1.33(6)$\hfill\cite{Laiho:2009eu}\\
$m_d(2\gev)=(4.73\pm0.12)\mev$	\hfill\cite{Laiho:2009eu}							& $\hat B_{B_s}/\hat B_{B_d}
= 1.05(7)$ \hfill \cite{Laiho:2009eu} \\
$m_s(2\gev)=(93.4\pm1.1) \mev$	\hfill\cite{Laiho:2009eu}				&
$F_{B_d} \sqrt{\hat
B_{B_d}} = 226(13)\mev$\hfill\cite{Laiho:2009eu} \\
$m_c(m_c) = (1.279\pm 0.013) \gev$ \hfill\cite{Chetyrkin:2009fv}					&
$F_{B_s} \sqrt{\hat B_{B_s}} =
279(13)\mev$\hfill\cite{Laiho:2009eu} \\
$m_b(m_b)=4.19^{+0.18}_{-0.06}\gev$\hfill\cite{Nakamura:2010zzi} 			& $\xi =
1.237(32)$\hfill\cite{Laiho:2009eu}
\\
$m_t(m_t) = 163(1)\gev$\hfill\cite{Laiho:2009eu,Allison:2008xk} &  $\eta_B=0.55(1)$\hfill\cite{Buras:1990fn,Urban:1997gw}  \\
$M_t=173.2\pm0.9 \gev$\hfill\cite{Aaltonen:2012ra}						&  $\Delta M_d = 0.507(4)
\,\text{ps}^{-1}$\hfill\cite{Amhis:2012bh}\\\cline{1-1}
$m_K= 497.614(24)\mev$	\hfill\cite{Nakamura:2010zzi}								&  $\Delta M_s = 17.72(4)
\,\text{ps}^{-1}$\hfill\cite{Amhis:2012bh}
\\	
$F_K = 156.1(11)\mev$\hfill\cite{Laiho:2009eu}												&
$S_{\psi K_S}= 0.679(20)$\hfill\cite{Nakamura:2010zzi}\\
$\hat B_K= 0.767(10)$\hfill\cite{Laiho:2009eu}												&
$S_{\psi\phi}= 0.001\pm 0.100$\hfill\cite{Raven:2012fb}\\
$\kappa_\epsilon=0.94(2)$\hfill\cite{Buras:2008nn,Buras:2010pza}				& $\Delta\Gamma_s=0.116\pm 0.019$
\cite{Raven:2012fb}
\\	
$\eta_1=1.87(76)$\hfill\cite{Brod:2011ty}												
	& $\tau(B_s)= 1.503(10)\,\text{ps}$\hfill\cite{Amhis:2012bh}\\		
$\eta_2=0.5765(65)$\hfill\cite{Buras:1990fn}												
& $\tau(B_d)= 1.519(7) \,\text{ps}$\hfill\cite{Amhis:2012bh}\\
$\eta_3= 0.496(47)$\hfill\cite{Brod:2010mj}												
& \\\cline{2-2}
$\Delta M_K= 0.5292(9)\times 10^{-2} \,\text{ps}^{-1}$\hfill\cite{Nakamura:2010zzi}	&
$|V_{us}|=0.2252(9)$\hfill\cite{Nakamura:2010zzi}\\
$|\eps_K|= 2.228(11)\times 10^{-3}$\hfill\cite{Nakamura:2010zzi}					& $|V_{cb}|=(40.9\pm1.1)\times
10^{-3}$\hfill\cite{Beringer:1900zz}\\\cline{1-1}
  $\mathcal{B}(B\to X_s\gamma)=(3.55\pm0.24\pm0.09) \times10^{-4}$\hfill\cite{Nakamura:2010zzi}                                             
                  &
$|V^\text{incl.}_{ub}|=(4.41\pm0.31)\times10^{-3}$\hfill\cite{Beringer:1900zz}\\
$\mathcal{B}(B^+\to\tau^+\nu)=(0.99\pm0.25)\times10^{-4}$\hfill\cite{Tarantino:2012mq}\	&
$|V^\text{excl.}_{ub}|=(3.23\pm0.31)\times10^{-3}$\hfill\cite{Beringer:1900zz}\\\cline{1-1}					
$\tau_{B^\pm}=(1641\pm8)\times10^{-3}\,\text{ps}$\hfill\cite{Amhis:2012bh}
														&
\\

\hline
\end{tabular}  }
\caption {\textit{Values of the experimental and theoretical
    quantities used as input parameters.}}
\label{tab:input}
\end{table}

Having fixed the three parameters of the CKM matrix to the values in (\ref{fixed}), for a given $\vub$  the {``true''} values
of the angle $\beta$  and
of the element $\vtd$
are  obtained from the unitarity of the CKM matrix:
\begin{equation} \label{eq:Rt_beta}
\vtd=\vus \vcb R_t,\quad
R_t=\sqrt{1+R_b^2-2 R_b\cos\gamma} ~,\quad
\cot\beta=\frac{1-R_b\cos\gamma}{R_b\sin\gamma}~,
\end{equation}
where
\be\label{Rb}
 R_b=\left(1-\frac{\lambda^2}{2}\right)\frac{1}{\lambda}\frac{|V_{ub}|}{\vcb}.
\ee

\begin{table}[!tb]
\centering
\begin{tabular}{|c||c|c|c|}
\hline
 & Scenario 1: & Scenario 2:   & Experiment\\
\hline
\hline
  \parbox[0pt][1.6em][c]{0cm}{} $|\varepsilon_K|$ & $1.72(22)  \cdot 10^{-3}$  & $2.15(32)\cdot 10^{-3}$ &$ 2.228(11)\times 10^{-3}$ \\
 \parbox[0pt][1.6em][c]{0cm}{}$(\sin2\beta)_\text{true}$ & 0.623(25) &0.770(23)  & $0.679(20)$\\
 \parbox[0pt][1.6em][c]{0cm}{}$\Delta M_s\, [\text{ps}^{-1}]$ &19.0(21)&  19.0(21) &$17.73(5)$ \\
 \parbox[0pt][1.6em][c]{0cm}{} $\Delta M_d\, [\text{ps}^{-1}]$ &0.56(6) &0.56(6)   &  $0.507(4)$\\
\parbox[0pt][1.6em][c]{0cm}{}$\mathcal{B}(B^+\to \tau^+\nu_\tau)$&  $0.62(14) \cdot 10^{-4}$&$1.02(20)\cdot 10^{-4}$ & $0.99(25) \times
10^{-4}$\\
\hline
\end{tabular}
\caption{\it SM prediction for various observables for  $|V_{ub}|=3.1\cdot 10^{-3}$ and $|V_{ub}|=4.0\cdot 10^{-3}$ and $\gamma =
68^\circ$ compared to experiment.
}\label{tab:SMpred}~\\[-2mm]\hrule
\end{table}

In Table~\ref{tab:SMpred} we
summarize for completeness the SM results for $|\varepsilon_K|$,  $\Delta M_{s,d}$,
$\left(\sin 2\beta\right)_\text{true}$ and $\mathcal{B}(B^+\to \tau^+\nu_\tau)$, obtained from (\ref{eq:Rt_beta}),
setting
$\gamma = 68^\circ$ and  choosing the two values for $\vub$ in (\ref{Vubrange}).
We observe that for both choices of $\vub$ the data show significant deviations from the SM predictions but
the character of the NP which could cure these tensions depends on the
choice of $\vub$ as already discussed in detail in \cite{Buras:2012ts} and in
the previous section.

What is
striking in this table is that
the predicted central values of $\Delta M_s$  and $\Delta M_d$, although
slightly above the data,  are both in  good agreement with the latter
when hadronic uncertainties are taken into account. In particular
the central value of the ratio $\Delta M_s/\Delta M_d$ is
 very close to  the data:
\be\label{Ratio}
\left(\frac{\Delta M_s}{\Delta M_d}\right)_{\rm SM}= 34.5\pm 3.0\qquad {\rm exp:~~ 35.0\pm 0.3}\,.
\ee
These results depend on the lattice input and in the case
of $\Delta M_d$ on the value of $\gamma$. Therefore to get a better insight
both lattice input and the tree level determination of $\gamma$
have to improve.

Similarly to the anatomy of $Z'$ models in \cite{Buras:2012jb}
we will deal
with two scenarios for $\vub$ and four scenarios LHS, RHS, LRS, ALRS
for flavour violating couplings of $H$ to quarks. Thus for a given scalar or
pseudoscalar we will deal with
eight scenarios of flavour violating $H$-physics to be denoted by
\be\label{generalS}
{\rm LHS1,\quad LHS2,\quad RHS1,\quad RHS2,\quad LRS1, \quad
LRS2, \quad ALRS1, \quad ALRS2}
\ee
with S1 and S2 indicating the $\vub$ scenarios. With the help of scalar,
pseudoscalar and mixed scenarios for leptonic couplings introduced in
Subsection~\ref{SPscenarios} in each
case, we will be able to get the full picture of various possibilities.

We should emphasize that in each of the scenarios listed in (\ref{generalS}), except for leptonic couplings,
 we have only two
free parameters describing
the $H$-quark couplings in each meson system except for the universal
$M_{H}$. Therefore, as in the case of $Z'$ models it is possible
to determine these couplings from flavour observables (see Section~\ref{sec:2})
provided flavour conserving $H$ couplings to  muons and $M_H$ are known.
While in the SM and some specific models scalar couplings are known, in the
present analysis  we
want to be more model independent.
While we will get some insight about them from $B_s\to\mu^+\mu^-$,
 determining them in purely leptonic processes increases the predictive
power of the theory.

Following Step 2 of our general strategy of Section~\ref{sec:2}, in what follows we will assume that
$\Delta_P^{\mu\bar\mu}(H)$ and $\Delta_S^{\mu\bar\mu}(H)$ have been determined
in purely leptonic processes. For definiteness
we set the lepton couplings
at the  following values
\be\label{leptonicset}
\tilde\Delta_P^{\mu\bar\mu}(H)=\pm 0.020\frac{m_b(M_H)}{m_b(m_b)}, \qquad
\Delta_S^{\mu\bar\mu}(H)=0.040\frac{m_b(M_H)}{m_b(m_b)}
\ee
with the latter factor being $0.61$ for $M_H=1\tev$. We show this factor 
explicitly to indicate how the correct scale for $m_b$ affects the allowed 
range for the lepton couplings.
As we will demonstrate in the course of our presentation these values are consistent with the allowed range for
$\overline{\mathcal{B}}(B_{s}\to\mu^+\mu^-)$
when the constraints on the quark couplings from $B_s^0-\bar B_s^0$ are
taken into account and $M_H=1\tev$.
The reason for choosing the scalar couplings to be larger than the pseudoscalar
ones is that they are weaker constrained than the latter because the
scalar contributions do not interfere with SM contributions. Note that because of
the lack of this interference, the values of $S$ are simply proportional to
$\Delta_S^{\mu\bar\mu}(H)$ and it is straightforward to obtain $S$ contributions
for different values of this coupling.

These couplings should be compared with SM Higgs couplings
\be 
[\tilde\Delta_P^{\mu\bar\mu}(H)]_{\rm SM}=0, \qquad [\Delta_S^{\mu\bar\mu}(H)]_{\rm SM}=1.2\times 10^{-3}.
\ee
{ As discussed in Section~\ref{sec:ZSM}
the smallness of these couplings precludes any visible SM Higgs effects in 
rare $B_d$ and $K$ decays after the constraints from $\Delta F=2$ processes 
have been taken into account. On the other hand SM Higgs effects
in $B_s\to\mu^+\mu^-$, although significantly smaller than in the case 
of heavy scalars, could enhance the branching ratio up to $8\%$ over the 
SM value and could also be seen in the asymmetry $S^s_{\mu^+\mu^-}$.}

Concerning the signs in (\ref{leptonicset}), the one of $\Delta_S^{\mu\bar\mu}(H)$
is irrelevant as only the square of this coupling enters various observables.
The sign of $\tilde\Delta_P^{\mu\bar\mu}(H)$ has an impact on the interference
of pseudoscalar and SM contributions and
 is thereby crucial for the identification of various enhancements
and suppressions with respect to SM branching ratios and CP asymmetries.
Consequently it plays a role
 of our search for successful oases in the space of parameters.

\subsection{Simplified Analysis}
As in \cite{Buras:2012jb} we will perform a simplified analysis of $\varepsilon_K$,
$\Delta M_{d,s}$, $S_{\psi K_S}$
and $S_{\psi\phi}$
in order to identify oases in the space of new parameters (see Section~\ref{sec:2})
for which these five observables are consistent with experiment.
To this end we set all other input parameters at their central values
but in order to take partially hadronic
and experimental uncertainties into account we require the theory in
each of the eight scenarios in (\ref{generalS})
to reproduce the data for $\varepsilon_K$ within $\pm 10\%$, $\Delta M_{s,d}$ within $\pm 5\%$ and the
data on $S_{\psi K_S}$ and $S_{\psi\phi}$ within experimental
$2\sigma$. We choose larger uncertainty for  $\varepsilon_K$  than  $\Delta M_{s,d}$ because of its strong $\vcb^4$ dependence. For $\Delta M_K$ we will
only require the agreement within $\pm 25\%$ because of potential long
distance uncertainties.

Specifically, our search is governed by the following allowed ranges\footnote{When using the constraint from $S_{\psi\phi}$ we take into
account that only
$B_s$ mixing phase close to its  SM value is allowed thereby removing some discrete ambiguities. The same is done for $S_{\psi
K_S}$.}:
\be\label{C1}
16.9/{\rm ps}\le \Delta M_s\le 18.7/{\rm ps},
\quad  -0.20\le S_{\psi\phi}\le 0.20,
\ee
\be\label{C2}
0.48/{\rm ps}\le \Delta M_d\le 0.53/{\rm ps},\quad
0.64\le S_{\psi K_S}\le 0.72 .
\ee

\be\label{C3}
0.75\le \frac{\Delta M_K}{(\Delta M_K)_{\rm SM}}\le 1.25,\qquad
2.0\times 10^{-3}\le |\varepsilon_K|\le 2.5 \times 10^{-3}.
\ee

The search for these oases in each of the scenarios in (\ref{generalS})
is simplified by the fact that for fixed $M_{H}$  each of the pairs
$(\Delta M_s,S_{\psi\phi})$,
$(\Delta M_d,S_{\psi K_S})$  and $(\Delta M_K,|\varepsilon_K|)$ depend only
on two variables. The fact that in the $K$ system we have only one
powerful constraint at present is rather unfortunate. Moreover, in the
models considered the decays $\kpn$ and $\klpn$ cannot help unless
charged Higgs contributions are considered, which is beyond the scope
of the present paper. While the constraint (\ref{eq:KLmm-bound})
on $K_L\to\mu^+\mu^-$ could have in principle an impact on our search for
oases, we have checked that this is not the case.

In what follows we will first for each scenario identify the allowed
oases. As in the case of $Z'$ models there will be in principle four  oases
allowed by the constraints in (\ref{C1})-(\ref{C2}). However, when
one takes into account that the data imply the phases in
$S_{\psi\phi}$ and $S_{\psi K_S}$ to be  close to the SM phases, only two big oases
are left in each case. Similarly the sign of $\varepsilon_K$ selects two allowed oases. In order to identify the final oasis
we will have to
invoke other observables, which are experimentally only weakly bounded
at present. Yet, our plots
will show that once these observables will be measured precisely one day
not only a unique oasis in the parameter space will be identified but the
specific correlations in this oasis will provide a powerful test of the
flavour violating $H$ scenarios.

\section{An Excursion through $H$ Scenarios}\label{sec:Excursion}
\subsection{The LHS1 and LHS2 Scenarios}\label{LHS12}
\boldmath
\subsubsection{The $B_s$ Meson System}
\unboldmath
We begin the search for the oases with the $B_s$ system as here the choice
of $\vub$ is immaterial and the results for LHS1 and LHS2 scenarios are almost identical. Basically only the asymmetry $S_{\psi\phi}$ within the
SM and $\vts$ are slightly modified because of the unitarity of the CKM
matrix. But this changes  $S_{\psi\phi}$ in the SM from $0.032$ to  $0.042$
and can be neglected.

\begin{figure}[!tb]
\begin{center}
\includegraphics[width=0.45\textwidth] {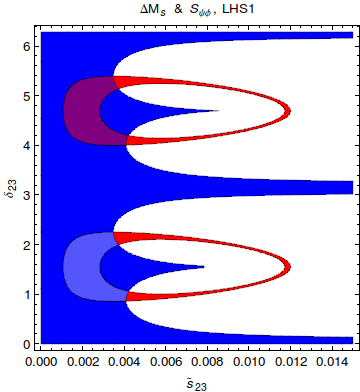}
\caption{\it  Ranges for $\Delta M_s$ (red region) and $S_{\psi \phi}$ (blue region) for $M_{H}=1$~TeV in LHS1 satisfying the bounds
in Eq.~(\ref{C1}).
}\label{fig:oasesBsLHS1}
~\\[-2mm]\hrule
\end{center}
\end{figure}

The result of this search for $M_{H}=1\tev$ is shown in Fig.~\ref{fig:oasesBsLHS1},
where we show the allowed ranges
for $(\tilde s_{23},\delta_{23})$.
The {\it red} regions correspond to the allowed ranges for $\Delta M_{s}$,
while the {\it blue} ones to the corresponding ranges for  $S_{\psi\phi}$. The overlap between red and blue regions identifies the
oases we were looking for. We observe that the requirement of suppression
of $\Delta M_s$ implies $\tilde s_{23}\not=0$.

Comparing Fig.~\ref{fig:oasesBsLHS1} with the corresponding $Z'$ result of Fig.~2 in \cite{Buras:2012jb} we observe that the
phase structure is identical to the one found in the case of $Z'$ but the
values of $\tilde s_{23}$ are smaller. This
behaviour is easy to understand. While the tree diagram with scalar exchange
has the overall sign opposite to the one of a gauge boson exchange, this difference is canceled by the opposite signs of the matrix element
of the leading
operator  $Q_1^\text{SLL}$ and $ Q_1^\text{VLL}$ in the case of $H$ and $Z'$
exchange, respectively. But the absolute value of  $\langle Q_1^\text{SLL}\rangle$  is
larger than of $\langle Q_1^\text{VLL}\rangle$ and consequently
$\tilde s_{23}$ in the Higgs case has to be smaller than in the $Z'$ case
in order to fit data.
We find that this suppression of $\tilde s_{23}$  that enters quadratically in $\Delta M_s$  amounts roughly to a factor of $1.5$.

In view of this simple change we do not show the table for the allowed
ranges for $\delta_{23}$ and $\tilde s_{23}$. They are obtained from Table 5 in
\cite{Buras:2012jb} by leaving  $\delta_{23}$ unchanged and rescaling
$\tilde s_{23}$ down by a factor of $1.5$.

Inspecting Fig.~\ref{fig:oasesBsLHS1} we observe the following pattern:
\begin{itemize}
\item
For each oasis with a given $\delta_{23}$ there is another oasis with $\delta_{23}$ shifted
by $180^\circ$ but the range for  $\tilde s_{23}$ is unchanged. This discrete
ambiguity results from the fact that $\Delta M_s$ and $S_{\psi\phi}$ are
governed by  $2\delta_{23}$. However, as we will see below
this ambiguity can be resolved by other observables.
Without the additional information on phases mentioned in connection with
constraints (\ref{C1})-(\ref{C3}) one would find two additional small oases
corresponding roughly to NP contribution to
$M^s_{12}$ twice as large as the SM one but carrying opposite sign. But taking
these constraints on the phases into account removes these oases from
our analysis.
\item
The increase of $M_{H}$ by a given factor allows to increase
 $\tilde s_{23}$ by the same factor. This structure is evident from the formulae
for $\Delta S(B_s)$. However, the inspection of the formulae for
$\Delta F=1$ transitions shows that this
change will have impact on rare decays, making the NP
effects in them with increased  $M_{H}$ smaller. This is evident
from the correlations derived in Subsection~\ref{CORR}.
\end{itemize}

We will next confine our
numerical analysis to these oases, investigating whether some of them can be
excluded by other constraints and studying correlations between various
observables. To this end we  consider in parallel {\it pseudoscalar} and
{\it scalar} scenarios setting the lepton couplings as given in
(\ref{leptonicset}). In addition to the general case corresponding to the
oases just discussed we will present in plots the results obtained when the
$U(2)^3$ symmetry is imposed on $B_s$ and $B_d$ systems. This case will be discussed in detail at the end  of this subsection but to avoid too many plots and to show
the impact of this symmetry we will already include the results in discussing
the results without this symmetry. Our colour coding will be as follows:
\begin{itemize}
\item
In the general case {\it blue} and {\it purple} allowed regions correspond to
oases with small and
large $\delta_{23}$, respectively. However, one should keep in mind the
next comment.
\item
In the $U(2)^3$ symmetry case, the allowed region will be in {\it magenta} and
and {\it cyan} for LHS1 and LHS2, respectively, as in this case even in the
$B_s$ system there is dependence on $\vub$ scenario.
These  regions are subregions of the general  blue or purple regions so that
they  cover some parts of them.
\end{itemize}

\begin{figure}[!tb]
\centering
\includegraphics[width = 0.45\textwidth]{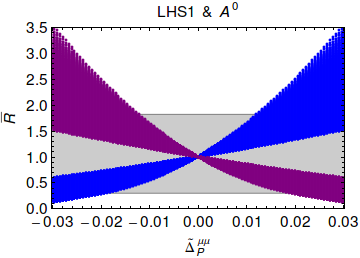}
\includegraphics[width = 0.46\textwidth]{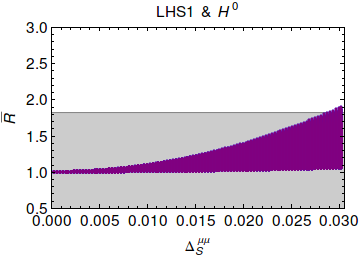}
\caption{\it  $\overline{R}$ as a function of $\tilde\Delta_P^{\mu\bar\mu}$ and $\Delta_S^{\mu\bar\mu}$ in LHS1. Left: $A^0$ case; right:
$H^0$ case.
Gray region: 95\% CL of $R$.}
 \label{fig:RvsDel}~\\[-2mm]\hrule
\end{figure}

In order to justify the values for the leptonic couplings in
(\ref{leptonicset}) we show in
 Fig.~\ref{fig:RvsDel}  $\overline{R}$ as  function of $\tilde\Delta_P^{\mu\bar\mu}$ and $\Delta_S^{\mu\bar\mu}$  in LHS1 for the
pseudoscalar
and scalar scenario, respectively. In ALRS effects are smaller and in LRS $\overline{R}$ does not depend on $\tilde\Delta_P^{\mu\bar\mu}$
and
$\Delta_S^{\mu\bar\mu}$. We observe
that for equal scalar and pseudoscalar couplings, the effects are significantly larger in the $A^0$ case and this is the reason why we have
chosen the scalar couplings to be larger.

 There are two striking differences between $A^0$ and $H^0$ cases originating in the fact that pseudoscalar contributions interfere with
the SM contribution, while this is not the case for a scalar:
\begin{itemize}
\item
While in the $H^0$ case $\overline{R}$ can
only be enhanced, it can also be suppressed in the $A^0$ case. This
difference could play an important role one day.
\item
In the $A^0$ case the result depends on the oasis considered and the sign
of $\tilde\Delta_P$. However changing simultaneously the sign of
 $\tilde\Delta_P$ and the oasis leaves $\overline{R}$ invariant. In the $H^0$ case
$\overline{R}$ is independent of the oasis considered and of the sign of  $\tilde\Delta_S$.
\end{itemize}

\begin{figure}[!tb]
\centering
\includegraphics[width = 0.45\textwidth]{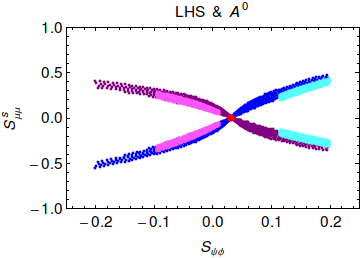}
\includegraphics[width = 0.45\textwidth]{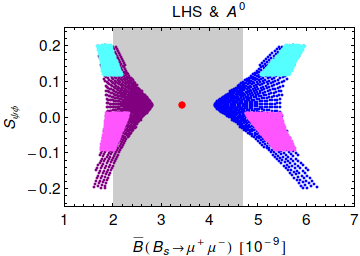}
\caption{\it$ S^s_{\mu^+\mu^-}$  versus $S_{\psi\phi}$ (left) and $S_{\psi\phi}$ versus $\overline{\mathcal{B}}(B_s\to\mu^+\mu^-)$
(right)
for
$M_{H} = 1~$TeV in LHS for two oases as explained in the text and $A^0$ case. The blue and purple regions are
almost identical for LHS1 and LHS2. The magenta region corresponds to the $U(2)^3$ limit for LHS1 and the cyan region for LHS2
(see Sect.~\ref{sec:U23}). Gray region: exp 1$\sigma$ range
$\overline{\mathcal{B}}(B_s\to\mu^+\mu^-) = (3.2^{+1.5}_{-1.2})\cdot 10^{-9}$.  Red point: SM central value.}
 \label{fig:SmusvsSphiLHS1P}~\\[-2mm]\hrule
\end{figure}

In Fig.~\ref{fig:SmusvsSphiLHS1P} (left) we show $S^s_{\mu^+\mu^-}$  vs $S_{\psi\phi}$ in the $A^0$ case.
In the same figure (right) we show the correlation between $\overline{\mathcal{B}}(B_{s}\to\mu^+\mu^-)$
and
$S_{\psi\phi}$.\footnote{The central values for
$\mathcal{B}(B_{d}\to\mu^+\mu^-)_{\rm SM}=1.0\times 10^{-10}$ and
$\overline{\mathcal{B}}(B_{s}\to\mu^+\mu^-)_{\rm SM}=3.45\times 10^{-9}$ shown in the plots
correspond to fixed CKM parameters chosen by us and differ from the ones
listed in~(\ref{LHCb3}) and (\ref{FleischerSM}) but are fully consistent
with them.}
We observe that for largest allowed values of  $S_{\psi\phi}$ the asymmetry
 $S^s_{\mu^+\mu^-}$ can be as large as $\pm 0.5$. Also the effects in 
$\overline{\mathcal{B}}(B_{s}\to\mu^+\mu^-)$ 
are expected to be sizable for the chosen
value of muon coupling.

Comparing the plots in  Fig.~\ref{fig:SmusvsSphiLHS1P} with the corresponding
results for $Z'$ in Fig.~3 of \cite{Buras:2012jb} we observe striking differences which allow to distinguish the case of tree-level pseudoscalar exchange
from the heavy gauge boson exchange:
\begin{itemize}
\item
In the $A^0$ case the asymmetry  $S^s_{\mu^+\mu^-}$ can be zero while this
 was not the case in the $Z'$ case where
the
requirement of suppression of $\Delta M_s$ directly translated in
$ S^s_{\mu^+\mu^-}$  being non-zero. Consequently in the $Z'$ case the sign of
$ S^s_{\mu^+\mu^-}$  could be used to identify the right oasis. The left plot
in  Fig.~\ref{fig:SmusvsSphiLHS1P} clearly shows that this is not possible
in the $A^0$ case. We also find that while in the $Z'$ case the
asymmetry $ S^s_{\mu^+\mu^-}$ could reach values as high as $\pm 0.9$, in
the $A^0$ case $|S^s_{\mu^+\mu^-}|$ can hardly be larger than 0.5.
\item
On the other hand we observe that in the $A^0$ case
the measurement of $\overline{\mathcal{B}}(B_{s}\to\mu^+\mu^-)$
uniquely chooses the
right oasis. The enhancement of this branching ratio relatively to the
SM chooses the blue oasis
while the suppression 
the purple one. This was not possible in the $Z'$ case. The maximal enhancements and suppressions are comparable in both cases but finding 
$\overline{\mathcal{B}}(B_{s}\to\mu^+\mu^-)$
close to SM value would require in the $A^0$ case either larger
$M_H$ or smaller muon coupling.
\end{itemize}

We observe that the roles of $S^s_{\mu^+\mu^-}$ and $\overline{\mathcal{B}}(B_{s}\to\mu^+\mu^-)$
in searching for optimal oasis have been interchanged
when going from the $Z'$ case to the $A^0$ case. While $\overline{\mathcal{B}}(B_{s}\to\mu^+\mu^-)$ 
identifies the
oasis the correlation of $S^s_{\mu^+\mu^-}$ vs. $S_{\psi\phi}$ constitutes
an important test of the model.
While in the blue
oasis
$S_{\psi\phi}$ increases (decreases) uniquely with increasing (decreasing)
 $S^s_{\mu^+\mu^-}$ , in the purple oasis, the increase of
$S_{\psi\phi}$ implies uniquely a decrease of  $S^s_{\mu^+\mu^-}$.
Therefore, while  $S^s_{\mu^+\mu^-}$ alone cannot uniquely
determine the optimal oasis, it can do in collaboration with $S_{\psi\phi}$.

If the favoured oasis will be found to  differ from the one found by means
of  $\overline{\mathcal{B}}(B_s\to\mu^+\mu^-)$ one day
the
 model in question will be in trouble. Indeed, let us
assume that  $\overline{\mathcal{B}}(B_s\to\mu^+\mu^-)$ will be found above its SM
value selecting  thereby blue oasis. Then the measurement of  $S_{\psi\phi}$
will uniquely predict
the sign of $S^s_{\mu^+\mu^-}$. Moreover, in the
case of $S^s_{\psi\phi}$ sufficiently different from zero, we will be able
to determine not only the sign but also the magnitude of  $S^s_{\mu^+\mu^-}$.

These striking differences between the $A^0$-scenario and $Z'$-scenario
can be traced back to the difference between the phase of the NP correction
to $P$ in these two NP scenarios. As the oasis structure as far as the phase
$\delta_{23}$ is concerned is the same in both scenarios the difference enters
through the muon couplings which are imaginary in the case of $A^0$-scenario but
real in the case of $Z'$.  This is in fact the main reason why the structure
of correlations in both scenarios is so different. Taking in addition into
account the sign difference between $Z'$ and pseudoscalar propagator in the
the $b\to s \mu^+\mu^-$ amplitude, which is now not compensated by a hadronic
matrix element, we find that
\be\label{PP1}
P(Z')=1+ r_{Z'} e^{i \delta_{Z'}}, \qquad P(A^0)=1 +  r_{A^0} e^{i\delta_{A^0}}
\ee
with
\be\label{SHIFT}
 r_{Z'}\approx  r_{A^0}, \qquad    \delta_{Z'}=\delta_{23}-\beta_s, \qquad \delta_{A^0}=\delta_{Z'}-\frac{\pi}{2}.
\ee

Therefore with $\delta_{23}$ of Fig.~\ref{fig:oasesBsLHS1} the phase
$\delta_{Z'}$ is around $90^\circ$ and
$270^\circ$ for the blue and purple oasis, respectively. Correspondingly
$\delta_{A^0}$ is around
$0^\circ$ and $180^0$. This
difference in the phases is at the origin of the differences listed above.
In particular, we understand now why the CP asymmetry $ S^s_{\mu^+\mu^-}$
can vanish in the $A^0$ case, while it was always different from zero in the $Z'$-case.
What is interesting is that this difference is just related to the different
particle exchanged: gauge boson and pseudoscalar. We summarize the ranges of
 $\delta_{Z'}$ and  $\delta_{A^0}$ in Table~\ref{tab:PZ}.

\begin{table}[!tb]
\centering
\begin{tabular}{|c||c|c|c|}
\hline
 Oasis  & $\delta_{Z'}$ & $\delta_{A^0}$   \\
\hline
\hline
  \parbox[0pt][1.6em][c]{0cm}{} $B_s$ (blue) & $50^\circ-130^\circ$  & $-40^\circ-(+40^\circ)$   \\
 \parbox[0pt][1.6em][c]{0cm}{}$B_s$ (purple) & $230^\circ-310^\circ$ & $140^\circ-220^\circ$  \\
\hline
 \parbox[0pt][1.6em][c]{0cm}{}$B_d$ (S1) (yellow) & $57^\circ-86^\circ$&  $-33^\circ-(+4^\circ)$   \\
 \parbox[0pt][1.6em][c]{0cm}{} $B_d$ (S1) (green) & $237^\circ-266^\circ$ & $147^\circ-176^\circ$   \\
\parbox[0pt][1.6em][c]{0cm}{}$B_d$ (S2) (yellow) & $103^\circ-125^\circ$&  $13^\circ-35^\circ$   \\
 \parbox[0pt][1.6em][c]{0cm}{} $B_d$ (S2) (green) & $283^\circ-305^\circ$ & $193^\circ-215^\circ$   \\
\hline
 \parbox[0pt][1.6em][c]{0cm}{}$U(2)^3$ (S1) (blue, magenta) & $55^\circ-84^\circ$&  $-35^\circ-(-6^\circ)$   \\
 \parbox[0pt][1.6em][c]{0cm}{} $U(2)^3$ (S1) (purple, magenta) & $235^\circ-264^\circ$ & $145^\circ-174^\circ$   \\
\parbox[0pt][1.6em][c]{0cm}{} $U(2)^3$ (S2) (blue, cyan) & $101^\circ-121^\circ$&  $11^\circ-31^\circ$  \\
 \parbox[0pt][1.6em][c]{0cm}{} $U(2)^3$ (S2) (purple, cyan) & $291^\circ-301^\circ$ & $201^\circ-211^\circ$   \\
\hline
\end{tabular}
\caption{\it Ranges for the values of $\delta_{Z'}$ and $\delta_{A^0}$ as defined in (\ref{PP1}) for the $B_s$ and $B_d$ systems
and various
cases discussed in the text. Also the result for $U(2)^3$ models is shown.  }\label{tab:PZ}~\\[-2mm]\hrule
\end{table}

\begin{figure}[!tb]
\centering
\includegraphics[width = 0.45\textwidth]{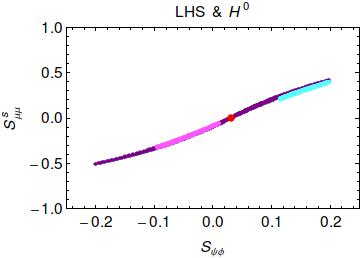}
\includegraphics[width = 0.45\textwidth]{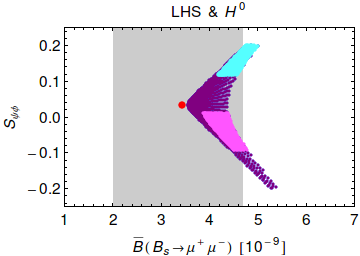}
\caption{\it$ S^s_{\mu^+\mu^-}$  versus $S_{\psi\phi}$ (left) and $S_{\psi\phi}$ versus $\overline{\mathcal{B}}(B_s\to\mu^+\mu^-)$
(right)
for
$M_{H} = 1~$TeV in LHS1 for two oases as explained in the text and $H^0$ case (the two oases overlap here).The magenta
region corresponds to the $U(2)^3$ limit for LHS1 and the cyan region for LHS2.  Gray region: exp 1$\sigma$
range
$\overline{\mathcal{B}}(B_s\to\mu^+\mu^-) = (3.2^{+1.5}_{-1.2})\cdot 10^{-9}$.  Red point: SM central value.}
 \label{fig:SmusvsSphiLHS1S}~\\[-2mm]\hrule
\end{figure}

The power of the correlations in question in distinguishing between various
scenarios is further demonstrated  when we consider the
case of a scalar in which there is no interference with the SM contribution.
 In Fig.~\ref{fig:SmusvsSphiLHS1S} we show the corresponding results in the
 $H^0$ case. We observe the following differences with respect to
Fig.~\ref{fig:SmusvsSphiLHS1P}:
\begin{itemize}
\item
$\overline{\mathcal{B}}(B_{s}\to\mu^+\mu^-)$
 can only be enhanced in this scenario and
this result is independent of the oasis considered. Thus
finding this branching ratio below its SM value would favour the
pseudoscalar scenario over scalar one. But the enhancement is not as
pronounced as in the pseudoscalar case because the correction to the
branching ratio is governed here by the square of the muon coupling while
in the pseudoscalar case the correction was proportional to this coupling
due to the interference with the SM contribution which is absent here.
\item
Concerning CP-asymmetries similarly to the branching ratio there is no dependence on the oasis considered but more importantly
$S^s_{\mu^+\mu^-}$  can
only increase with increasing $S_{\psi\phi}$.
\end{itemize}

It is instructive to understand better the results in the scalar scenario.
Inspecting the formulae for the Wilson coefficients we arrive at an important
relation:
\be\label{SZP}
\varphi_S=\delta_{Z'}-\pi,
\ee
where the shift is related to the minus sign difference in the $Z'$ and scalar
propagators.

But as seen in (\ref{SB}) the three observables given there, all depend
on $2\varphi_S$, implying that from the point of view of these quantities
this shift is irrelevant. As different oases correspond to phases shifted by
$\pi$ this also explains  why in the scalar case the results in different
oases are the same. That the branching ratio can only be enhanced follows
just from the absence of the interference with the SM contributions. In order
to understand the signs in $S_{\mu\mu}^s$ one should note the minus sign in
front of sine in the corresponding formula. Rest follows from (\ref{SZP})
and Table~\ref{tab:PZ}.

\begin{figure}[!tb]
\centering
\includegraphics[width = 0.45\textwidth]{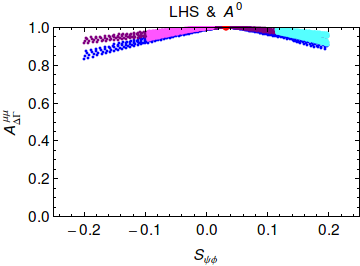}
\includegraphics[width = 0.45\textwidth]{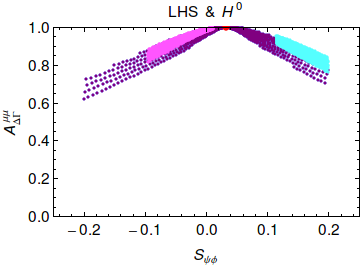}
\caption{\it $\mathcal{A}^\lambda_{\Delta\Gamma}$ versus $S_{\psi\phi}$ for
$M_{H} = 1~$TeV in LHS1, blue and purple oases and $A^0$ case (left) and
$H^0$ case (right). In $H^0$ case the two oases overlap.  The magenta
region corresponds to the $U(2)^3$ limit for LHS1 and the cyan region for LHS2. Red point: SM central value.}
 \label{fig:AGammavsSphiLHS1}~\\[-2mm]\hrule
\end{figure}

In Fig.~\ref{fig:AGammavsSphiLHS1}  we plot $\mathcal{A}^\lambda_{\Delta\Gamma}$ vs  $S_{\psi\phi}$ for $A^0$ and $H^0$ cases.
We observe that
for  $M_{H}=1\tev$, even for $S_{\psi\phi}$ significantly different from
zero, $\mathcal{A}^\lambda_{\Delta\Gamma}$ does not defer significantly
 from unity in both scenarios. Larger effects have been found in the
$Z'$ case as seen Fig.~4 of \cite{Buras:2012jb}.

\begin{figure}[!tb]
\centering
\includegraphics[width = 0.45\textwidth]{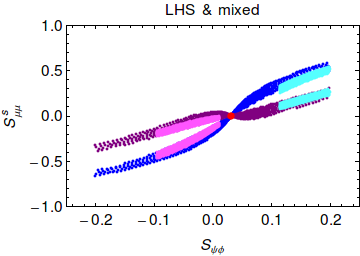}
\includegraphics[width = 0.45\textwidth]{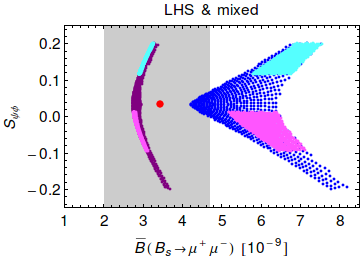}
\caption{\it$ S^s_{\mu^+\mu^-}$  versus $S_{\psi\phi}$ (left) and $S_{\psi\phi}$ versus $\overline{\mathcal{B}}(B_s\to\mu^+\mu^-)$
(right)
for
$M_{H} = 1~$TeV in LHS1 for two oases and the mixed $H^0$ and $A^0$ case with $\Delta_S^{\mu\bar\mu}=
2\tilde\Delta_P^{\mu\bar\mu}$.  The magenta
region corresponds to the $U(2)^3$ limit for LHS1 and the cyan region for LHS2. Gray
region: exp 1$\sigma$
range
$\overline{\mathcal{B}}(B_s\to\mu^+\mu^-) = (3.2^{+1.5}_{-1.2})\cdot 10^{-9}$.  Red point: SM central value.}
 \label{fig:SmusvsSphiLHS1mix}~\\[-2mm]\hrule
\end{figure}

In Fig.~\ref{fig:SmusvsSphiLHS1mix} we show how the plots in
Figs.~\ref{fig:SmusvsSphiLHS1P} and \ref{fig:SmusvsSphiLHS1S} change
when the exchanged particle has both scalar and pseudoscalar couplings to muons
with 
\be
\Delta_S^{\mu\bar\mu}=
2\tilde\Delta_P^{\mu\bar\mu}=0.4 \frac{m_b(M_H)}{m_b(m_b)}.
\ee
We observe that while the correlation between $ S^s_{\mu^+\mu^-}$  and $S_{\psi\phi}$ is relative to $A^0$ case practically
unmodified, the correlation between
 $S_{\psi\phi}$ and $\overline{\mathcal{B}}(B_{s}\to\mu^+\mu^-)$
is visibly modified
for $\overline{\mathcal{B}}(B_{s}\to\mu^+\mu^-)$  below the SM value while less if an enhancement is present. 

\begin{figure}[!tb]
\centering
\includegraphics[width = 0.45\textwidth]{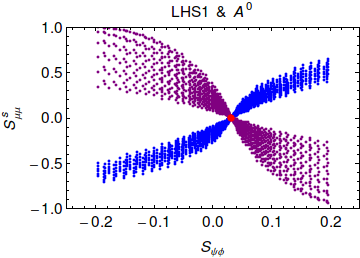}
\includegraphics[width = 0.45\textwidth]{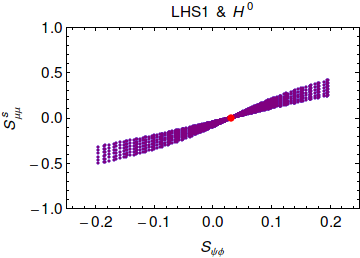}

\includegraphics[width = 0.45\textwidth]{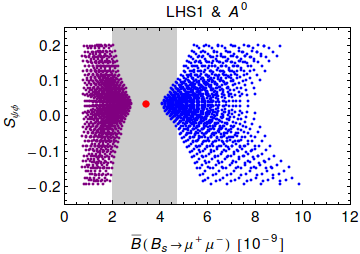}
\includegraphics[width = 0.45\textwidth]{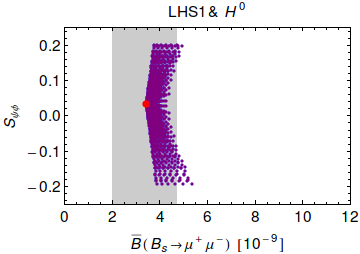}

\includegraphics[width = 0.45\textwidth]{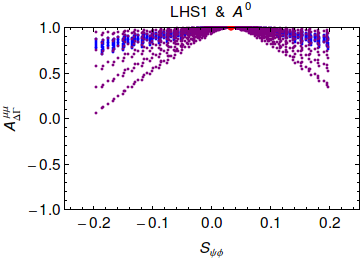}
\includegraphics[width = 0.45\textwidth]{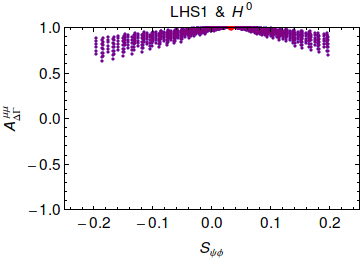}

 \caption{\it Correlation plots as in Fig.~\ref{fig:SmusvsSphiLHS1P}, \ref{fig:SmusvsSphiLHS1S},
and \ref{fig:AGammavsSphiLHS1} but now with lepton coupling $\tilde
\Delta_P\in[0.02,0.04]$, $\Delta_S = 0$
(left) and $\Delta_S\in[0.02,0.04]$, $\tilde
\Delta_P = 0$ (right) in LHS1. Gray
 region: exp 1$\sigma$
 range
 $\overline{\mathcal{B}}(B_s\to\mu^+\mu^-) = (3.2^{+1.5}_{-1.2})\cdot 10^{-9}$.  Red point: SM central value.}
  \label{fig:PandSscan}~\\[-2mm]\hrule
\end{figure}

\begin{figure}[!tb]
\centering

\includegraphics[width = 0.45\textwidth]{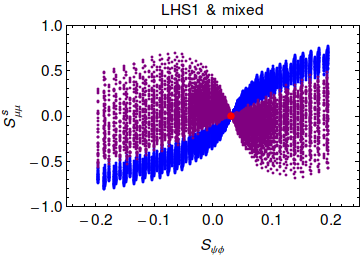}
\includegraphics[width = 0.45\textwidth]{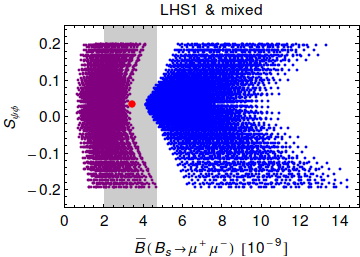}
 \caption{\it Correlation plots as in Fig.~\ref{fig:SmusvsSphiLHS1mix} but now
with $\tilde
\Delta_P\in[0.02,0.04]$ and $\Delta_S\in[0.02,0.04]$. Gray
 region: exp 1$\sigma$
 range
 $\overline{\mathcal{B}}(B_{s}\to\mu^+\mu^-) = (3.2^{+1.5}_{-1.2})\cdot 10^{-9}$.  Red point: SM central value.}
  \label{fig:mixscan}~\\[-2mm]\hrule
\end{figure}

 Clearly for a fixed $M_H$ the results presented so far  depend on the 
choice on muon couplings made by us. 
In Figs.~\ref{fig:PandSscan} and \ref{fig:mixscan} we show the corresponding 
plots when  the lepton couplings  $\tilde\Delta_P$ and $\Delta_S$ are varied independently in the range $0.02-0.04$. Evidently,
the allowed regions are now larger 
but the general pattern of correlations remains.
These results are presented here only for
illustration and we will not discuss this mixed scenario for other meson
systems.

\boldmath
\subsubsection{The $B_d$ Meson System}
\unboldmath

\begin{figure}[!tb]
\begin{center}
\includegraphics[width=0.45\textwidth] {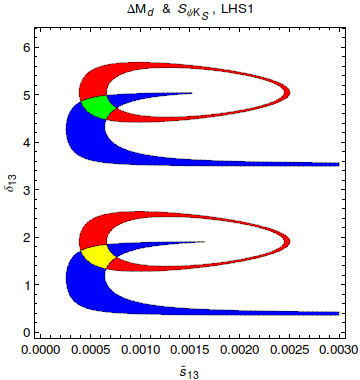}
\includegraphics[width=0.45\textwidth] {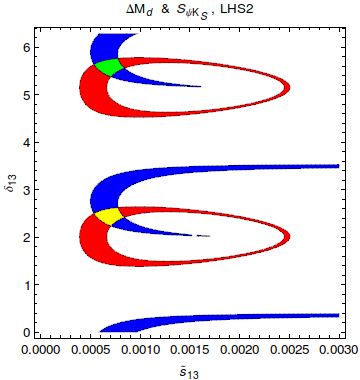}
\caption{\it  Ranges for $\Delta M_d$ (red region) and $S_{\psi K_S}$ (blue region) for $M_{H}=1$ TeV in LHS1 (left) and
LHS2 (right) satisfying the bounds in Eq.~(\ref{C2}).
}\label{fig:oasesBdLHS}~\\[-2mm]\hrule
\end{center}
\end{figure}

We begin by searching for the allowed oases in this case. The result is shown
in Fig.~\ref{fig:oasesBdLHS}. The general structure
of the discrete
ambiguities is as in the $B_s$ case but now as expected the selected
oases in S1 and S2 differ significantly from each other. In fact this figure
has the same phase structure as Fig.~6 in \cite{Buras:2012jb} except that
the allowed values of $\tilde s_{13}$ are reduced with respect to the $Z'$
case for the same reason as in the $B_s$ system: the relevant hadronic matrix
elements are larger.

\begin{figure}[!tb]
\begin{center}
\includegraphics[width=0.45\textwidth] {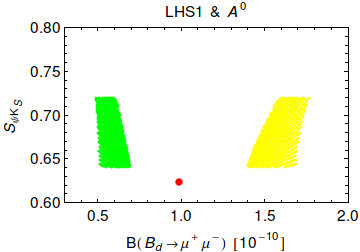}
\includegraphics[width=0.45\textwidth] {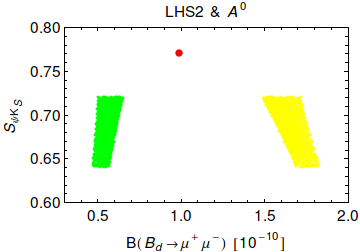}
\caption{\it  $S_{\psi K_S}$ versus   $\mathcal{B}(B_d\to\mu^+\mu^-)$  in $A^0$
scenario for
$M_{H}=1$ TeV in LHS1 (left) and
LHS2 (right) in the yellow and green oases as discussed in the text.   Red point:
SM central
value.}\label{fig:BdmuvsSKSLHSP}~\\[-2mm]\hrule
\end{center}
\end{figure}

Let us first concentrate on S2 scenario and the $A^0$ case.
 Our colour coding is such that
\begin{itemize}
\item
In the general case {\it yellow} and {\it green} allowed regions correspond to
oases with small and
large $\delta_{13}$, respectively.
\item
In principle in  the $U(2)^3$ symmetry case we could again show the reduced
regions with {\it magenta} and {\it cyan} for LHS1 and LHS2, respectively but
this reduction amounts typically to $5-10\%$ at most and it is more
transparent not to show it. This small impact of $U(2)^3$ symmetry in
the $B_d$ system is evident from Table~\ref{tab:PZ}.
\end{itemize}

In the right panel of Fig.~\ref{fig:BdmuvsSKSLHSP} we show  $S_{\psi K_S}$ vs   $\mathcal{B}(B_d\to\mu^+\mu^-)$. This result
should be compared with the one for
$Z'$ case shown
in the right panel of Fig.~7 in \cite{Buras:2012jb}.

 In order to understand the differences between these two scenarios of NP we
again look at the phase of the correction to $P$ which now is given as follows:
\be\label{SHIFTBd}
 r_{Z'}\approx  r_{A^0}, \qquad    \delta_{Z'}=\delta_{13}-\beta, \qquad 
\delta_{A^0}=\delta_{Z'}-\frac{\pi}{2}.
\ee
Note that this time the phase of $V_{td}$ enters the analysis with $\beta\approx 19^\circ$ and $\beta\approx 25^\circ$ for S1 and
S2
scenario of $\vub$, respectively. We find then that in scenario S2 the phase $\delta_{Z'}$ is around
$115^\circ$ and $295^\circ$ for yellow and green oases, respectively.
Correspondingly $\delta_{A^0}$  is around $25^\circ$ and $205^\circ$.
 We summarize the ranges of
 $\delta_{Z'}$ and  $\delta_{A^0}$ in Table~\ref{tab:PZ}.

With
this insight at hand we can easily understand the plots in question
noting that the enhancements and suppressions of $\mathcal{B}(B_d\to\mu^+\mu^-)$ are governed by the cosine of the phase of the
correction:
\begin{itemize}
\item
In the $A^0$ case $\mathcal{B}(B_d\to\mu^+\mu^-)$ is enhanced in the yellow
oasis but suppressed in the green oasis.
\item
In the $Z'$-case the behaviour is opposite:
$\mathcal{B}(B_d\to\mu^+\mu^-)$ is suppressed  in the yellow
oasis but enhanced in the green oasis.
\item
For the choice of parameters NP effects are a bit larger in the $A^0$ case.
\end{itemize}

Note that in both cases the
requirement on $S_{\psi K_S}$ and $\Delta M_d$ forces
$\mathcal{B}(B_d\to\mu^+\mu^-)$ to differ from the SM value. In the $A^0$ case
these enhancements and suppressions amount up to  $\pm 70\%$
for $M_{H}=1\tev$. They increase with decreasing  $S_{\psi K_S}$.

Note that because of the correlation between
$\mathcal{B}(B_d\to\mu^+\mu^-)$
and $S_{\psi K_S}$ and the fact that the latter is already
well determined, the range of $\delta_{13}$ cannot be large and consequently
also  the ranges for the phases $\delta_{Z'}$ and $\delta_{A^0}$ are small allowing
thereby the identification of the right oasis by
 $\mathcal{B}(B_d\to\mu^+\mu^-)$  only.

\begin{figure}[!tb]
\begin{center}
\includegraphics[width=0.45\textwidth] {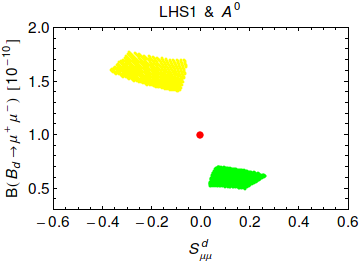}
\includegraphics[width=0.45\textwidth] {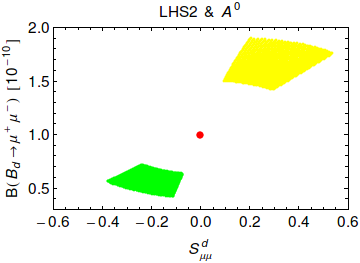}
\caption{\it $\mathcal{B}(B_d\to\mu\bar\mu)$ versus $S_{\mu^+\mu^-}^d$  in $A^0$ case  for $M_{Z^\prime}=1$ TeV in LHS1 (left)
and
LHS2 (right) for the green and yellow oases as discussed in the text.  Red point: SM central
value.}\label{fig:BdmuvsSmudLHSP}~\\[-2mm]\hrule
\end{center}
\end{figure}

While the correlation between $\mathcal{B}(B_d\to\mu^+\mu^-)$
and $S_{\psi K_S}$  offers a distinction between $Z'$ and pseudoscalar scenario
even more interesting in this respect is the correlation between
 $\mathcal{B}(B_d\to\mu^+\mu^-)$ and $S_{\mu^+\mu^-}^d$. We show this correlation
in Fig.~\ref{fig:BdmuvsSmudLHSP} (right panel).
Comparing this result with right panel of Fig.~8 in \cite{Buras:2012jb} we
observe that in LHS2 these two observables are correlated within $A^0$ case
but anticorrelated in the $Z'$ scenario. Also this behaviour follows directly
from the phase structure of NP contributions in these two cases. Measuring
only the signs of shifts in these two observables relative to the SM values
will uniquely tell us which of these two NP scenarios could be at work and
which not.

We next turn to LHS1 scenario for $\vub$.
We observe that the phase $\delta_{13}$
is lower than in  the case of scenario S2 but $\tilde s_{13}$ is basically
the same.  Using (\ref{SHIFTBd}) we can again calculate the phases of NP physics contributions to $P$. We find
that now
 $\delta_{Z'}$ is around
$70^\circ$ and $250^\circ$ for yellow and green oases, respectively.
Correspondingly $\delta_{A^0}$  is around  $-20^\circ$ and $160^\circ$.
 We summarize the ranges of
 $\delta_{Z'}$ and  $\delta_{A^0}$ in Table~\ref{tab:PZ}.

With
this insight at hand we can easily understand the plots
in the left panels in Figs.~\ref{fig:BdmuvsSKSLHSP} and
\ref{fig:BdmuvsSmudLHSP} and analogous plots in the left panels of Figs.~7 and
8 in \cite{Buras:2012jb}. In particular as seen in Fig.~\ref{fig:BdmuvsSmudLHSP}
there is a flip in sign of $S^d_{\mu^+\mu^-}$  when moving from LHS1 to LHS2 while there is no qualitative change in the case of
$\mathcal{B}(B_d\to\mu^+\mu^-)$.
The opposite behaviour is found in the $Z'$ case.

Therefore what distinguishes
LHS1 from LHS2 in both NP scenarios is the sign of the correlation between $S^d_{\mu^+\mu^-}$ and
$\mathcal{B}(B_d\to\mu^+\mu^-)$. In the $A^0$ case a positive $S^d_{\mu^+\mu^-}$ implies
suppression of $\mathcal{B}(B_d\to\mu^+\mu^-)$ in LHS1 but enhancement in
LHS2. Note that this pattern is independent of the sign of $\tilde\Delta_P^{\mu\bar\mu}$
coupling as this coupling enters both observables. On the other hand the
flip of this sign would interchange colours
in Figs.~\ref{fig:BdmuvsSKSLHSP} and
\ref{fig:BdmuvsSmudLHSP}.   As seen in  Figs.~7 and 8 in \cite{Buras:2012jb}
in the $Z'$-case the behaviour is opposite to the one found in the $A^0$ case:
anti-correlation in LHS1 and correlation in LHS2 between
$S^d_{\mu^+\mu^-}$ and
$\mathcal{B}(B_d\to\mu^+\mu^-)$ in the $A^0$ case is changed respectively to correlation
and anti-correlation in the $Z'$-case. This means again that once we
will know whether LHS1 or LHS2 is chosen by nature the measurements of
$S^d_{\mu^+\mu^-}$ and $\mathcal{B}(B_d\to\mu^+\mu^-)$ will tell us which
of the two NP scenarios are favoured. However, we are aware of the fact
that while the measurement of $S^s_{\mu^+\mu^-}$ is extremely difficult,
the measurement of $S^d_{\mu^+\mu^-}$ will require heroic efforts and it may
take decades to realize such a measurement.

\begin{figure}[!tb]
\begin{center}
\includegraphics[width=0.45\textwidth] {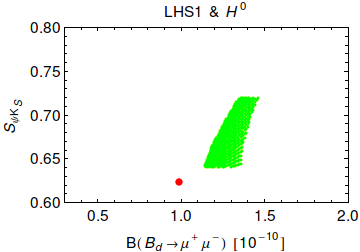}
\includegraphics[width=0.45\textwidth] {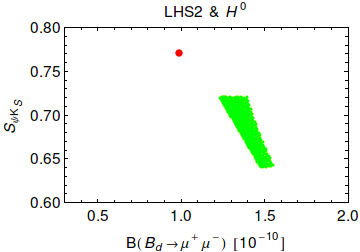}
\caption{\it  $S_{\psi K_S}$ versus   $\mathcal{B}(B_d\to\mu^+\mu^-)$  in $H^0$
scenario for
$M_{H}=1$ TeV in LHS1 (left) and
LHS2 (right) in the yellow and green oases that overlap here. Red point: SM central
value.}\label{fig:BdmuvsSKSLHSS}~\\[-2mm]\hrule
\end{center}
\end{figure}

\begin{figure}[!tb]
\begin{center}
\includegraphics[width=0.45\textwidth] {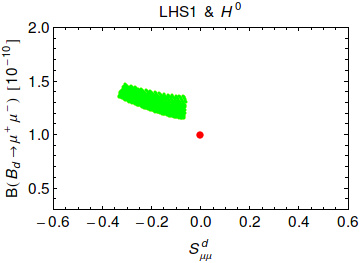}
\includegraphics[width=0.45\textwidth] {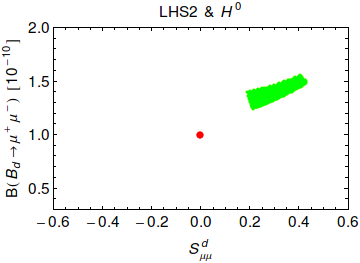}
\caption{\it $\mathcal{B}(B_d\to\mu\bar\mu)$ versus $S_{\mu^+\mu^-}^d$  in $H^0$ case  for $M_{H}=1$ TeV in LHS1 (left) and
LHS2 (right) for the green and yellow oases (they overlap here) as discussed in the text.  Red point: SM central
value.}\label{fig:BdmuvsSmudLHSS}~\\[-2mm]\hrule
\end{center}
\end{figure}

We next move to consider the $H^0$ case and show the results in this case
in Figs.~\ref{fig:BdmuvsSKSLHSS} and
\ref{fig:BdmuvsSmudLHSS} that correspond to
Figs.~\ref{fig:BdmuvsSKSLHSP} and
\ref{fig:BdmuvsSmudLHSP} in the $A^0$ case, respectively. We observe the following  differences between $A^0$ and $H^0$ cases:
\begin{itemize}
\item
As in the case of $B_s\to\mu^+\mu^-$ there is no dependence on the oasis considered and in all plots in
Figs.~\ref{fig:BdmuvsSKSLHSS} and
\ref{fig:BdmuvsSmudLHSS} $\mathcal{B}(B_d\to\mu^+\mu^-)$ is always enhanced as
opposed to the $A^0$ case where it could also be suppressed.
\item
The asymmetry $S^d_{\mu^+\mu^-}$ is negative and positive in LHS1 and LHS2,
respectively, while in the $A^0$ case both signs were possible in LHS1 and
LHS2.
\end{itemize}

In order to understand the signs of $S^d_{\mu^+\mu^-}$ in this case one
should just use Table~\ref{tab:PZ} and the relation (\ref{SZP}) in the $B_d$
system. Effectively these signs in
 Figs.~\ref{fig:BdmuvsSKSLHSS} and \ref{fig:BdmuvsSmudLHSS} can be obtained from the corresponding plots
in Figs.~\ref{fig:BdmuvsSKSLHSP} and \ref{fig:BdmuvsSmudLHSP} by simply
removing the regions with suppression of  $\mathcal{B}(B_d\to\mu^+\mu^-)$.
Therefore the distinction between $S$ and $P$ will only be easy if this
branching ratio will turn out to be suppressed with respect to its SM value.

\boldmath
\subsubsection{The $U(2)^3$ Limit}\label{sec:U23}
\unboldmath
We have investigated how the results presented until now are modified when
the flavour $U(2)^3$ symmetry
\cite{Barbieri:2011ci,Barbieri:2011fc,Barbieri:2012uh,Barbieri:2012bh,Crivellin:2011fb,Crivellin:2011sj,Crivellin:2008mq}
is imposed on the $H$ couplings.
 As pointed out in \cite{Buras:2012sd} in this case $\varphi_{B_d}=\varphi_{B_s}$
which in turn implies
not only  the correlation between CP asymmetries
$S_{\psi K_S}$ and $S_{\psi\phi}$
but also  a  triple $S_{\psi K_S}-S_{\psi\phi}-|V_{ub}|$ correlation.

Usually, when considering
the case of $U(2)^3$
broken by the minimal set of spurions, the $MU(2)^3$ case, only SM operators
are involved. Yet, as mentioned in \cite{Buras:2012sd} in the case of
${\rm 2HDM_{\overline{\rm MFV}}}$~\cite{Buras:2010mh} with flavour blind phases dominantly in the Higgs potential and the
dominance of scalar left-handed currents also the
 $U(2)^3$ structure of scalar contributions to FCNC transitions is obtained.
Thus  only the
LHS1 and LHS2 scenarios are involved in this case.
In what follows we will confine our discussion to $B_s$ and $B_d$
systems as NP effects in the $K$ system are much less interesting in the
NP scenarios considered in this paper. General discussion can be found in
\cite{Buras:2012sd}.

Until now NP effects in the observables in $B_d$ and $B_s$ systems where
uncorrelated but now they are correlated with each other
due to the relations:
\be\label{equ:U23relation}
\frac{\tilde s_{13}}{\vtd}=\frac{\tilde s_{23}}{\vts}, \qquad
\delta_{13}-\delta_{23}=\beta-\beta_s.
\ee
Thus, once the allowed oases in the $B_d$ system are fixed, the oases in $B_s$
system are determined. Moreover, all observables in both systems are described
by only one real positive parameter and one phase, e.g. $({\tilde s}_{23},\delta_{23})$.

We also have in this case \cite{Buras:2012sd}
\be\label{Fleischer}
S^s_{\mu^+\mu^-}=S^d_{\mu^+\mu^-}
\ee
 for which formulae in the $H^0$ and $A^0$ case can be found in 
Subsection~\ref{SPscenarios}.

\begin{figure}[!tb]
\centering
\includegraphics[width = 0.45\textwidth]{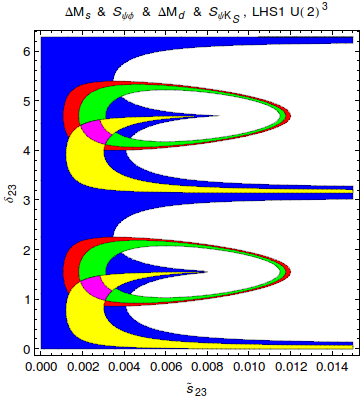}
\includegraphics[width = 0.45\textwidth]{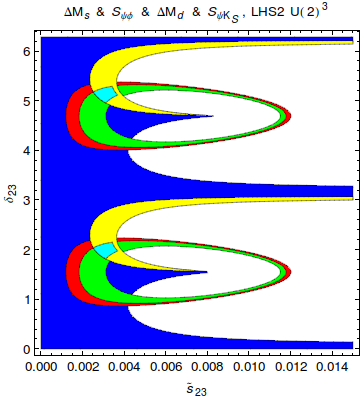}
\caption{\it Ranges for $\Delta M_s$ (red region), $S_{\psi \phi}$ (blue region), $\Delta M_d$ (green region) and $S_{\psi K_S}$
(yellow region)for $M_{H}=1$~TeV in LHS1 (left) and LHS2 (right) in the $U(2)^3$ limit satisfying the bounds
in Eq.~(\ref{C1}) and ~(\ref{C2}). The overlap region of all four regions is shown in magenta in LHS1 and in cyan in LHS2. }
 \label{fig:oasesU2}~\\[-2mm]\hrule
\end{figure}

In Fig.~\ref{fig:oasesU2} we combine Figs.~\ref{fig:oasesBsLHS1} and~\ref{fig:oasesBdLHS} using the $U(2)^3$ symmetry
relations in (\ref{equ:U23relation}). In the $U(2)^3$ limit the allowed oases get smaller. This decrease turns out to be
not very pronounced in the case of $({\tilde s}_{13},\delta_{13})$ oases
as they were already small as seen in Fig.~\ref{fig:oasesBdLHS} but has
a significant impact on $({\tilde s}_{23},\delta_{23})$ oases which where
much larger as seen in  Fig.~\ref{fig:oasesBsLHS1}. Moreover the fact that
the results in the $B_d$ system depend on whether LHS1 or LHS2 is considered
is now transfered through the relations in (\ref{equ:U23relation}) into
the $B_s$ system. This is clearly seen in Fig.~\ref{fig:oasesU2}, in particular
the final oases in {\it cyan} in LHS2 are  visibly smaller than the {\it magenta} oases in LHS1 due to the required
shift of $S_{\psi K_S}$.

Inspecting the ranges for the phases in the last four rows of
Table~\ref{tab:PZ}, which now apply to both $B_s$ and $B_d$ systems and
comparing them with the remaining rows of these tables we can predict
the impact of the imposition of $U(2)^3$ symmetry on the results in both systems presented so far:
\begin{itemize}
\item
In the case of $B_s$ system in which previously there was no distinction
between LHS1 and LHS2 the changes are as follows. The plots in Figs.~\ref{fig:SmusvsSphiLHS1P} and \ref{fig:SmusvsSphiLHS1S} still
apply but the allowed
regions get smaller and some of them are valid only for LHS1 ({\it magenta}) and other for
LHS2 ({\it cyan}).  We also note that the imposition of $U(2)^3$ symmetry 
favours regions away from the SM point.
\item
In the $B_d$ system the plots in Figs.~\ref{fig:BdmuvsSKSLHSP}--\ref{fig:BdmuvsSmudLHSS} have the same structure as previously and
as stated previously the
effect of the imposition of $U(2)^3$ symmetry is so small that we do not
show {\it magenta} and {\it cyan} areas in this case. They would cover almost
completely the yellow and green areas, respectively.
\end{itemize}

These expectations could be already made by comparing the right panel of Fig.~3
and Fig.~25 in \cite{Buras:2012jb} in the case of $Z'$ scenario. Indeed also
there the imposition of $U(2)^3$ symmetry reduces the allowed ranges significantly in the $B_s$ system and
make a clear distinction between LHS1 and LHS2 scenario which was absent
previously. We see again how important the determination of $\vub$ is.
Knowing future precise values
of $\vub$ as well as $S_{\psi\phi}$ and $\mathcal{B}(B_{s,d}\to\mu^+\mu^-)$
will confirm or rule out this scenario of NP.
These correlations are particular
examples of the correlations in $MU(2)^3$ models pointed out in
\cite{Buras:2012sd}. What
is new here is that in a specific model considered by us the $\vub-S_{\psi\phi}$
correlation has now also implications for $\overline{\mathcal{B}}(B_{s}\to\mu^+\mu^-)$
and $S_{\mu\mu}^s$.

\begin{figure}[!tb]
\centering
\includegraphics[width = 0.45\textwidth]{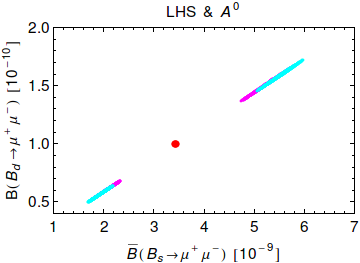}
\includegraphics[width = 0.45\textwidth]{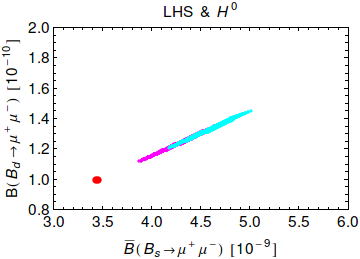}
\caption{\it $\mathcal{B}(B_d\to \mu^+\mu^-)$ versus $\overline{\mathcal{B}}(B_s\to \mu^+\mu^-)$
for $M_{H}=1$~TeV in $A^0$ case (left) and $H^0$ case (right) in the $U(2)^3$ limit satisfying the bounds
in Eq.~(\ref{C1}) and ~(\ref{C2}).}
 \label{fig:BdBs}~\\[-2mm]\hrule
\end{figure}

In Fig.~\ref{fig:BdBs} we show $\mathcal{B}(B_d\to \mu^+\mu^-)$ versus  
$\overline{\mathcal{B}}(B_{s}\to\mu^+\mu^-)$
for the $A^0$ and $H^0$
cases. As expected on the basis of a general 
discussion in \cite{Buras:2012sd} there is a very strong correlation between 
these two branching ratios. Again, while in the $A^0$ case both branching ratios 
can be enhanced or suppressed  with respect to the SM, they can be only 
enhanced in the $H^0$ case.

\subsection{The RHS1 and RHS2 Scenarios}

\subsubsection{First Observations}

We will now investigate $H$ scenario with exclusively RH couplings to quarks.
Now in the RHS1 and RHS2 scenarios only RH couplings to quarks
are present in $H$ contributions. As QCD
is parity conserving, the hadronic matrix elements for operators with
RH currents as well as QCD corrections remain unchanged. The expressions
for $\Delta F=2$ observables in RHS1 and RHS2 scenarios as well as the
corresponding constraints have precisely the same structure as in the
LHS1 and LHS2 cases just discussed. Therefore the oases in the space of
parameters related to RH currents are precisely the same as in LHS1 and LHS2
scenarios,
except that the parameters $\tilde s_{ij}$
and $\delta_{ij}$ parametrize now RH and not LH currents. Anticipating this
result we have not introduced separate description of LH and RH oases.
Yet, in the case of $\Delta F=1$ observables several changes are present which
allow in principle to distinguish the RHS1 and RHS2 scenarios from the
corresponding LHS1 and LHS2 scenarios.

\boldmath
\subsubsection{The $B_s$ Meson System}
\unboldmath

On the left in Figs.~\ref{fig:SmusvsSphiLHS1P} and ~\ref{fig:SmusvsSphiLHS1S} we have shown  $S^s_{\mu^+\mu^-}$  vs $S_{\psi\phi}$
in the LHS1
scenario for $A^0$ and $H^0$ cases, respectively.
Analogous plots for the correlation of $S_{\psi\phi}$ vs $\overline{\mathcal{B}}(B_{s}\to\mu^+\mu^-)$  in
the LHS1 scenario are shown in
Figs.~\ref{fig:SmusvsSphiLHS1P} and \ref{fig:SmusvsSphiLHS1S}. Inspecting
the related formulae for RHS scenario we conclude similarly to the $Z'$
case that in the case of $A^0$  these plots are
also valid for RHS1 scenario except that the colours should be
interchanged. Therefore on the basis of observables considered here
it is not
possible to distinguish between LHS1 and RHS1 scenarios because in the
RHS1 scenario one can simply interchange the two oases
 to obtain the same
physical results as in LHS1 scenario. We also note that reversing simultaneously the sign of $\tilde\Delta_P$ would keep also the
oases unchanged.

The situation is even simpler in the case of $H^0$ case. As the plots 
in question did not depend on oasis considered, the correlations in this
case are identical in RHS1 and LHS1 independently of oasis considered.

Clearly as in the LHS1 scenario this result represents a test of the
RHS1 scenario but if one day we will have precise measurements of
 $S^s_{\mu^+\mu^-}$,  $S_{\psi\phi}$ and $\overline{\mathcal{B}}(B_{s}\to\mu^+\mu^-)$ 
we will still not be able to distinguish for instance whether we deal
with LHS1 scenario in the blue oasis or RHS1 scenario in purple oasis.

In principle, one could make a  distinction between LHS and RHS scenarios by considering model independent bounds
from $B\to K \mu^+\mu^-$ and $B\to K^* \mu^+\mu^-$ on the Wilson coefficients
of the scalar operators. However, as discussed in  Subsection~\ref{bsllc},
this is presently not the case. This should be contrasted with $Z'$ analysis
in \cite{Buras:2012jb} where in fact such a distinction could be made.

 \boldmath
\subsubsection{The $B_d$ Meson System}
\unboldmath

Similarly to the $B_s$ case the structure of oases is as
in Fig.~\ref{fig:oasesBdLHS}. Moreover, the results in
in Figs.~\ref{fig:BdmuvsSKSLHSP} --\ref{fig:BdmuvsSmudLHSS} are valid
for RHS1 and RHS2 scenarios except that in the $A^0$ case the colours should
be interchanged, while there is no modification in the $H^0$ case.
Thus we cannot distinguish
between LHS and RHS scenarios  on the basis of considered observables. Clearly
the study of $b\to d \mu^+\bar\mu^-$ transitions could help in this context
but they are more challenging both theoretically and experimentally.

\subsection{The LRS1 and LRS2 Scenarios}

\subsubsection{First Observations}

If both LH and RH currents are present in NP contributions, the pattern
of flavour violation can differ from the scenarios considered until now
in a profound manner. If the LH and RH couplings differ from each other,
the number of parameters increases and it is harder to get clear cut
conclusions without some underlying fundamental theory. On the other hand
some of the ``symmetries'' between LHS and RHS scenarios identified above
are broken and
the effect of RH currents in certain cases could in principle be better
visible.

Here in order to keep
the same number of parameters as in previous scenarios we will assume
a left-right symmetry in the $H$-couplings to quarks. That is the
LH couplings $\Delta_L$ are equal in magnitudes and phases to the corresponding
RH couplings $\Delta_R$. In this manner we can also keep the same
parametrization of couplings as in previous scenarios.

Before entering the details let us emphasize two new features relative
to the cases in which either LH or RH couplings in NP contributions
were present:
\begin{itemize}
\item
NP contributions to $\Delta F=2$ observables receive now new LR
operators, whose contributions are enhanced through renormalization group
effects relative to SM operators, however as scalar
LL and RR operators are also enhanced by such effects the difference between
LL (RR) scenarios and LR scenario in the scalar case is much smaller than in the
$Z'$ case.
\item
NP contributions to $B_{d,s}\to\mu^+\mu^-$ and $K_L\to\mu^+\mu^-$ vanish
eliminating in this manner $S^{s,d}_{\mu^+\mu^-}$ and $\mathcal{B}(B_{s,d}\to\mu^+\mu^-)$ as basic observables in the
identification of acceptable oases.
On the other hand $B\to K^*\mu^+\mu^-$ and $B\to K\mu^+\mu^-$ receive
still NP contributions and can help in this context.
\end{itemize}

While  $S^{s,d}_{\mu^+\mu^-}$ cannot help in the identification of the
optimal
oasis in the LR scenarios they are non-vanishing:

\be\label{SmumuLR}
S_{\mu^+\mu^-}^q=-\sin(2\varphi_{B_q}).
\ee
While rather small they offer a clean test of the LR scenarios.

\boldmath
\subsubsection{The $B_s$ Meson System}
\begin{figure}[!tb]
\begin{center}
\includegraphics[width=0.45\textwidth] {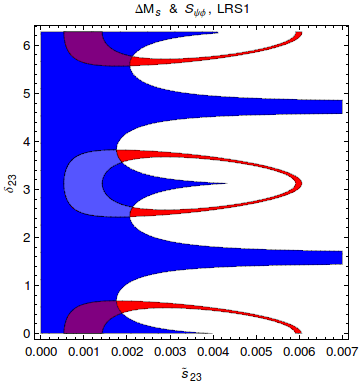}
\caption{\it  Ranges for $\Delta M_s$ (red region) and $S_{\psi \phi}$ (blue region) for $M_{H}=1$~TeV in LRS1 satisfying the
bounds
in
Eq.~(\ref{C1}).
}\label{fig:oasesBsLRS1}~\\[-2mm]\hrule
\end{center}
\end{figure}

\unboldmath
We begin the search for the oases with the $B_s$ system proceeding
with input parameters as in the previous scenarios.
The result of this search for $M_{H}=1\tev$ is shown in Fig.~\ref{fig:oasesBsLRS1},
where we show the allowed ranges
for $(\tilde s_{23},\delta_{23})$.
The {\it red} regions correspond to the allowed ranges for $\Delta M_{s}$,
while the {\it blue} ones to the corresponding ranges for  $S_{\psi\phi}$. The overlap between red and blue regions identifies the
oases we were looking for.

The notations are as in previous cases but it
should be kept in mind that the parameters  $(\tilde s_{23},\delta_{23})$
describe both LH and RH couplings.

In order to understand the structure of oases in Fig.~\ref{fig:oasesBsLRS1}, that differs from the ones
found so far, we note that the matrix element of the dominant $Q_2^{\rm LR}$
operator has the sign opposite to the dominant $Q_1^{\rm SLL}$ operator.
Therefore, this operator
naturally suppresses $\Delta M_s$ with the phase $\delta_{23}$ centered in the
ballpark of $0^\circ$ and $180^\circ$, that is shifted down by roughly
$90^\circ$ relatively to the LHS scenarios. As the matrix element of
$Q_2^{\rm LR}$ is larger than that of $Q_1^{\rm SLL}$  operator in LHS and RHS scenarios, $\tilde s_{23}$ has to be sufficiently
smaller to agree with data.

The crucial role in the $B_s$ meson system in this scenario,
in the absence of NP contributions to $B_{s,d}\to\mu^+\mu^-$ decays, is
now played by  $B\to K^*\mu^+\mu^-$ and  $B\to K\mu^+\mu^-$.
We
will discuss the latter decays at the end of this section.

 \boldmath
\subsubsection{The $B_d$ Meson System}
\unboldmath

\begin{figure}[!tb]
\begin{center}
\includegraphics[width=0.45\textwidth] {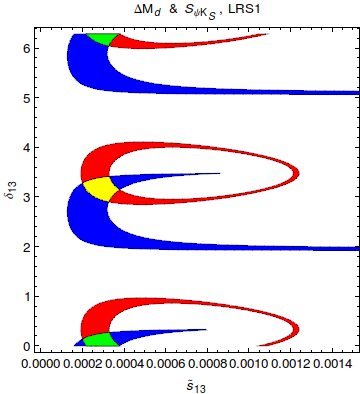}
\includegraphics[width=0.45\textwidth] {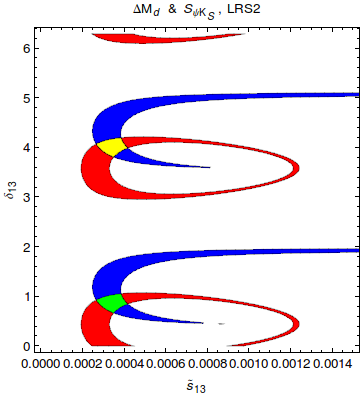}

\caption{\it  Ranges for $\Delta M_d$ (red region) and $S_{\psi K_S}$ (blue region) for $M_{H}=1$~TeV in LRS1 (left) and
LRS2 (right) satisfying the bounds
in
Eq.~(\ref{C2}).
}\label{fig:oasesBdLRS1}~\\[-2mm]\hrule
\end{center}
\end{figure}

The structure of oases in this case is given
in Fig.~\ref{fig:oasesBdLRS1}. As we do not have
$\mathcal{B}(B_d\to\mu^+\mu^-)$ to our disposal and $b\to d\ell^+\ell^-$
decays are challenging this system is not very useful to provide
tests of LRS scenarios without some fundamental theory.

Due to the sign
of the matrix element of the dominant $Q_2^{\rm LR}$ operator in both
LRS1 and LRS2 the mass difference $\Delta M_d$ is naturally suppressed.
The requested size of this suppression together with significant
suppression of $S_{\psi K_S}$ in LRS2 and slight enhancement of it in LRS1
governs the structure of the phases.

\subsection{The ALRS1 and ALRS2 Scenarios}
We include this case as well because it has not been discussed in
the literature but it is an interesting NP scenario for the following
reasons:
\begin{itemize}
\item
NP contributions to $\Delta F=2$ observables are dominated as in
LRS scenarios by new LR
operators but as the sign of LR interference is flipped some differences
arise.
\item
NP contributions to $B_{d,s}\to\mu^+\mu^-$ enter again with full power.
Therefore these decays together with
$S^q_{\mu^+\mu^-}$ offer as in the LHS and RHS scenarios some help
in the identification of acceptable oases and to study differences
between scalars, pseudoscalars and $Z'$ bosons.
\item
The phase structure of the oases is as in LHS scenario but due to enhanced
hadronic matrix elements of LR operators the mixing parameters $\tilde s_{ij}$
are decreased.
\item
NP contributions to $K_L\to \pi^0\ell^+\ell^-$
vanish in this scenario.
\end{itemize}

In view of this simple structure of modifications with respect to LHS
scenario, all plots have the same structure as LH scenarios but NP effects are
smaller. Therefore we will  not show these plots.

\subsection{Implications of $b\to s \ell^+\ell^-$ Constraints}\label{bsllc}
Presently the NP effects found by us are consistent with the experimental
data on $B_{s,d}\to\mu^+\mu^-$. However,
also the data on $B\to X_s \ell^+\ell^-$, $B\to   K^*\ell^+\ell^-$  and
 $B\to   K\ell^+\ell^-$ recently improved a lot and it is of interest to see whether this has 
an impact on our results. It should be emphasized that $B\to X_s \ell^+\ell^-$, $B\to   K^*\ell^+\ell^-$
 are not as theoretically clean as $B_s\to\mu^+\mu^-$ because of the
presence of form factors. However in the case of  $B\to   K\ell^+\ell^-$
 progress in  lattice calculations of the relevant form factors is expected
soon and as stressed in particular in \cite{Becirevic:2012fy} a simultaneous
consideration of this decay together with $B_s\to\mu^+\mu^-$  provides
useful tests of extensions of the SM. Indeed, while $B_s\to\mu^+\mu^-$
is sensitive only to the differences  $C_P-C_P'$ and  $C_S-C_S'$, the
decay $B\to   K\ell^+\ell^-$ is sensitive to their sums  $C_P+C_P'$ and  $C_S+C_S'$. A very extensive model independent
analysis of $C_P(C_P')$ and  $C_S(C_S')$ in the context of the data on  $B_s\to\mu^+\mu^-$ and $B\to  K\ell^+\ell^-$ has been
performed in
\cite{Becirevic:2012fy}. Adjusting their normalization of Wilson coefficients to ours the final result of this paper reads:
\be
m_b|C_S^{(\prime)}|\le 0.7,\qquad   m_b|C_P^{(\prime)}|\le 1.0,
\ee
which implies
\begin{equation}
 \left|\frac{1}{g_{\text{SM}}^2\sin^2\theta_W}\frac{1}{ M_H^2}\frac{\Delta_{L,R}^{sb}(H)\Delta_S^{\mu\bar\mu}(H)}{V_{ts}^*
V_{tb}}\right|\le 0.70
\end{equation}
\begin{equation}
 \left|\frac{1}{g_{\text{SM}}^2\sin^2\theta_W}\frac{1}{ M_H^2}\frac{\Delta_{L,R}^{sb}(H)\Delta_P^{\mu\bar\mu}(H)}{V_{ts}^*
V_{tb}}\right|\le 1.0
\end{equation}
or equivalently for $M_H=1\tev$
\be
|\tilde s_{23}\Delta_S^{\mu\bar\mu}(H)|\le 0.00115,
\ee
\be
|\tilde s_{23}\Delta_P^{\mu\bar\mu}(H)|\le 0.00164.
\ee

The largest values of $\tilde s_{23}$ used in our analysis are $0.0041$ in LHS
and RHS scenarios with smaller values for LR and ALR scenarios. We find then
$0.0001$ and $0.00005$ for the two products respectively. 
 Therefore, these bounds do not have any impact on our results.

\boldmath
\section{The $K$ Meson System}\label{sec:U(2)}
\unboldmath

As we already stated previously we do not expect any visible
effects in $\kpn$ and
$\klpn$ and our discussion will concentrate on $\varepsilon_K$, $K_L\to\mu^+\mu^-$ and $K_L\to\pi^0\ell^+\ell^-$.

\begin{figure}[!tb]
\begin{center}
\includegraphics[width=0.45\textwidth] {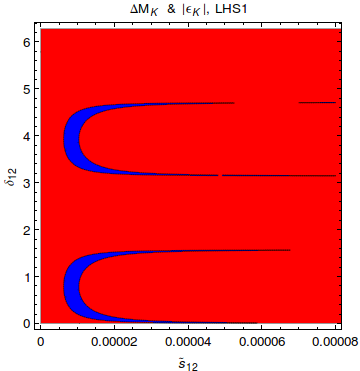}
\includegraphics[width=0.45\textwidth] {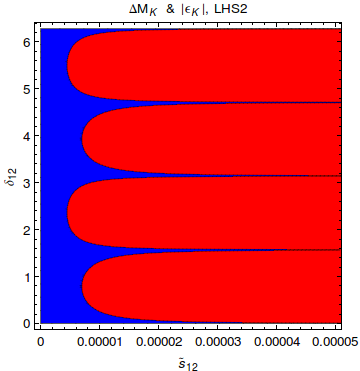}
\caption{\it  Ranges for $\Delta M_K$ (red region) and $\varepsilon_K$ (blue region) (LHS1: left, LHS2: right) for
$M_{H}=1$ TeV  satisfying the bounds in Eq.~(\ref{C3}).
}\label{fig:oasesKLHS}~\\[-2mm]\hrule
\end{center}
\end{figure}

As seen in (\ref{C3}) the constraints from $\Delta F=2$ observables are
weaker than in previous cases. Yet as seen in Fig.~\ref{fig:oasesKLHS}, obtained within LHS1 and LHS2 scenarios,
it is possible to identify the allowed oases. These plots have the same phase
structure
as the plot in Fig.~9 of \cite{Buras:2012jb} for $Z'$ scenario
except that $\tilde s_{12}$ is by a factor of five smaller because of
the enhanced matrix element of the relevant scalar operator.

Due to weaker constraints in the $K$ system the oases are rather large.
We have two oases in S1:
\be\label{KoasesS1}
C_1(S1):~~0^\circ \le \delta_{12} \le 90^\circ,\qquad C_2(S1):~~180^\circ \le \delta_{12} \le 270^\circ
\ee
 and only one oasis in S2:
\be\label{KoasesS2}
C_1(S2):~~0^\circ \le \delta_{12} \le 360^\circ.
\ee

\begin{figure}[!tb]
\begin{center}
\includegraphics[width=0.45\textwidth] {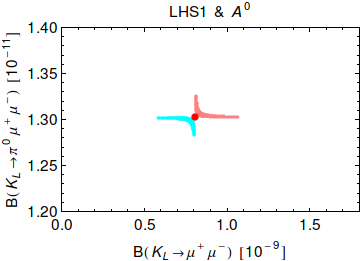}
\includegraphics[width=0.45\textwidth] {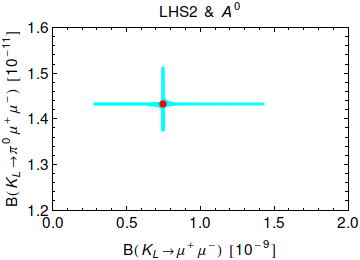}
\caption{\it$\mathcal{B}(K_L \to \pi^0 \mu^+\mu^-)$ versus $\mathcal{B}(K_L\to\mu^+\mu^-)$ for P
scenario and LHS1 (left), LHS2 (right). Red  point: SM central value.
}\label{fig:eevsmumuPLHS}~\\[-2mm]\hrule
\end{center}
\end{figure}

With these constraints at hand we have calculated the branching ratios
for $K_L\to\mu^+\mu^-$ and $K_L\to\pi^0\ell^+\ell^-$  decays. We can summarize
our results as follows:
\begin{itemize}
\item
 NP effects in $K_L\to\pi^0e^+e^-$ are totally negligible both for $H^0$ and $A^0$ cases.
\item
 NP effects in $K_L\to\pi^0\mu^+\mu^-$ are larger but amount to at most
 $\pm 5\%$ at the level of the branching ratio which is also negligibly
 small in view of large theoretical uncertainties.
\item
 The short distance branching ratio for $K_L\to\mu^+\mu⁻$ can a be modified
 up to $\pm 50\%$ in the $A^0$ case. Still such effects are fully consistent with the upper bound on this  branching ratio.
NP effects in the $H^0$ case are much
 smaller.
\end{itemize}

These results are rather disappointing but allow to distinguish scenarios
discussed here from $Z'$ scenario, where effects have been found to be larger.
As an example we show in Fig.~\ref{fig:eevsmumuPLHS}
the correlation
between $\mathcal{B}(K_L\to\pi^0\mu^+\mu^-)$ and
$\mathcal{B}(K_L\to\mu^+\mu^-)$ in LHS1 and LHS2.

The effects in other scenarios are rather uninteresting as well and
we will not present any results for rare $K$ decays in them.

\begin{figure}[!tb]
\begin{center}
\includegraphics[width=0.45\textwidth] {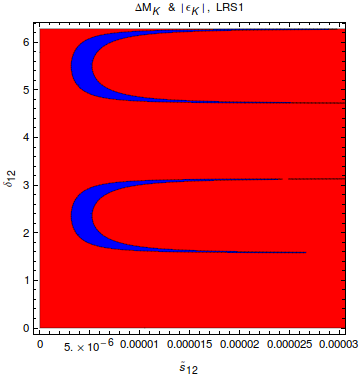}
\includegraphics[width=0.45\textwidth] {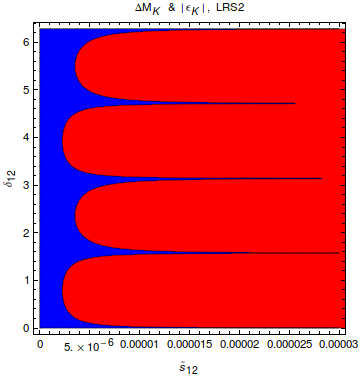}
\caption{\it  Ranges for $\Delta M_K$ (red region) and $\varepsilon_K$ (LRS1: left, LRS2: right) for
$M_{H}=1$ TeV  satisfying the bounds in Eq.~(\ref{C3}).
}\label{fig:oasesKLRS}~\\[-2mm]\hrule
\end{center}
\end{figure}

However, it is of interest to see how the oases change in the presence of LR
operators. This we show in
Fig.~\ref{fig:oasesKLRS}. Due to the presence of
LR operators the structure of oases is different than in LHS1 and LHS2 scenarios. While
the shape of the single oasis in the LRS2 case is similar to the LHS2, for
LRS1 the oases are shifted by $90^\circ$:
\be\label{LRKoasesS1}
C_1(S1):~~90^\circ \le \delta_{12} \le 180^\circ,\qquad C_2(S1):~~270^\circ \le \delta_{12} \le 360^\circ.
\ee

NP effects in $K_L\to \mu^+\mu^-$ vanish in these scenarios and  in 
$K_L\to\pi^0\ell^+\ell^-$ they are negligible. Therefore we do not show any plots.

Concerning  NP contributions to $K\to\pi\nu\bar\nu$ decays all scalar
scenarios could turn out one day to be interesting if
the data on observables in $B_s$ and $B_d$ systems will show the presence
of NP but negligible NP effects in $K\to\pi\nu\bar\nu$.

\section{Flavour Violating SM $H$ Boson}\label{sec:ZSM}

We will next  turn our attention to flavour violating couplings of the SM
Higgs ($h$) that
can be generated in the presence of other scalar particles and or
new heavy vectorial fermions with $+2/3$ and $-1/3$ electric charges.
 In the case considered by us, new quarks with  $-1/3$ charges are essential for generating
flavour violating couplings to SM down-quarks but the presence of  heavy
quarks with $+2/3$ charges could be relevant for charm physics. Moreover,
such heavy fermions could contribute to rare $K$ and $B$ decays through
loop diagrams. In what follows we will not consider these loop contributions
as they would lead us beyond the scope of our paper.

\begin{figure}[!tb]
\centering
 \includegraphics[width= 0.45\textwidth]{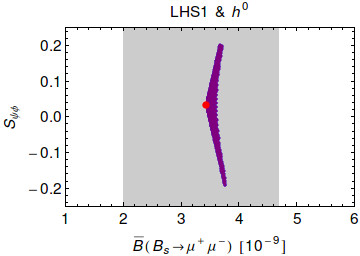}
 \includegraphics[width= 0.45\textwidth]{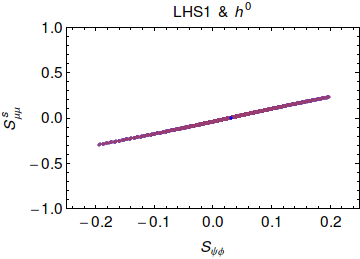}
\includegraphics[width = 0.45\textwidth]{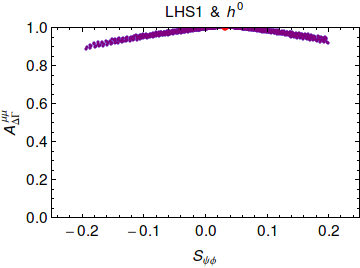}
\caption{\it  $S_{\psi\phi}$ versus $\overline{\mathcal{B}}(B_s\to\mu^+\mu^-)$, $S^s_{\mu^+\mu^-}$  versus $S_{\psi\phi}$  and 
$\mathcal{A}^\lambda_{\Delta\Gamma}$ versus $S_{\psi\phi}$ for scalar $h^0$ case with
$M_{h} = 125~$GeV in LHS1. The two oases (blue and purple) overlap.   Red point: SM central
value.}\label{fig:lightHiggsBs}
~\\[-2mm]\hrule
\end{figure}

The strategy and formalism developed in the previous sections can be used
in a straightforward manner  for the case of $h$ flavour-violating couplings to
quarks. Using the general relations of Section~\ref{sec:3a} one then finds that these couplings have to be significantly smaller,
by roughly an order of magnitude relatively to the corresponding couplings
of a heavy scalar with a mass in the ballpark of 1~TeV. Still the presence
of such contributions can remove all possible tensions within the SM in
$\Delta F=2$ transitions without being in conflict with constraints from
rare decays, where the SM Higgs contributions, in spite of a low Higgs mass,
 turn out to be small.

The reason for the smallness of SM Higgs contributions to rare decays
originates from the smallness of scalar Higgs coupling to $\mu^+\mu^-$ with
$\Delta^{\mu\mu}_S(h)=1.2\times 10^{-3}$. It is roughly a factor of 25 smaller than
the largest scalar coupling allowed by $B_s\to\mu^+\mu^-$ for a scalar with
$M_H=1\tev$. As the correlations between $\Delta F=2$ and $\Delta F=1$
transitions in Section~\ref{sec:3a} show, this smallness of the muon couplings
of SM Higgs can be compensated partly by the smallness of its mass. {  But it turns out that this
compensation is insufficient to make the SM Higgs contributions to
rare $B_{d}$ and $K$ decays relevant.} We recall that these contributions do not
interfere with the SM contribution from $Z$-penguins and box diagrams and 
 are suppressed by the square of $\Delta_S^{\mu\mu}(h)$.

On the other hand small but visible effects in $B_s\to\mu^+\mu^-$ decay 
are still possible. We illustrate this  in Figs.~\ref{fig:lightHiggsBs} which have been obtained 
using the technology developed for heavy scalars. We note that 
$\overline{\mathcal{B}}(B_{s}\to\mu^+\mu^-)$ can be enhanced up to $8\%$ and 
$|S^s_{\mu^+\mu^-}|$ can be as large as $0.3$ but only for the maximal allowed 
values of $S_{\psi\phi}$.

\section{Summary and Conclusions}\label{sec:5}

In this paper we exhibited the pattern of flavour violation in models in which
NP effects are dominated by tree-level heavy pseudoscalar (A) or scalar (H)
 exchanges under the assumption that the theoretical and experimental errors on
various input parameters will
decrease with time. In particular we have identified a number of correlations
between $\Delta F=2$ and $\Delta F=1$ processes that will enable in due time
to test this NP scenario.
Our detailed
analysis of these correlations in Section~\ref{sec:Excursion} shows that  in the $B_s$ and $B_d$ systems a very rich pattern of NP effects is present while
this is not the case in $K$ decays. This is partly opposite to $Z'$ and $Z$
scenarios considered in \cite{Buras:2012jb} where the largest effects
have been found in the $K$ system, even for masses of $M_{Z'}$ outside
the LHC reach.

Our results are summarized in a number of plots that have been obtained
in various scenarios for the $H$ couplings and for inclusive and
exclusive values of $\vub$. We list here only
few highlights:
\begin{itemize}
\item
For each scenario we have identified allowed oases in the parameter space
of the model. In each oasis particular structure of correlations between
various observables will in the future either favour  or exclude a given oasis.
\item
For the near future the correlations involving $S_{\psi K_S}$, $S_{\psi\phi}$ and
$\mathcal{B}(B_{d}\to\mu^+\mu^-)$  and $\overline{\mathcal{B}}(B_{s}\to\mu^+\mu^-)$  will be the most interesting as the
data on these four observables will be improved in the coming years,
sharpening the outcome of our analysis and possibly ruling out some oases
and scenarios of the couplings.
\item
Most importantly we have found that various correlations involving
$S_{\psi\phi}$,  $\overline{\mathcal{B}}(B_{s}\to\mu^+\mu^-)$ and the CP asymmetry
$ S^s_{\mu^+\mu^-}$ show profound differences between the scenario with
a pseudoscalar tree-level exchange and $Z'$ exchange. As we explained in
detail, these differences
are directly related to the difference in the fundamental properties of the particles involved: their spin and CP-parity. As far as the last property is
concerned also differences between the implications of the pseudoscalar and scalar exchanges
have been identified. In particular the scalar contributions can only enhance
$\overline{\mathcal{B}}(B_{s}\to\mu^+\mu^-)$  and are invariant under the interchange
of two oases in parameter space involved, which is not the case of pseudoscalar
exchanges where the branching ratio can also be suppressed.
The symphony of plots in Subsection~\ref{LHS12}, where the most important
results of our paper are shown
will be helpful in monitoring further  developments in the measurements
of the observables in question.
\item
Analogous comments apply to the correlations involving
$S_{\psi K_S}$,  $\mathcal{B}(B_{d}\to\mu^+\mu^-)$ and the CP asymmetry
$ S^d_{\mu^+\mu^-}$ for which the symphony of plots, also presented in
Subsection~\ref{LHS12},
 is doubled in view
of the dependence on the value of $\vub$.
\item
We have demonstrated that the imposition of $U(2)^3$ symmetry on pseudoscalar
and scalar quark
couplings has a profound impact on the correlations in the $B_s$ system
with much smaller effects in the $B_d$ system.
\item
We have also pointed out additional differences between $Z'$ and pseudoscalar
or scalar tree-level contributions related
to channels with
neutrinos in the final state, where in the $Z'$ case these contributions
could be very large but are expected to be negligible in NP scenarios
considered here.
\item
{ Our short analysis of flavour-violating SM Higgs-couplings shows that in the case of rare $B_{d}$ and $K$ decays, the SM Higgs 
contributions are irrelevant
due to the smallness of  the Higgs coupling to muons after corresponding 
constraints from $\Delta F=2$ transitions have been taken into account.
However, small but visible effects in $B_s\to\mu^+\mu^-$ are still allowed.}
 On the other hand such
contributions could in principle remove all tensions within $\Delta F=2$
observables observed within the SM.
\end{itemize}

\begin{figure}[!tb]
\begin{center}
\includegraphics[width=0.46\textwidth]{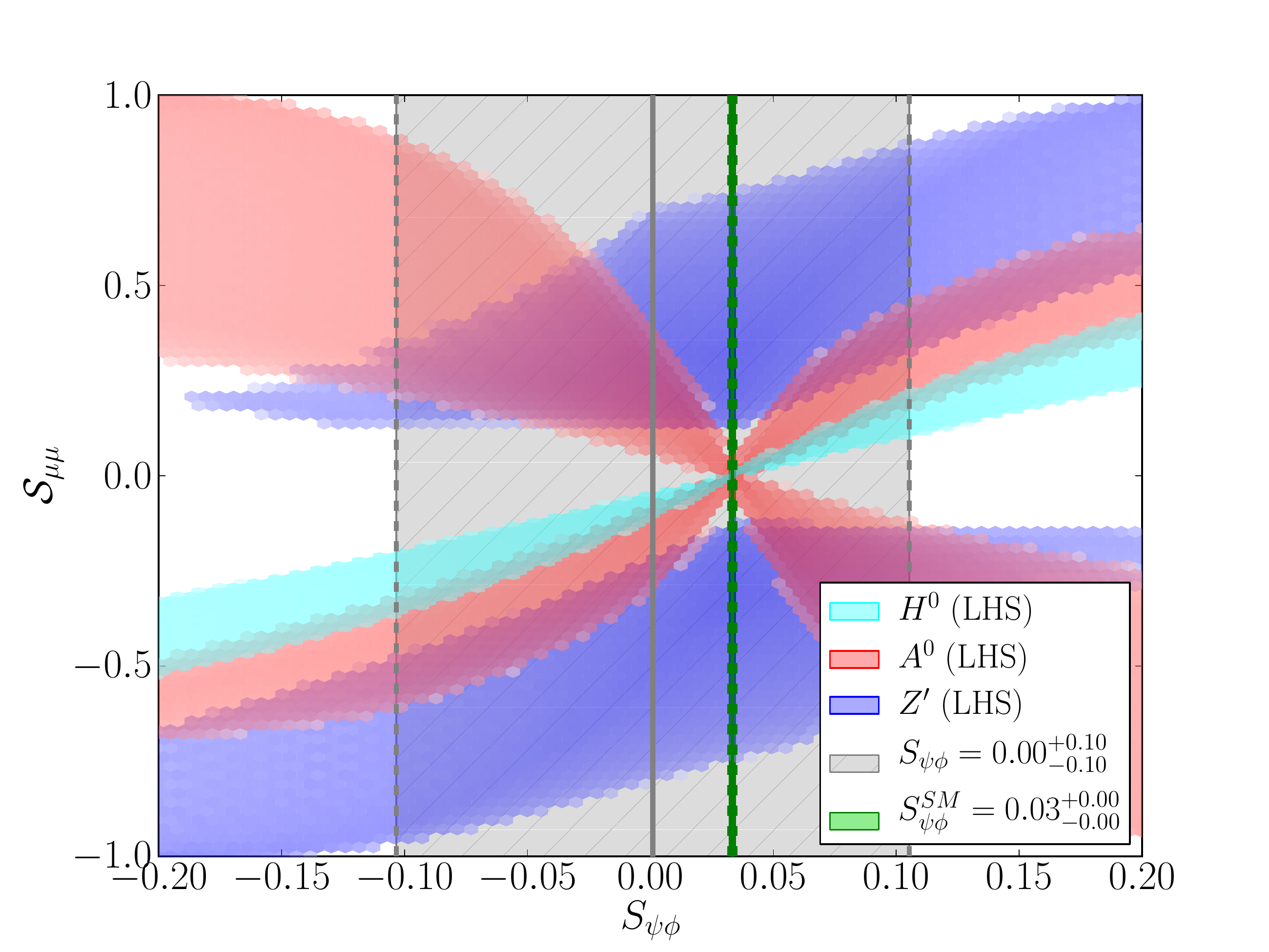}
\includegraphics[width=0.46\textwidth]{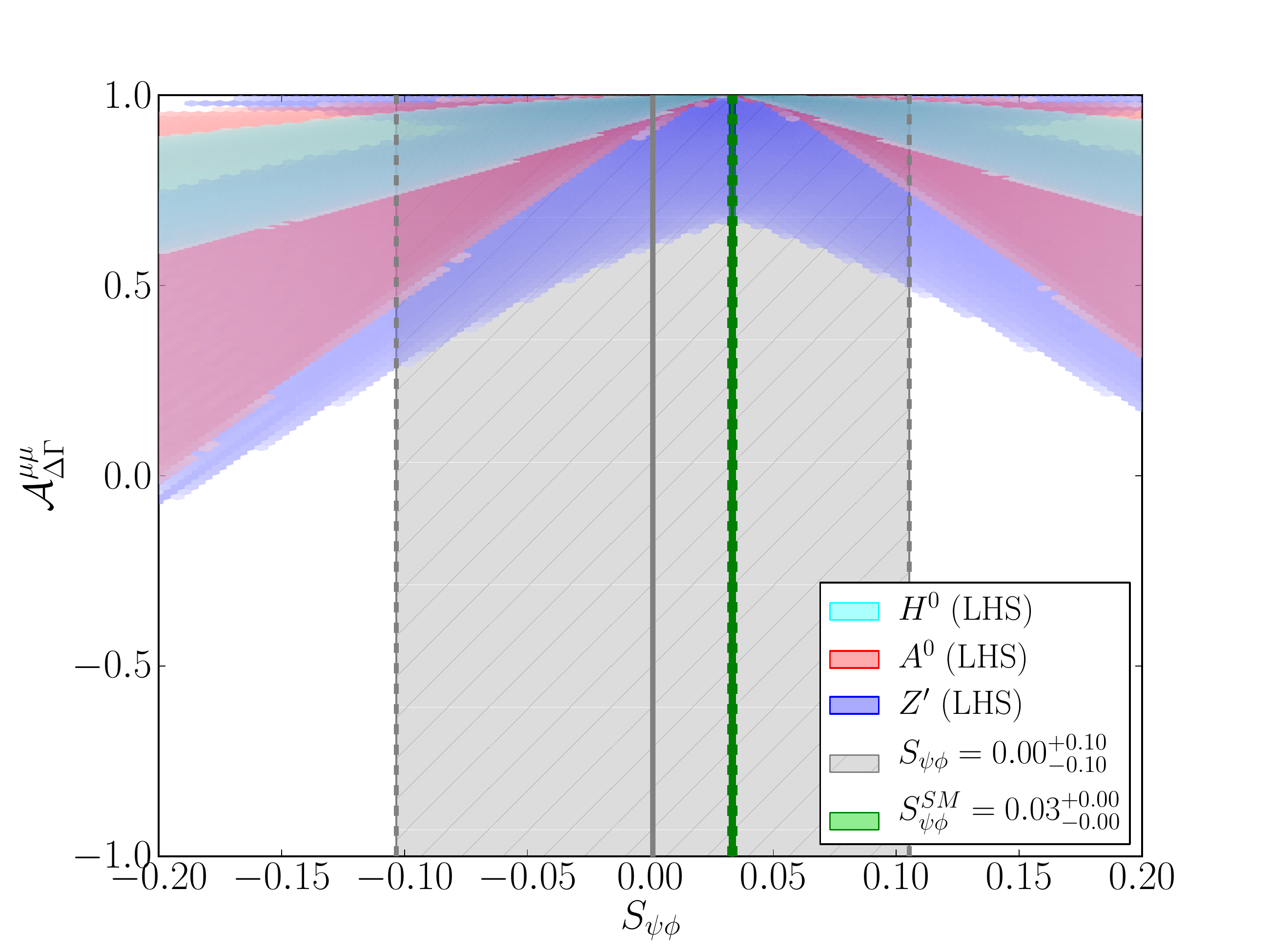}
\includegraphics[width=0.55\textwidth]{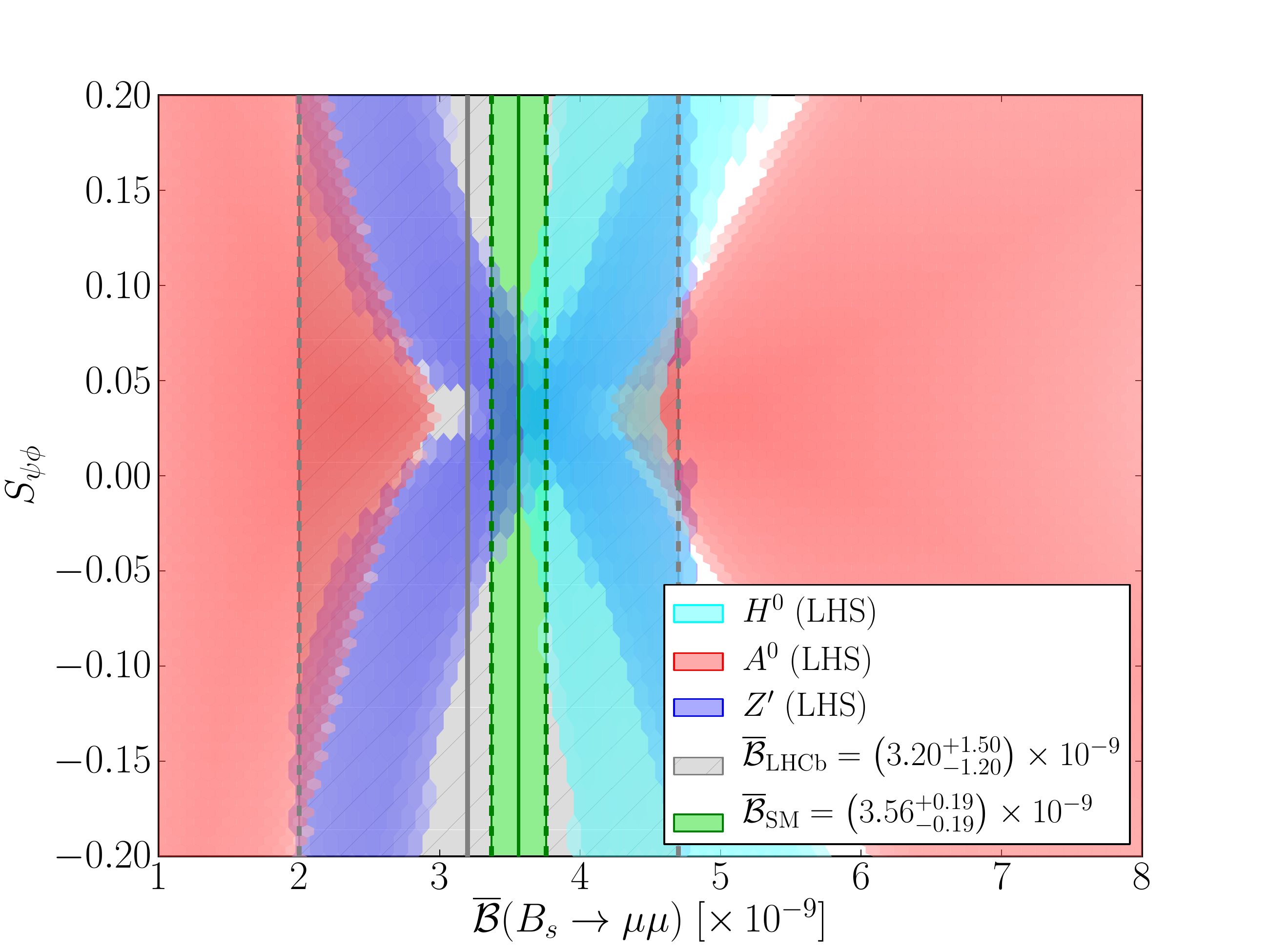}
\caption{\it  Overlay of the correlations for $S_{\mu\mu}^s$ versus $S_{\psi\phi}$ (top left), $A_{\Delta\Gamma}^{\mu\mu}$ versus
$S_{\psi\phi}$ (top right) and $S_{\psi\phi}$ versus $\overline{\mathcal{B}}(B_s\to\mu^+\mu^-)$ (bottom) for tree
level scalar (cyan), pseudoscalar (red) and $Z^\prime$ (blue) exchange (both oases in same colour respectively) in LHS. The lepton
couplings are varied in the ranges $|\Delta_{S,P}^{\mu\mu}(H)| \in [0.012,0.024]$ and $\Delta_A^{\mu\mu}(Z')\in [0.3,0.7]$.
}\label{fig:grandplot}~\\[-2mm]\hrule
\end{center}
\end{figure}

We close our paper by  Fig.~\ref{fig:grandplot} in which 
we  show the correlations involving  $S_{\mu\mu}^s$,  $S_{\psi\phi}$ 
and $\overline{\mathcal{B}}(B_{s}\to\mu^+\mu^-)$
combining information of Fig.~\ref{fig:PandSscan} for the tree-level scalar and pseudoscalar exchange and include
also 
tree-level $Z^\prime$ exchange. The lepton couplings are not fixed but varied in the following ranges: $|\Delta_{S,P}^{\mu\mu}(H)| \in
[0.012,0.024]$ and $\Delta_A^{\mu\mu}(Z')\in [0.3,0.7]$. Further we do not distinguish between the two different oases here. In 
the
$Z^\prime$ case we also take into account the bounds from $b\to s\ell^+\ell^-$ transitions from \cite{Altmannshofer:2012ir}. The patterns
already identified 
previously and summarized above are clearly visible in these plots.

 We are aware of the fact that some of the correlations presented
by us would be washed out if we included all existing uncertainties. Yet, our
simplified numerical analysis had as the main goal to illustrate how the
decrease of theoretical, parametric and experimental uncertainties in the
coming years might allow to exhibit certain features of NP, even if
deviations from the SM will be only moderate.
In this manner we have uncovered a world of correlations present
in NP scenarios, where new effects are dominated by flavour-violating couplings
of a heavy neutral pseudoscalar and scalar.
In fact, within the
coming years the size of the assumed uncertainties in our analysis
 could likely become reality not only because of improved experimental
data but also improvements in theory, in particular lattice calculations of hadronic matrix
elements, $B_i$ parameters, form factors and weak decay constants.

We are looking forward to improved experimental data  and improved lattice
calculations. The correlations identified in this paper will allow to
monitor how  simple NP scenarios discussed by us
face the future precision flavour data.

{\bf Acknowledgements}\\
We thank Robert Fleischer and Robert Ziegler for discussions.
This research was financially supported by the ERC Advanced Grant project ``FLAVOUR'' (267104) and the Foundation for Fundamental Research
on Matter (FOM). It was also  partially supported by the DFG cluster
of excellence ``Origin and Structure of the Universe''.

\bibliographystyle{JHEP}
\bibliography{allrefs}
\end{document}